\pdfoutput=1
\documentclass[prd,showpacs,letterpaper,10pt,aps,notitlepage]{revtex4}

\usepackage{amsfonts,amsmath,graphicx,bm}
\usepackage{multirow}

\newcommand{\nb}{\phantom{0}}
\newcommand{\bs}[1]{\ensuremath{{\boldsymbol{#1}}}}

\begin{document}

\title{Bottomonium spectrum at order $\bs{v^6}$ from domain-wall lattice QCD: \\ precise results for hyperfine splittings}

\author{Stefan Meinel}
\affiliation{Department of Physics, College of William \&  Mary, Williamsburg,
  VA 23187-8795, USA}

\date{October 26, 2010}


\begin{abstract}
The bottomonium spectrum is computed in dynamical 2+1 flavor lattice QCD, using NRQCD for the $b$ quarks.
The main calculations in this work are based on gauge field ensembles generated by the RBC and UKQCD collaborations
with the Iwasaki action for the gluons and a domain-wall action for the sea quarks. Lattice spacing values of approximately $0.08$ fm
and $0.11$ fm are used, and simultaneous chiral extrapolations to the physical pion mass are performed.
As a test for gluon discretization errors, the calculations are repeated on two ensembles generated by the MILC collaboration
with the L\"uscher-Weisz gauge action. Gluon discretization errors are also studied in a lattice potential model
using perturbation theory for four different gauge actions.
The nonperturbative lattice QCD results for the radial and orbital bottomonium energy splittings
obtained from the RBC/UKQCD ensembles are found to be in excellent agreement with experiment.
To get accurate results for spin splittings, the spin-dependent order-$v^6$ terms are included in the NRQCD action,
and suitable ratios are calculated such that most of the unknown radiative corrections cancel. The cancellation of radiative corrections
is verified explicitly by repeating the calculations with different values of the couplings in the NRQCD action.
Using the lattice ratios of the $S$-wave hyperfine and the $1P$ tensor splitting,
and the experimental result for the $1P$ tensor splitting, the $1S$ hyperfine splitting
is found to be $60.3\pm5.5_{\rm\:stat } \pm5.0_{\rm\: syst } \pm2.1_{\rm\: exp }$ MeV, and the $2S$
hyperfine splitting is predicted to be $23.5\pm4.1_{\rm\: stat } \pm2.1_{\rm\: syst } \pm0.8_{\rm\: exp }$ MeV.
\end{abstract}

\maketitle

\section{Introduction}
The low-lying radial and orbital energy splittings in bottomonium are well known from experiment and well understood theoretically.
Spin-dependent energy splittings pose a significantly greater challenge. On the experimental side, the observation
of the $S=0$ states is very difficult. So far, only the $\eta_b(1S)$ has been found. The weighted average of the results
from \cite{Aubert:2008vj,Aubert:2009pz,Bonvicini:2009hs} gives a value of $69.3 \pm 2.9$ MeV for the $\Upsilon(1S)-\eta_b(1S)$ hyperfine splitting.

Calculations of this hyperfine splitting using perturbative QCD gave significantly lower values around $40$ MeV
\cite{Recksiegel:2003fm, Kniehl:2003ap, Penin:2009wf}. Mixing of the $\eta_b$ with a light $CP$-odd Higgs boson has been suggested as a possible
explanation of this discrepancy \cite{Domingo:2009tb}. To definitely answer the question whether QCD alone is
able to correctly predict the $\Upsilon(1S)-\eta_b(1S)$ mass difference, precise non-perturbative calculations from lattice QCD are required.

Presently, lattice calculations with dynamical light quarks are performed at lattice spacings that are too coarse
to resolve the Compton wave length of the $b$ quark, and therefore special heavy-quark techniques are required.
Two such techniques are the Fermilab method \cite{ElKhadra:1996mp} and
lattice nonrelativistic QCD (NRQCD) \cite{Thacker:1990bm, Lepage:1992tx}. The bottomonium spectrum
has also been calculated using relativistic actions, on anisotropic lattices \cite{Liao:2001yh} and on
very fine, small lattices \cite{Chiu:2007km}, but so far without dynamical light quarks.

With the Fermilab method, the heavy quark is implemented by an improved Wilson-like action, where the parameters are tuned such
that heavy-quark discretization errors are reduced. The Fermilab method has the advantage over NRQCD
that continuum extrapolations can be performed safely. In the simplest case, only the mass parameter in the action
is adjusted such that the kinetic mass of a heavy meson agrees with experiment.
This method, in combination with MILC gauge field configurations generated with the L\"{u}scher-Weisz gluon action
and 2+1 flavors of rooted staggered sea quarks \cite{Bazavov:2009bb}, has been employed in \cite{Burch:2009az}
to calculate the bottomonium and charmonium spectra. In the continuum limit, the $\Upsilon(1S)-\eta_b(1S)$ splitting was found to be $54 \pm 12$ MeV.

A version of the Fermilab action with three tuned parameters was used in Ref.~\cite{Li:2008kb} to calculate bottomonium
masses at one lattice spacing using gauge field ensembles generated by the RBC and UKQCD collaborations using the Iwasaki gluon action and 2+1 flavors
of domain-wall sea quarks \cite{Allton:2008pn}.

In contrast to the Fermilab method, lattice NRQCD is based on the direct discretization of an
effective field theory for heavy quarks, in which an expansion in powers of the heavy-quark velocity $v$
is performed \cite{Thacker:1990bm, Lepage:1992tx}. For bottomonium, one has
$v^2\approx0.1$ \cite{Eichten:1979ms}. With NRQCD, it is required that the UV cut-off provided by the lattice is lower
than the heavy-quark mass: one must have $a m_b \gtrsim 1$, where $a$ is the lattice spacing. Discretization errors can be removed
through Symanzik improvement.
The bottomonium spectrum was calculated using improved lattice NRQCD of order $v^4$ on the MILC gauge field ensembles in \cite{Gray:2005ur}.
Good agreement with experiment was seen for the radial and orbital energy splittings. However, the results for the spin-dependent energy splittings
suffered from large uncertainties due to missing radiative and higher-order relativistic corrections in the NRQCD action used there.
Spin-dependent splittings are an effect of order $v^4$, and therefore $v^6$ corrections are significant. These $v^6$ corrections
were included in earlier calculations \cite{Trottier:1996ce, Manke:1997gt, Eicker:1998vx, Manke:2000dg}
and found to reduce the bottomonium spin splittings by 10-30\%.
Radiative corrections to the spin-dependent couplings in the NRQCD action are expected to be of order $\alpha_s$ ($\approx$20-30\%),
and are still unknown. However, as will be demonstrated in this work, these radiative corrections largely cancel in
suitable ratios of spin splittings.

In the following, a new calculation of the bottomonium spectrum in lattice QCD with 2+1 flavors of dynamical sea quarks is presented.
For the $b$ quarks, an improved NRQCD action including the spin-dependent order-$v^6$ terms is used (for comparison, results obtained without these terms are also shown).
By varying the couplings of the leading-order spin-dependent terms in the action, the cancellation of radiative corrections is demonstrated nonperturbatively for the
ratio of the $S$-wave hyperfine and $1P$ tensor splittings, as well as the ratio of the $2S$ and $1S$ hyperfine splittings. For these quantities,
results with unprecedented precision are obtained here.

The main calculations in this paper are done on RBC/UKQCD gauge field ensembles, which were generated with the Iwasaki gluon action and
a domain-wall sea quark action. The calculations are performed at lattice spacing values of approximately $0.08$ fm (with spatial lattice size $L=32$) and
$0.11$ fm (for both $L=24$ and $L=16$). This work is an extension of the first calculation by the author
that was using only the $L=24$ ensembles and only the $v^4$ action \cite{Meinel:2009rd}.
By including the finer $L=32$ ensembles, discretization errors can now be studied directly,
and by including the $L=16$ ensembles with their smaller box size, the size of finite-volume effects can be estimated.
The data at $L=24$ are also reanalyzed with improved methods leading to smaller statistical errors.
With the better accuracy, chiral extrapolations to the physical pion mass are now possible.

Systematic errors caused by the lattice NRQCD action are estimated using power counting. In order
to study discretization errors caused by the lattice gluon action, two approaches are used here. First,
the radial and orbital energy splittings are calculated in a lattice potential model, using the static
quark-antiquark potential derived from the gluon action in lattice perturbation theory. A comprehensive study
of the scaling behavior is presented for four different gluon actions (Plaquette, L\"uscher-Weisz, Iwasaki, and DBW2).
This model is however limited to the tree level, and to radial and orbital energy splittings only.
In order to go beyond tree-level and include spin splittings, the nonperturbative lattice QCD calculations
of the bottomonium spectrum are repeated on gauge field ensembles generated with the L\"uscher-Weisz gluon action by the MILC
collaboration, and a detailed comparison to the results from the RBC/UKQCD ensembles (with the Iwasaki action)
is made.

This paper is organized as follows: the lattice methods and parameters are described in Sec.~\ref{sec:methods}.
The ``speed of light'' is studied in Sec.~\ref{sec:speed_of_light}. In
Sec.~\ref{sec:radial_orbital_lattice_spacing}, the results for the
radial and orbital energy splittings are presented, followed by the spin-dependent
energy splittings in Sec.~\ref{sec:spindep}. The conclusions
are given in Sec.~\ref{sec:conclusions}.
A simple analysis of autocorrelations is described in Appendix \ref{sec:autocorr},
the tuning of the $b$-quark mass is discussed in Appendix \ref{sec:mb_tuning}, and tables with results in lattice units
can be found in Appendix \ref{sec:results_lattice_units}. The lattice potential model calculations
of gluon discretization errors and the comparison of nonperturbative results from the MILC and RBC/UKQCD ensembles
are presented in Appendix \ref{sec:gluon_errors}.

\section{\label{sec:methods}Methods}

\subsection{Lattice actions and parameters}

The calculations in this work are based on gauge field ensembles that include the effects of
dynamical up- down- and strange sea quarks (with $m_u=m_d$, in the following denoted as $m_l$).
The ensembles used for the main part of the calculations
were generated by the RBC and UKQCD collaborations \cite{Allton:2007hx,Allton:2008pn}.
The sea quarks are implemented with a domain wall action \cite{Kaplan:1992bt,Shamir:1993zy,Furman:1994ky},
which yields an exact chiral symmetry when the extent $L_5$ of the auxiliary fifth dimension is taken to infinity.
The gluons are implemented with the Iwasaki action \cite{Iwasaki:1983ck,Iwasaki:1984cj, Iwasaki:1996sn}, which
suppresses the residual chiral symmetry breaking at finite $L_5$ \cite{Aoki:2002vt}. The form of the Iwasaki action
can be found in Appendix \ref{sec:gluon_errors}, where the discretization errors associated with this action
are analyzed.

The parameters of the RBC/UKQCD gauge field configurations used here are given in Table \ref{tab:lattices}. All ensembles
have $L_5=16$. There are ensembles with two different values of the bare gauge coupling, here given as $\beta=6/g^2$.
The two values $\beta=2.13$ and $\beta=2.25$ correspond to lattice spacings of $a\approx 0.11$ fm and $a\approx0.08$ fm,
respectively (see Sec.~\ref{sec:radial_orbital_results}). The box sizes at the coarser lattice spacing are about $1.8$ fm ($L=16$) and $2.7$ fm ($L=24$);
for the finer lattice spacing the box size is about $2.7$ fm ($L=32$).

\begin{table*}[h!]
\begin{ruledtabular}
\begin{tabular}{cccclccccccccccclcc}
$L^3 \times T$ &  & $\beta$ &  & $a m_l$ &  & $a m_s$  &  & $a m_b$ &  & $u_{0L}$ &  & MD range, step  &  & $a$ (fm) &  & $m_\pi$ (GeV) &  & $a m_b^{\rm (phys.)}$ \\
\hline
$16^3 \times 32$ && $2.13$  && $0.01$  && $0.04$  && $2.536$                 && $0.8439$ && $500$ - $4010$, $10$  && $0.1117(33)$ && $0.436(14)$  && $2.469(72)$ \\
$16^3 \times 32$ && $2.13$  && $0.02$  && $0.04$  && $2.536$                 && $0.8433$ && $500$ - $4040$, $10$  && $0.1170(32)$ && $0.548(16)$  && $2.604(75)$ \\ 
$16^3 \times 32$ && $2.13$  && $0.03$  && $0.04$  && $2.536$                 && $0.8428$ && $500$ - $7600$, $10$  && $0.1195(24)$ && $0.639(14)$  && $2.689(56)$ \\
\\[-1ex]
$24^3 \times 64$ && $2.13$  && $0.005$ && $0.04$  && $2.3$, $2.536$, $2.7$   && $0.8439$ && $915$ - $8665$,  $25$ && $0.1119(17)$ && $0.3377(54)$ && $2.487(39)$ \\ 
$24^3 \times 64$ && $2.13$  && $0.01$  && $0.04$  && $2.3$, $2.536$, $2.7$   && $0.8439$ && $1475$ - $8525$, $25$ && $0.1139(19)$ && $0.4194(70)$ && $2.522(42)$ \\ 
$24^3 \times 64$ && $2.13$  && $0.02$  && $0.04$  && $2.3$, $2.536$, $2.7$   && $0.8433$ && $1800$ - $3600$, $25$ && $0.1177(29)$ && $0.541(14)$  && $2.622(70)$ \\ 
$24^3 \times 64$ && $2.13$  && $0.03$  && $0.04$  && $2.3$, $2.536$, $2.7$   && $0.8428$ && $1275$ - $3050$, $25$ && $0.1196(29)$ && $0.641(15)$  && $2.691(66)$ \\ 
\\[-1ex]
$32^3 \times 64$ && $2.25$  && $0.004$ && $0.03$  && $1.75$, $1.87$, $2.05$  && $0.8609$ && $580$ - $6840$, $20$  && $0.0849(12)$ && $0.2950(40)$ && $1.831(25)$ \\ 
$32^3 \times 64$ && $2.25$  && $0.006$ && $0.03$  && $1.75$, $1.87$, $2.05$  && $0.8608$ && $552$ - $7632$, $24$  && $0.0848(17)$ && $0.3529(69)$ && $1.829(36)$ \\  
$32^3 \times 64$ && $2.25$  && $0.008$ && $0.03$  && $1.75$, $1.87$, $2.05$  && $0.8608$ && $540$ - $5920$, $20$  && $0.0864(12)$ && $0.3950(55)$ && $1.864(27)$ \\ 
\end{tabular}
\end{ruledtabular}
\caption{\label{tab:lattices}Summary of lattice parameters for the RBC/UKQCD ensembles.
The values for the lattice spacing are results of this work and are determined from the $\Upsilon(2S)-\Upsilon(1S)$
energy splitting (see Sec.~\ref{sec:radial_orbital_results}). The bare gauge coupling is given as $\beta=6/g^2$. The pion masses in lattice units were taken from
\cite{Allton:2007hx, Allton:2008pn, Syritsyn:2009np} and converted to physical units using the lattice spacings given here. The last column gives the value of the
bare $b$ quark mass that would yield agreement of the $\eta_b(1S)$ kinetic mass with experiment (see Appendix \ref{sec:mb_tuning}).}
\end{table*}

The bottom quark is implemented with lattice NRQCD \cite{Thacker:1990bm, Lepage:1992tx}. The Euclidean action has the form
\begin{equation}
S_{\psi}=a^3\sum_{\bs{x},t}\psi^\dagger(\bs{x},t)\big[{\psi}(\bs{x},t)
-K(t) \: {\psi}(\bs{x},t-a) \big], \label{eq:latact}
\end{equation}
where $\psi$ is the two-component bottom quark field, and $K(t)$ is given by
\begin{equation}
K(t)=\left(1-\frac{a\:\delta H|_t}{2}\right)
\left(1-\frac{a H_0|_t}{2n} \right)^n U_0^\dag(t-a)\left(1-\frac{a H_0|_{t-a}}{2n} \right)^n
\left(1-\frac{a\:\delta H|_{t-a}}{2}\right).\label{eq:mNRQCD_action_kernel}
\end{equation}
Here, $H_0$ is the order-$v^2$ term,
\begin{equation}
H_0 = -\frac{\Delta^{(2)}}{2 m_b}, \label{eq:H0}
\end{equation}
and $\delta H$ contains higher-order corrections,
\begin{eqnarray}
\nonumber\delta H&=&-c_1\:\frac{\left(\Delta^{(2)}\right)^2}{8 m_b^3}+c_2\:\frac{ig}{8 m_b^2}\:\Big(\bs{\nabla}\cdot\bs{\widetilde{E}}
-\bs{\widetilde{E}}\cdot\bs{\nabla}\Big)\\
\nonumber&&
\label{eq:adder1}\\
\nonumber&&-c_3\:\frac{g}{8 m_b^2}\:\bs{\sigma}\cdot
\left(\bs{\widetilde{\nabla}}\times\bs{\widetilde{E}}
-\bs{\widetilde{E}}\times\bs{\widetilde{\nabla}} \right)-c_4\:\frac{g}{2 m_b}\:\bs{\sigma}\cdot\bs{\widetilde{B}}\\
\nonumber && + c_5\:\frac{a^2\Delta^{(4)}}{24m_b}
-c_6\:\frac{a\left(\Delta^{(2)}\right)^2}{16n\:m_b^2}\\
&&-c_7\:\frac{g}{8 m_b^3}\Big\{ \Delta^{(2)}, \: \bs{\sigma}\cdot\bs{\widetilde{B}} \Big \}
-c_8\:\frac{3g}{64 m_b^4}\left\{ \Delta^{(2)}, \: 
\bs{\sigma}\cdot \left(\bs{\widetilde{\nabla}}\times\bs{\widetilde{E}}-\bs{\widetilde{E}}\times\bs{\widetilde{\nabla}} \right) \right\}
-c_9\:\frac{i g^2}{8 m_b^3}\:\bs{\sigma}\cdot(\bs{\widetilde{E}}\times\bs{\widetilde{E}}).
\label{eq:dH_full}
\end{eqnarray}
Note that antiquark propagators can be obtained from quark propagators calculated with the action (\ref{eq:latact}) through Hermitian conjugation.

Above, $\Delta^{(2)}=\sum_{j=1}^3 \nabla^{(2)}_j$ and $\Delta^{(4)}=\sum_{j=1}^3 \nabla^{(4)}_j$, where $\nabla^{(p)}_\mu$
denotes the $p$-th order symmetric and maximally local covariant lattice derivative in $\mu$-direction. All derivatives are understood to act
on all quantities to their right. The chromo-electric and chromo-magnetic
fields are defined as $E_j \equiv F_{j0}$, $B_j \equiv -\frac12 \epsilon_{jkl}F_{kl}$.
The terms with coefficients $c_1$ to $c_4$ in (\ref{eq:dH_full}) are the order-$v^4$ corrections,
while the terms with coefficients $c_7$ to $c_9$ are the spin-dependent
order-$v^6$ corrections (note that spin-\emph{independent} order-$v^6$ terms are not included).
The terms with coefficients $c_5$ and $c_6$ are spatial and temporal discretization corrections for $H_0$.
Quantities with a tilde also include discretization corrections:
\begin{equation}
\widetilde{\nabla}_\mu = \nabla_\mu - \frac{a^2}{6}\:\nabla^{(3)}_\mu , \label{eq:improved_D}
\end{equation}
\begin{equation}
 \widetilde{F}_{\mu\nu} = F_{\mu\nu} - \frac{a^2}{6} \left[ \nabla^{(2,{\rm ad})}_\mu + \nabla^{(2,{\rm ad})}_\nu \right] F_{\mu\nu}. \label{eq:improved_F}
\end{equation}
In (\ref{eq:improved_F}), $\nabla^{(2,{\rm ad})}_\mu$ is a second-order adjoint derivative (which acts only on $F_{\mu\nu}$) and $F_{\mu\nu}$
is the standard cloverleaf lattice gluon field strength.

At tree-level,
the coefficients $c_i$ are equal to 1. The action is tadpole-improved \cite{Lepage:1992xa} using the mean link in Landau gauge,
$u_{0L}$. Using the mean link instead of the plaquette for tadpole improvement leads to a better scaling behavior \cite{Shakespeare:1998dt}.

In this work, calculations were performed either with $c_7=c_8=c_9=0$ or with $c_7=c_8=c_9=1$. These two actions
will be referred to as the $v^4$ action and the $v^6$ action, respectively.  The $v^4$ action is identical to the action used in \cite{Gray:2005ur}.
The stability parameter $n$ in (\ref{eq:mNRQCD_action_kernel}) was always set to $n=2$. Calculations were performed for multiple values of
the bare $b$-quark mass $a m_b$, as shown in Table \ref{tab:lattices}. The spin-dependent energy splittings, which show significant $a m_b$-dependence
(see Appendix \ref{sec:spindep_splittings_mb_dep}) were then interpolated to $a m_b^{\rm (phys.)}$, where $a m_b^{\rm (phys.)}$
is the value of the bare $b$-quark mass that would yield agreement of the $\eta_b(1S)$ kinetic mass with experiment (see Appendix \ref{sec:mb_tuning}).

\subsection{Calculation and fitting of two-point functions}

The interpolating fields for the bottomonium two-point functions used here are the same as in \cite{Meinel:2009rd}, except that the cut-off radius
was chosen differently (equal to $L/2$). For $a\approx0.08$ fm the smearing parameters in lattice units were rescaled from those used
at $a\approx0.11$ fm so that the smearing functions in physical units remain the same.

The methods used here for fitting the two-point functions are also the same as in \cite{Meinel:2009rd}, i.e.~multi-exponential Bayesian matrix-correlator fitting
combined with statistical bootstrap (suitably modified for Bayesian fitting) \cite{Lepage:2001ym}. As in \cite{Meinel:2009rd}, bottomonium two-point functions were calculated for
32 different source locations on each gauge field configuration to increase statistics. Note that in \cite{Meinel:2009rd} the data were averaged over
those source locations prior to the analysis. However, the reduced sample size can lead to overestimates of errors due to poorly determined data correlation matrices.
As shown in Appendix \ref{sec:autocorr}, for the $L=24$ and $L=32$ ensembles, the bottomonium data from the 32 source locations are in fact sufficiently independent.
Therefore, in the present work the data correlation matrices are calculated with the unblocked data sets for
the $L=24$ and $L=32$ ensembles. For the $L=16$ ensembles, some autocorrelations between the data from different source locations were seen, and therefore binning over
source locations was performed.

\subsection{Chiral extrapolations}

The dependence of the bottomonium energy splittings on the light-quark masses is expected to be weak. Therefore, chiral extrapolations to the physical pion mass are
performed linearly in $m_\pi^2$.
Before chiral extrapolation, the energy splittings are converted to physical units using the
lattice spacing determinations on the individual ensembles.
In addition, the spin-splittings are interpolated to the physical $b$-quark mass on each individual ensemble
(see Appendices \ref{sec:spindep_splittings_mb_dep} and \ref{sec:mb_tuning}).

For a given energy splitting $E(m_\pi,\:a)$ that depends on the pion mass $m_\pi$ and the lattice spacing $a$,
the chiral extrapolation of the data from the $L=32$ ensembles with lattice spacing $a_1\approx0.08$ fm and the $L=24$ ensembles with lattice spacing $a_2\approx0.11$ fm
is performed simultaneously, using the functional form
\begin{eqnarray}
\nonumber E(m_\pi,\: a_1) &=& E(0, a_1)+A \:m_\pi^2,\\
E(m_\pi,\: a_2) &=& E(0, a_2)+A \:m_\pi^2. \label{eq:sim_chiral_extrap}
\end{eqnarray}
The three unconstrained fit parameters are $E(0, a_1)$, $E(0, a_2)$ and $A$. This allows for an arbitrary dependence of the energy splitting $E(m_\pi,\:a)$ on the lattice spacing $a$.
Higher-order terms depending on both $a$ and $m_\pi$ are neglected. As will be shown later, the $a$-dependence in most quantities is very weak, and therefore these higher-order
effects are small. No extrapolation in $a$ is performed here, since one can not take the continuum limit with NRQCD. Instead, discretization errors can be estimated from the difference
of the two results $E(0, a_1)$ and $E(0, a_2)$. The simultaneous chiral extrapolation was found to significantly improve the statistical accuracy of $E(0, a_1)$ from the
$L=32$ ensembles, due to the wider range in $m_\pi^2$ on the $L=24$ ensembles.

The data from the $L=16$ ensembles, which have a smaller volume, are extrapolated independently, to allow for
an arbitrary volume-dependence of the coefficients $A$ (the $L=24$ and $L=32$ ensembles have approximately
the same spatial volume in physical units).

When performing the extrapolations, the statistical uncertainties in the physical pion mass values on the individual ensembles (due to the scale uncertainty) are taken into account,
by making the pion masses themselves also parameters of the fit, constrained with Gaussian priors. The central values and widths of these priors were set equal to the values
and uncertainties of the pion masses given in Table \ref{tab:lattices}.

\section{Results}

\subsection{\label{sec:speed_of_light}Speed of light}

In the continuum, the energy of a particle with mass $M$ and three-momentum $\bs{p}$ satisfies the relationship
$E^2=M^2+\bs{p}^2$ (in the units used here). Equivalently, one has
\begin{equation}
\frac{E^2-M^2}{\bs{p}^2}=1. \label{sec:sol1}
\end{equation}
To study deviations from the relativistic continuum energy-momentum relationship on the lattice,
in the following the square of the ``speed of light'' (\ref{sec:sol1}) will be calculated for
the $\eta_b(1S)$ meson. Note that due to the use of NRQCD, energies extracted from fits
to bottomonium two-point functions are shifted by approximately $-2 m_b$. This shift
does not affect energy differences. To obtain the full mass of a
bottomonium state, one can calculate the \emph{kinetic mass}, defined as
\begin{equation}
M_{\rm kin} \equiv \frac{\bs{p}^2-\left[E(\bs{p})-E(0)\right]^2}{2\left[E(\bs{p})-E(0)\right]}. \label{eq:mkin}
\end{equation}
Equation (\ref{eq:mkin}) will be equal to the physical mass if the relativistic energy-momentum relationship
is satisfied up to a constant shift; with lattice NRQCD this is not exact and
$M_{\rm kin}$ will depend slightly on the momentum $\bs{p}$. On a lattice with $L$ points in the spatial
directions and periodic boundary conditions, the momentum can have the values
$\bs{p}=\bs{n}\cdot 2\pi/(aL)$ where $\bs{n}=(n_1,n_2,n_3)$ with $n_i\in\mathbb{Z}$ and $-L/2 < n_i \leq L/2$.
We therefore define the square of the speed of light as
\begin{equation}
 c^2 \equiv \frac{\left[E(\bs{p})-E(0)+M_{\rm kin,1}\right]^2-M_{\rm kin,1}^2}{\bs{p}^2}, \label{eq:speed_of_light}
\end{equation}
where $M_{\rm kin,1}$ denotes the kinetic mass calculated with $\bs{n}^2=1$.

Deviations of $c^2$ from 1 can be caused by missing relativistic and radiative corrections
in the NRQCD action (mainly in the coefficients $c_1, c_5$ and $c_6$) and by gluon discretization errors.

\begin{figure}[h!]
 \includegraphics[width=0.49\linewidth]{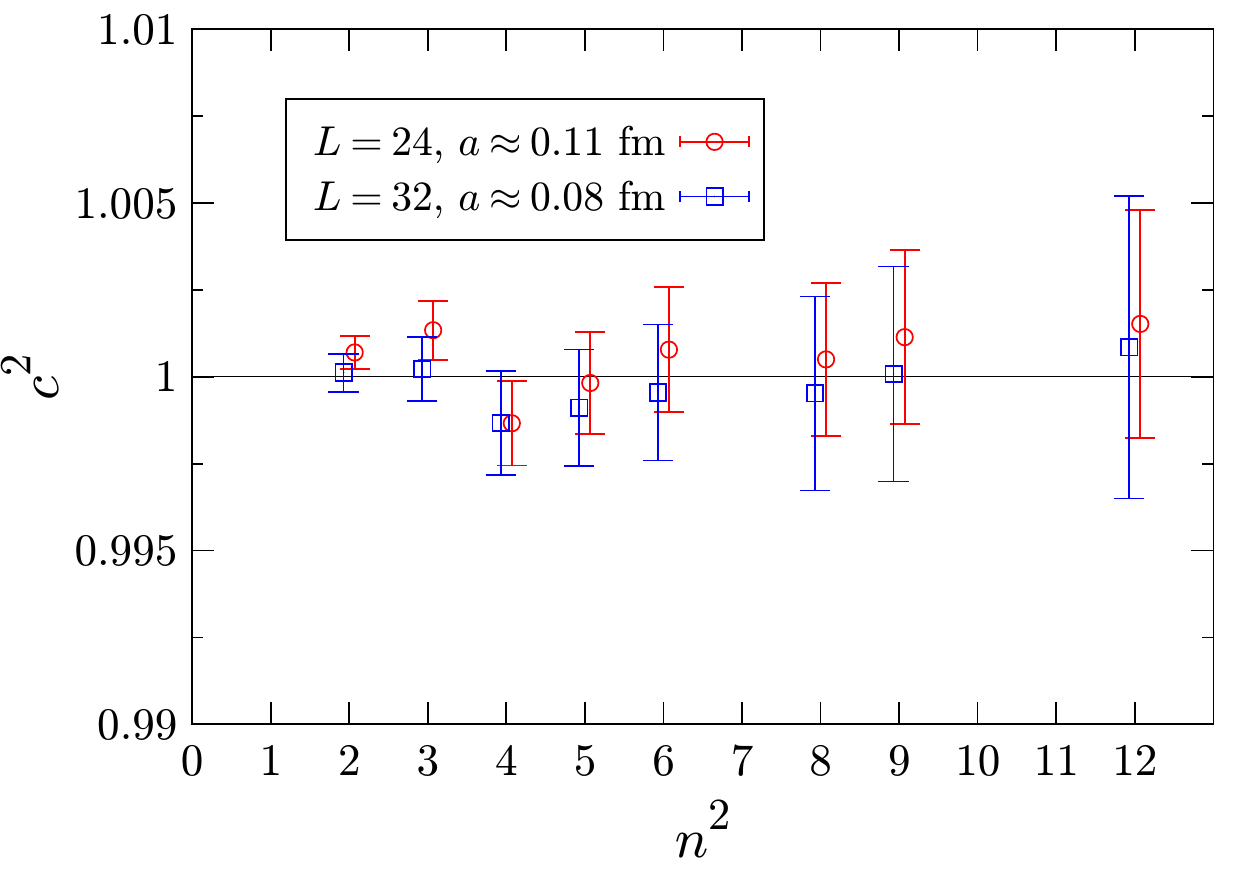}
\caption{\label{fig:speed_of_light}Square of the speed of light, calculated with the $v^4$ action
(points from the two different lattice spacings are offset horizontally for legibility). The
points at $\bs{n}^2=12$ correspond to a meson momentum of about 1.6 GeV.
Note that the very small deviation of $c^2$ from 1 at $a\approx0.11$ fm, seen for $\bs{n}^2=2$ and $\bs{n}^2=3$,
disappears at $a\approx0.08$ fm.}
\end{figure}

The numerical results for $c^2$,
calculated using the $v^4$ action from the $\eta_b(1S)$ energies, are given in Table \ref{tab:speed_of_light}
and plotted in Fig.~\ref{fig:speed_of_light}.
The results at $a\approx0.11$ fm given here have smaller statistical errors than the previous ones in \cite{Meinel:2009rd}.
At $a\approx0.11$ fm, a very small deviation of $c^2$ from 1, at the level of about 0.1\% (1.5 $\sigma$),
can now be resolved for $\bs{n}^2=2$ and $\bs{n}^2=3$. At $a\approx0.08$ fm, this deviation goes away,
indicating that discretization errors are indeed smaller at the finer lattice spacing.

The statistical error in $c^2$ grows with $\bs{n}^2$. Notice however that even at $\bs{n}^2=12$, which
corresponds to a meson momentum of about 1.6 GeV, the deviation of $c^2$ from 1 is found
to be less than 0.3\% at $a\approx0.11$ fm and less than 0.4\% at $a\approx0.08$ fm. This demonstrates
that the lattice NRQCD in combination with the Iwasaki gluon action works very well for bottomonium.

\begin{table}[h!]
\begin{tabular}{lllll}
\hline\hline
$\bs{n}^2$ & \hspace{4ex} & $c^2$ ($a\approx 0.11$ fm) & \hspace{4ex} & $c^2$ ($a \approx 0.08$ fm) \\
\hline
$2$       &&   $1.00070(47)$   &&   $1.00012(55)$       \\
$3$       &&   $1.00134(85)$   &&   $1.00022(92)$       \\
$4$       &&   $0.9987(12)$    &&   $0.9987(15)$        \\
$5$       &&   $0.9998(15)$    &&   $0.9991(17)$        \\
$6$       &&   $1.0008(18)$    &&   $0.9995(20)$        \\
$8$       &&   $1.0005(22)$    &&   $0.9995(28)$        \\
$9$       &&   $1.0011(25)$    &&   $1.0001(31)$        \\
$12$      &&   $1.0015(33)$    &&   $1.0009(44)$        \\
\hline\hline
\end{tabular}
\caption{\label{tab:speed_of_light}Square of the speed of light, calculated with the $v^4$ action.
For $\bs{n}^2=9$, the components are $\bs{n}=(2,2,1)$ and octahedral transformations thereof.
The data shown are from the $L=24$ ensemble ($a \approx 0.11$ fm)
with $a m_l=0.005$, $a m_b=2.536$ and from the $L=32$ ensemble ($a \approx 0.08$ fm) with $a m_l=0.004$, $a m_b=1.87$.}
\end{table}

\subsection{\label{sec:radial_orbital_lattice_spacing}Radial/orbital splittings and the lattice spacing}

\subsubsection{\label{sec:radial_orbital_lattice_spacing_sys_err}Estimates of systematic errors}

The radial and orbital bottomonium energy splittings are calculated in this work with the $v^4$ action on all ensembles.
Radial and orbital energy splittings are an effect of order $v^2$,
and therefore the relative error in the radial and orbital energy splittings due to the missing $v^6$
corrections is of order $v^4\approx 0.01$. Further systematic errors are introduced by the missing radiative
corrections to the $v^4$ terms, which leads to a relative error of order $\alpha_s v^2\approx0.02$ (here and in
the following, $\alpha_s\approx0.2$ is used).

The dominant discretization errors in radial and orbital energy splittings
are expected to be caused by missing radiative corrections to the couplings
$c_5$ and $c_6$ in (\ref{eq:dH_full}), and by discretization errors in
the gluon action.

The relative errors caused by the radiative corrections to $c_5$ and $c_6$ can be estimated
as follows: by replacing every derivative with a factor of $m_b v$,
we see that the terms with $c_5$ and $c_6$ in (\ref{eq:dH_full}) should be of order
$a^2 m_b^4 v^4/(24 m_b)$ and $a \:m_b^4 v^4 / (16 \:n\: m_b^2)$, respectively. The radiative
corrections should be of order $\alpha_s$ times these estimates. The relative error in radial and orbital
energy splittings is obtained through dividing by the estimate $m_b^2 v^2/(2 m_b)$ of $H_0$.
The relative error due to the missing radiative corrections to $c_5$ is then
\begin{equation}
 \frac{\alpha_s \:a^2\: m_b^2 \:v^2}{12}, \label{eq:c5_rad_cor}
\end{equation}
which is about 1\% at $a\approx0.11$ fm and 0.6\% at $a\approx0.08$ fm.
The relative error due to the missing radiative corrections to $c_6$ is
\begin{equation}
 \frac{\alpha_s \:a \:m_b\: v^2}{8n}, \label{eq:c6_rad_cor}
\end{equation}
which is about 0.3\% at $a\approx0.11$ fm and 0.2\% at $a\approx0.08$ fm. More sophisticated
estimates of these errors, using wave functions from a potential model, have been made in \cite{Gray:2005ur}.

The discretization errors caused by the gluon action are discussed in detail in Appendix \ref{sec:gluon_errors}.
Estimates using tree-level perturbation theory are derived in Sec.~\ref{sec:gluon_errors_tree}. The values for
the Iwasaki action at the lattice spacings used here can be found in Table \ref{tab:gluonic_discr_errors}. The
nonperturbative results presented in Secs.~\ref{sec:radial_orbital_results} and \ref{sec:gluon_errors_NP} indicate
that the errors may actually be smaller than the tree-level estimates.

\subsubsection{\label{sec:radial_orbital_results}Results}

The results for the radial and orbital energy splittings in lattice units at $a m_b=2.536$ (for $a\approx0.11$ fm) and
$a m_b=1.87$ (for $a\approx 0.08$ fm) are given in Appendix \ref{sec:radial_orbital_lattice}. The spin-averaged masses
are denoted with a bar, and are defined as
\begin{equation}
\overline{M}=\frac{\sum_J (2J+1)M_J}{\sum_J (2J+1)}. \label{eq:spinav}
\end{equation}
In Appendix \ref{sec:radial_orbital_mb_dep}, results for the radial and orbital energy splittings
for multiple values of $a m_b$, varying by about 15\%, are given. As can be seen there,
the dependence on $a m_b$ is very weak. The change when interpolating from $a m_b=2.536$ at $a\approx0.11$ fm
or $a m_b=1.87$ at $a\approx0.08$ fm to the physical values of $a m_b$ given in Table \ref{tab:lattices}
would be much smaller than the statistical errors.

The inverse lattice spacings of the gauge field ensembles are determined from the $\Upsilon(2S)-\Upsilon(1S)$ splitting,
dividing the experimental value from \cite{Amsler:2008zzb} by the dimensionless lattice value. The
$2S-1S$ splitting is expected to have smaller systematic errors than the
$1P-1S$ splitting \cite{Gray:2005ur}. In particular, as shown in Sec.~\ref{sec:gluon_errors_tree_results},
the gluonic discretization errors partially cancel in the $2S-1S$ splitting. For the Iwasaki action
and at tree-level, the remaining gluon errors in the
$2S-1S$ splitting are estimated to be about 2.6\% at $a\approx0.11$ fm and 1.6\% at $a\approx0.08$ fm.

Results for the lattice spacings of all ensembles are given in Tables \ref{tab:lattice_spacing_L16}, \ref{tab:lattice_spacing_L24}, and \ref{tab:lattice_spacing_L32}.
The errors shown there are statistical/fitting only.
For comparison, results from both the $\Upsilon(2S)-\Upsilon(1S)$ and $\overline{1^3P}-\Upsilon(1S)$ splitting are given.
They are found to be mostly consistent within the statistical errors here. Note that in the quenched approximation,
the ratio of the $2S-1S$ and $1P-1S$ splittings was previously found to be in disagreement with experiment \cite{Gray:2005ur}.

The lattice spacings obtained here are seen to become slightly finer as the sea quark mass is reduced (in \cite{Meinel:2009rd},
this dependence was hidden by the larger statistical errors). This behavior is in fact expected here, since in the RBC/UKQCD ensembles,
the bare gauge coupling is kept constant when varying the sea-quark masses (see Table \ref{tab:lattices}). In contrast to this,
the MILC collaboration decreases $\beta\propto 1/g^2$ when decreasing the sea-quark masses \cite{Bazavov:2009bb},
such that the lattice spacing remains approximately constant \cite{Gray:2005ur}.

\begin{table}[h!]
\begin{minipage}{.48\linewidth}
\begin{center}
\begin{tabular}{cccccc}
\hline\hline
                 & $am_l=0.01$    & & $am_l=0.02$    & & $am_l=0.03$     \\
\hline
$a^{-1}_{2S-1S}$ &  $1.766(52)$    && $1.687(46)$    &&  $1.651(33)$ \\
$a^{-1}_{1P-1S}$ &  $1.718(16)$    && $1.678(13)$    &&  $1.661(10)$   \\
\hline\hline
\end{tabular}
\caption{\label{tab:lattice_spacing_L16}Inverse lattice spacings of the $L=16$ ensembles in GeV.}
\end{center}
\end{minipage}
\hfill
\begin{minipage}{.48\linewidth}
\begin{center}
\begin{tabular}{cccccccc}
\hline\hline
                  & $am_l\!=\!0.005$    &  &  $am_l\!=\!0.01$  &  & $am_l\!=\!0.02$   &  & $am_l\!=\!0.03$     \\
\hline
$a^{-1}_{2S-1S}$  & $1.763(27)$   && $1.732(28)$  && $1.676(42)$  &&  $1.650(39)$   \\
$a^{-1}_{1P-1S}$  & $1.742(14)$   && $1.703(12)$  && $1.680(27)$  &&  $1.670(33)$   \\
\hline\hline
\end{tabular}
\caption{\label{tab:lattice_spacing_L24}Inverse lattice spacings of the $L=24$ ensembles in GeV.}
\end{center}
\end{minipage}
\end{table}

\begin{table}[h!]
\begin{tabular}{ccccccc}
\hline\hline
                  & \hspace{1ex} &  $am_l=0.004$  & \hspace{1ex} & $am_l=0.006$   & \hspace{1ex} & $am_l=0.008$     \\
\hline
$a^{-1}_{2S-1S}$  &&   $2.325(32)$  &&  $2.328(45)$    &&  $2.285(32)$  \\
$a^{-1}_{1P-1S}$  &&   $2.305(24)$  &&  $2.329(23)$    &&  $2.328(23)$  \\
\hline\hline
\end{tabular}
\caption{\label{tab:lattice_spacing_L32}Inverse lattice spacings of the $L=32$ ensembles in GeV.}
\end{table}

The lattice spacings from the $2S-1S$ splittings on the individual ensembles were then used to convert
the results for the other radial and orbital energy splittings to physical units. The conversion was performed
using the bootstrap method to take into account correlations between the $2S-1S$ splitting and the other splittings.

\begin{figure*}[h!]
 \includegraphics[width=0.43\linewidth]{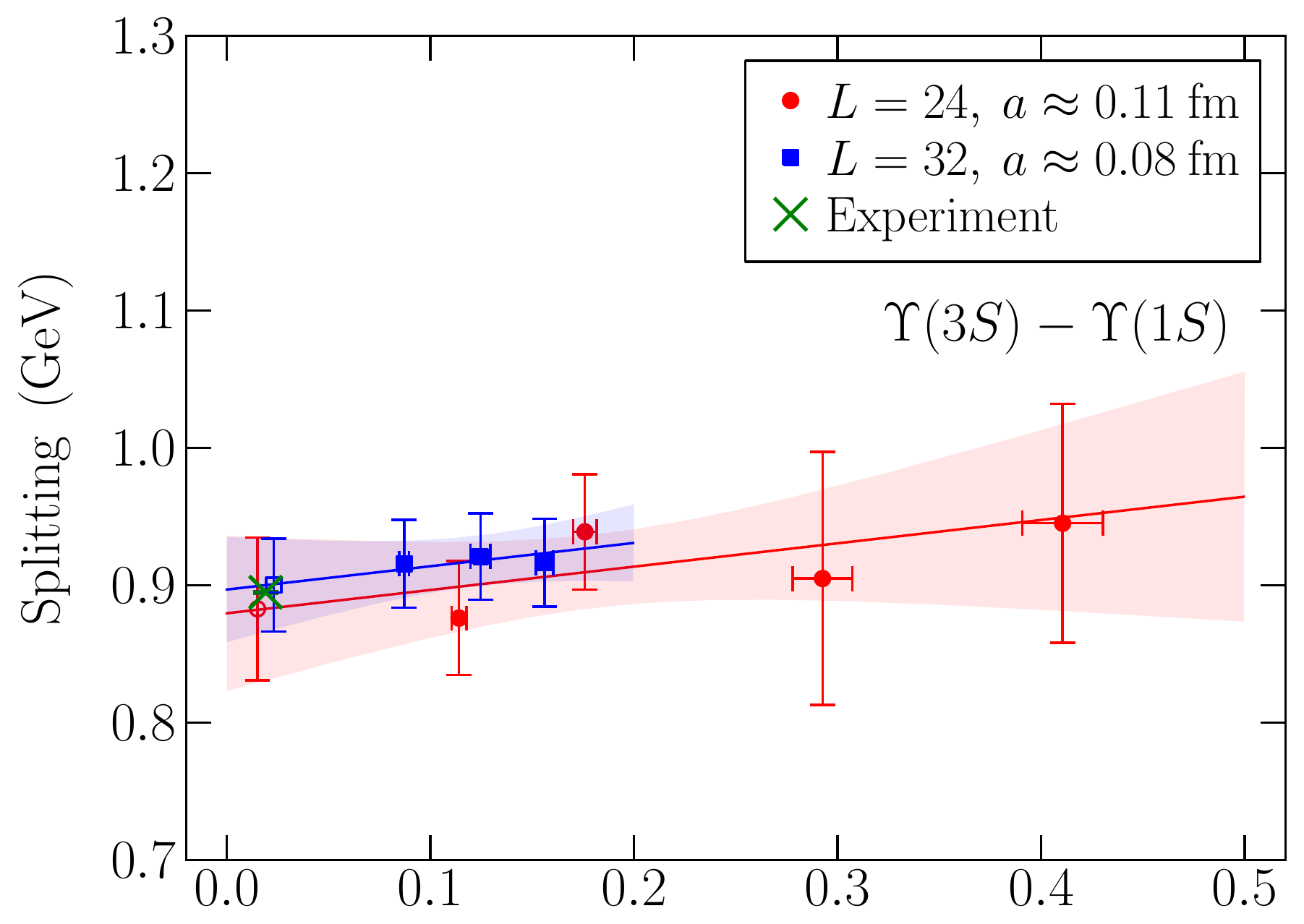} \hfill \includegraphics[width=0.43\linewidth]{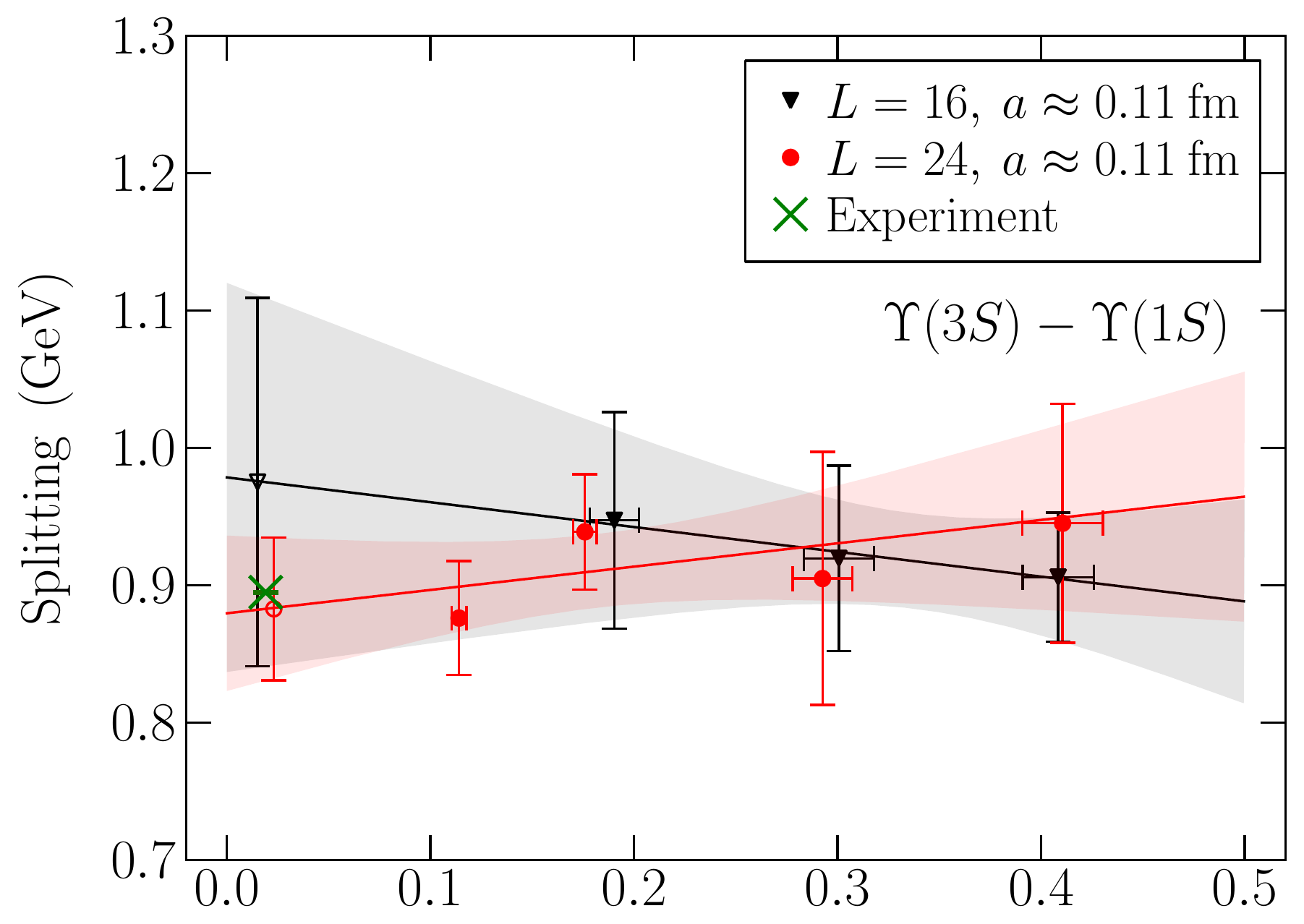}
 \includegraphics[width=0.43\linewidth]{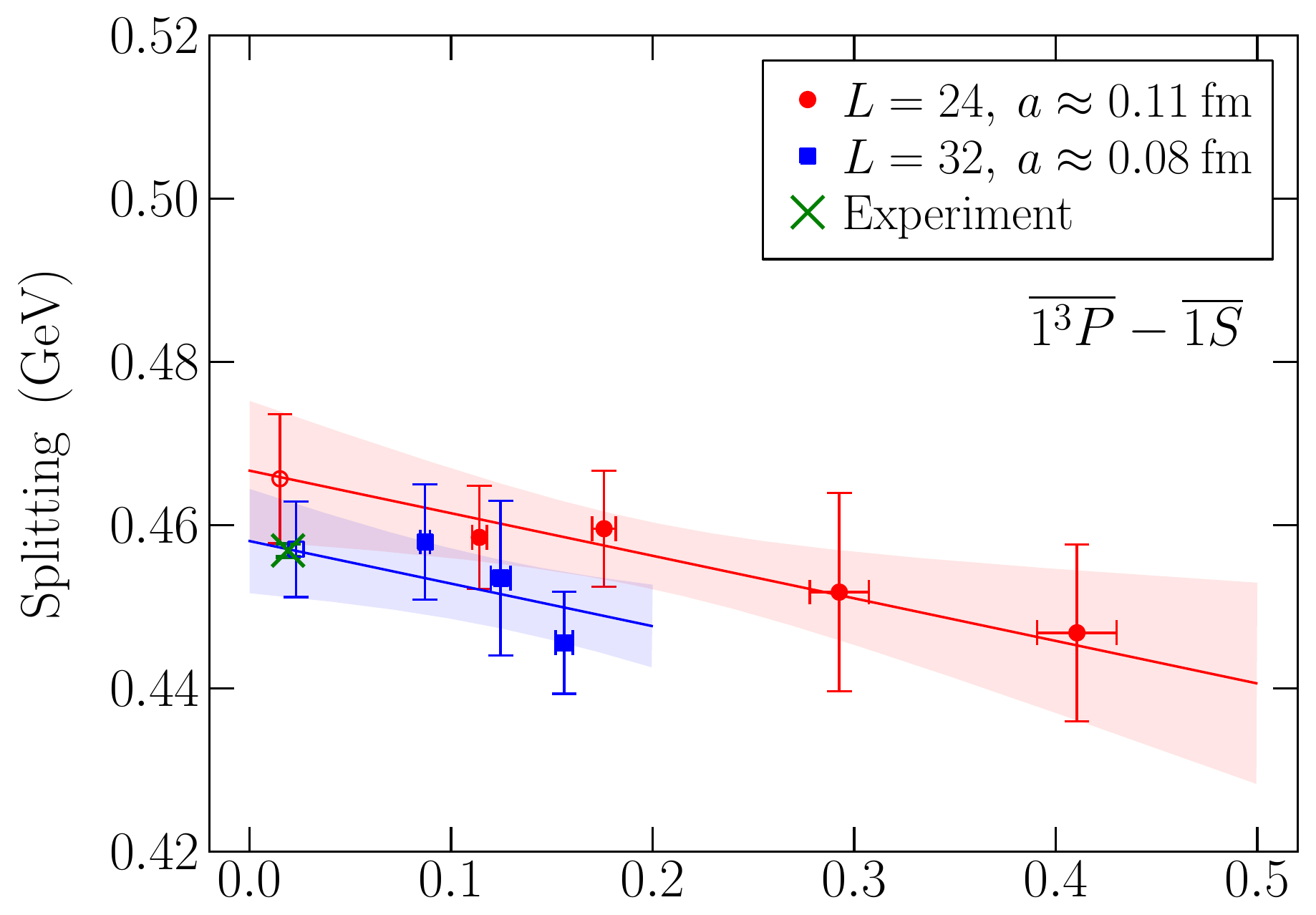}        \hfill \includegraphics[width=0.43\linewidth]{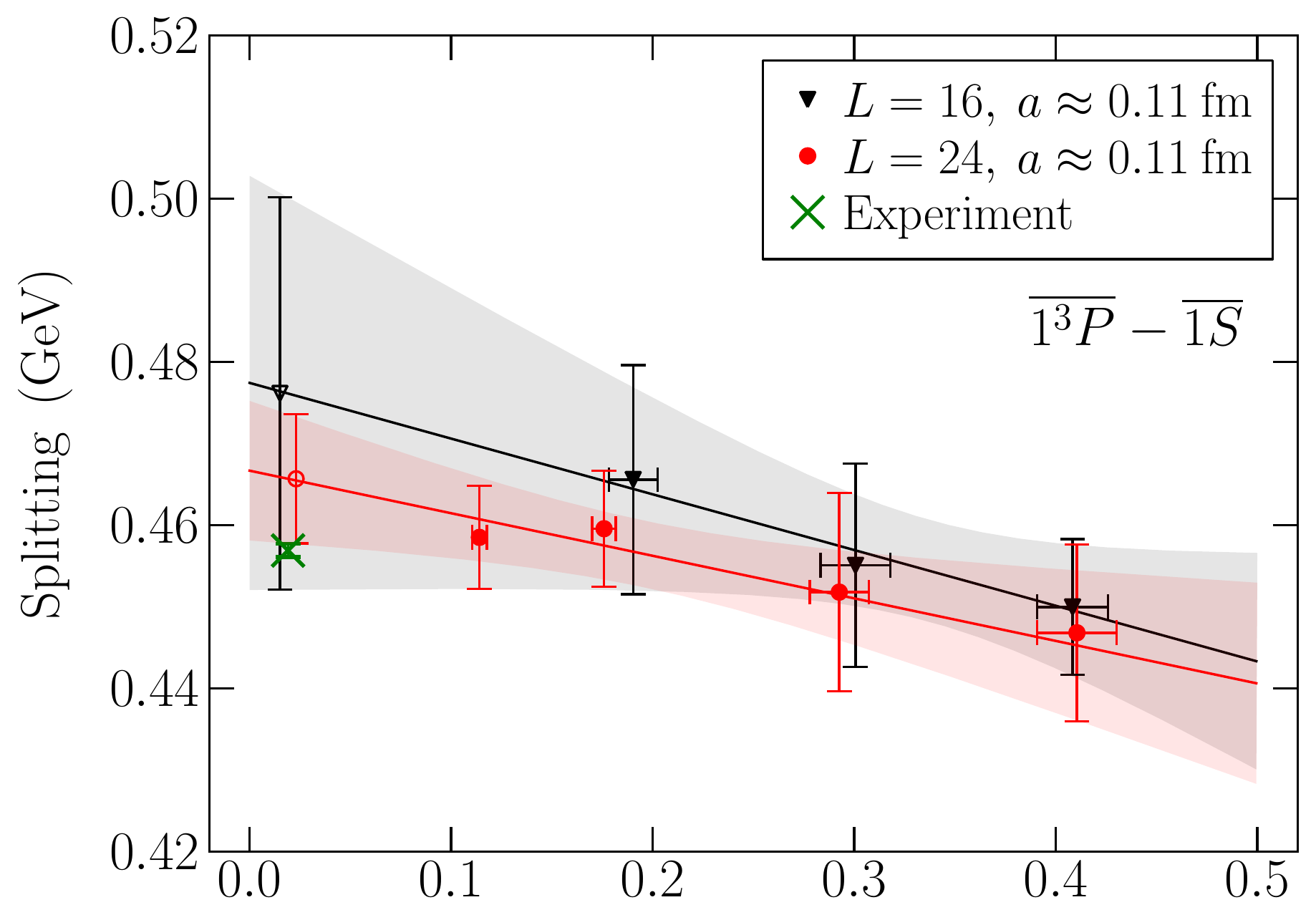}
 \includegraphics[width=0.43\linewidth]{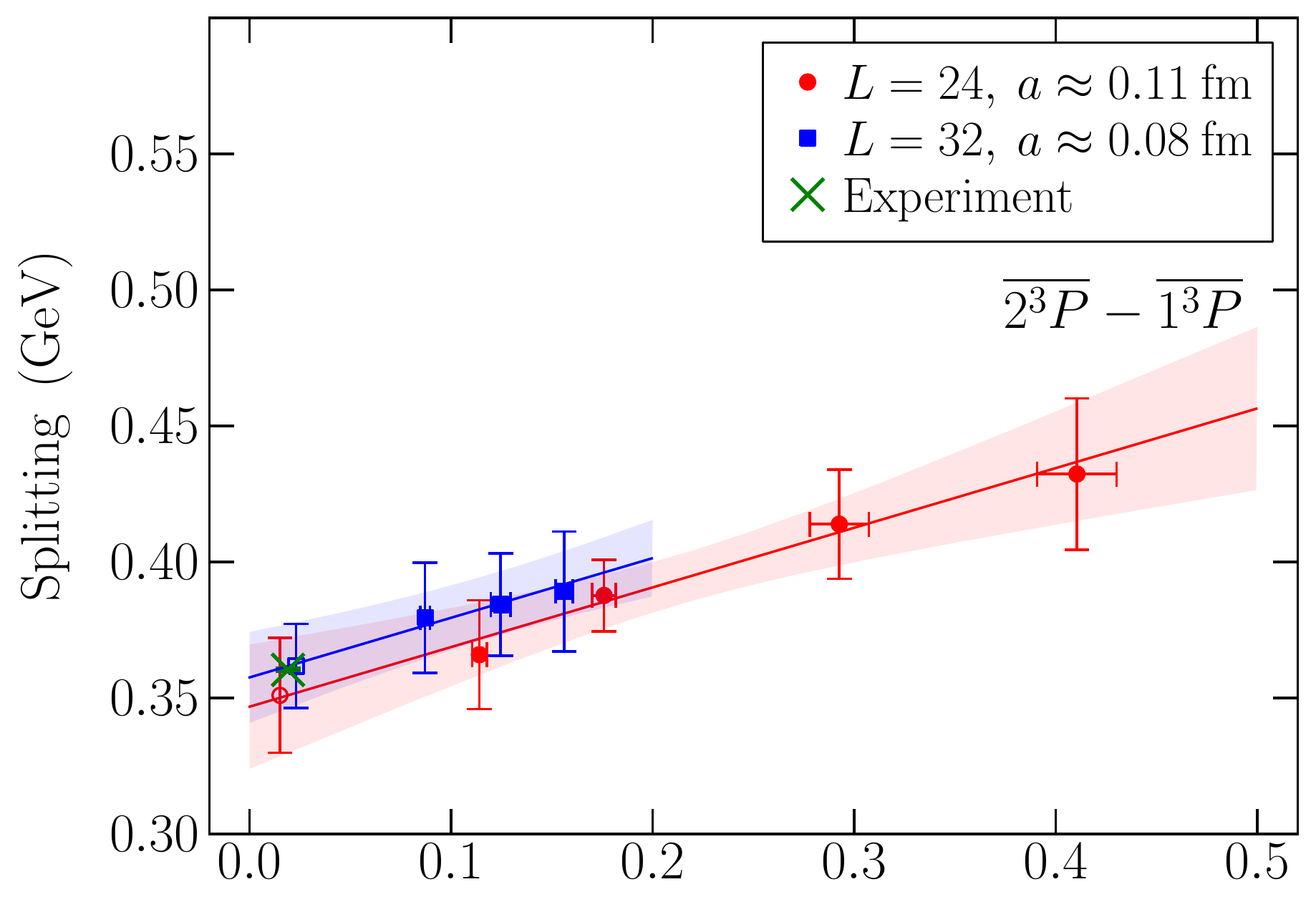}     \hfill \includegraphics[width=0.43\linewidth]{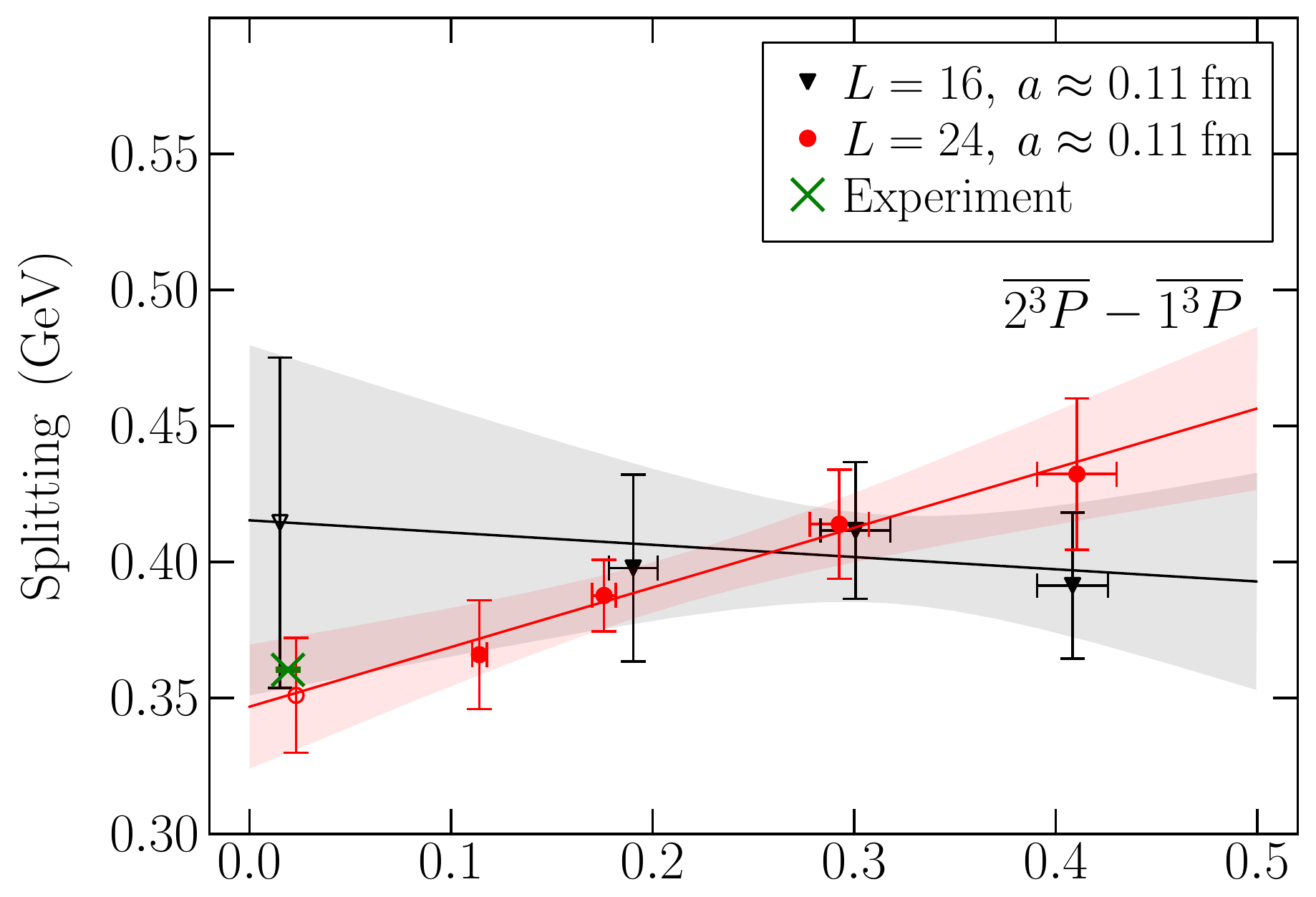}
 \includegraphics[width=0.43\linewidth]{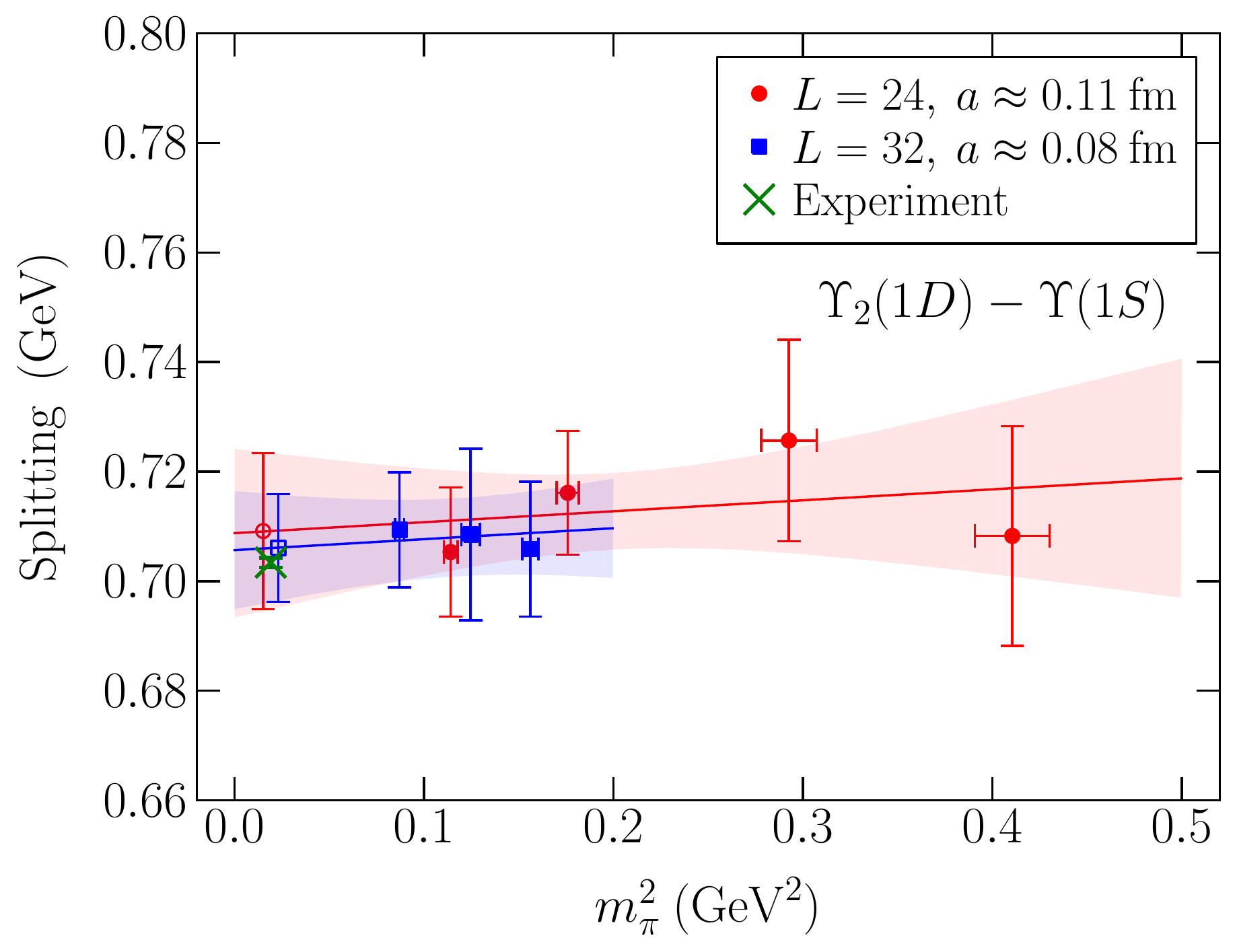} \hfill \includegraphics[width=0.43\linewidth]{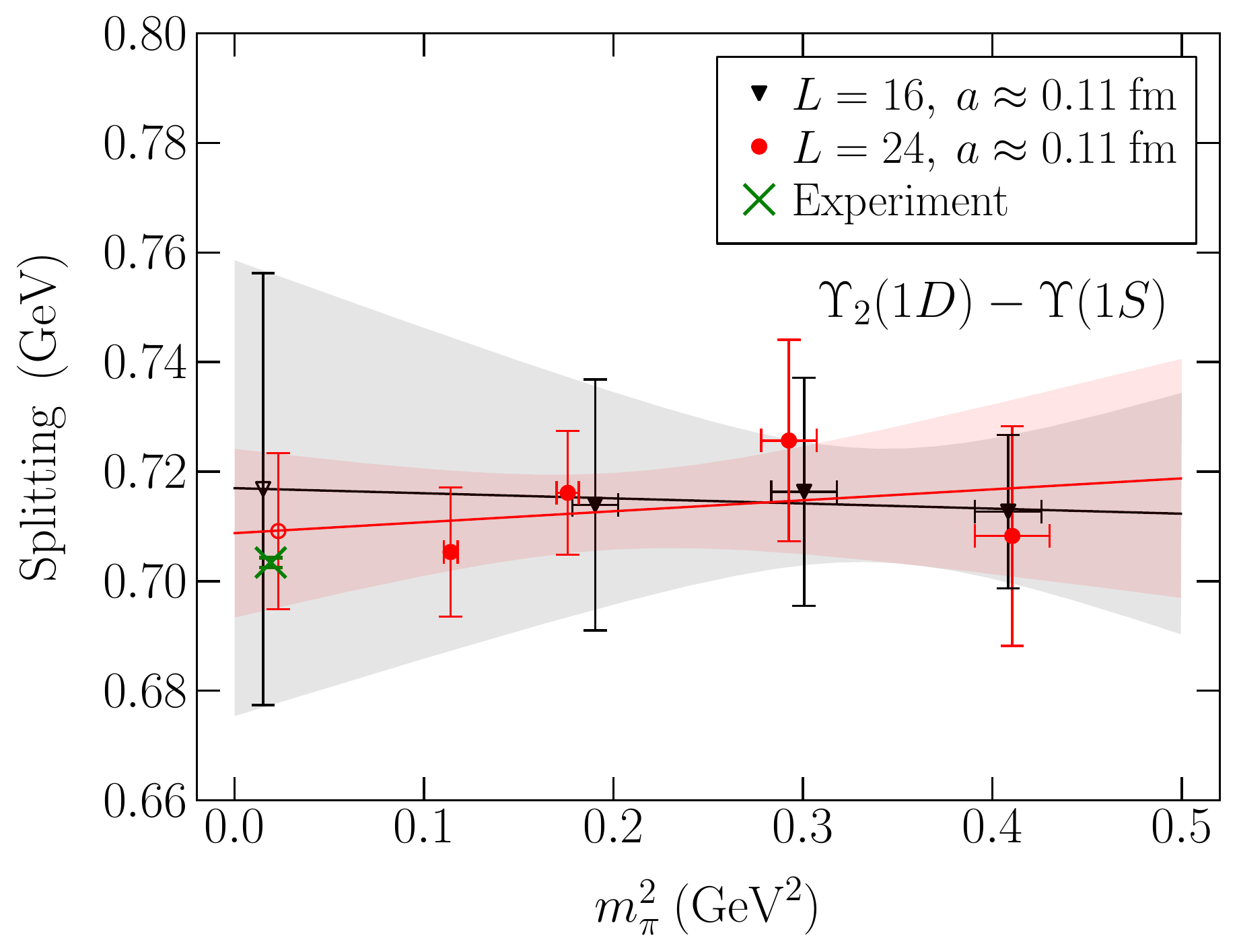}
\caption{\label{fig:radial_orbital_chiral_extrap}Chiral extrapolation of radial and orbital energy splittings (calculated with the $v^4$ action).
Extrapolated points are offset horizontally for legibility.}
\end{figure*}

Finally, the results in physical units were extrapolated to the physical pion mass, as shown in Fig.~\ref{fig:radial_orbital_chiral_extrap}.
The data from the $L=24$ and $L=32$ ensembles (both have a box size of about $2.7$ fm) were extrapolated simultaneously using the function (\ref{eq:sim_chiral_extrap}).
The data from the $L=16$ ensembles were treated independently, since the dependence on the pion mass may be different
for the smaller box size of about $1.8$ fm.

\clearpage

\begin{table*}[h!]
\begin{ruledtabular}
\begin{tabular}{lcccc}
                                  &   $L=16$, $a\approx 0.11$ fm   &  $L=24$, $a\approx 0.11$ fm        & $L=32$, $a\approx 0.08$ fm          & Experiment  \\
\hline
$\Upsilon(3S)-\Upsilon(1S)$       &  $0.98(13)\nb$ &  $0.883(52)\nb$ &  $0.900(34)\nb$ &  $0.89490(56)$ \\
$\overline{1^3P}-\Upsilon(1S)$    &  $0.462(24)$   &  $0.4524(78)$   &  $0.4428(57)$   &  $0.43957(37)$ \\
$\overline{1^3P}-\overline{1S}$   &  $0.476(24)$   &  $0.4657(79)$   &  $0.4571(59)$   &  $0.45690(79)$ \\
$\overline{2^3P}-\overline{1^3P}$ &  $0.414(61)$   &  $0.351(21)\nb$ &  $0.362(15)\nb$ &  $0.36033(45)$ \\
$\overline{2^3P}-\Upsilon(1S)$    &  $0.865(70)$   &  $0.802(23)\nb$ &  $0.803(17)\nb$ &  $0.79990(47)$ \\
$\overline{2^3P}-\overline{1S}$   &  $0.878(70)$   &  $0.816(23)\nb$ &  $0.817(17)\nb$ &  $0.81722(83)$ \\
$\Upsilon_2(1D)-\Upsilon(1S)$     &  $0.717(39)$   &  $0.709(14)\nb$ &  $0.7060(98)$   &  $0.70340(87)$ \\
\end{tabular}
\end{ruledtabular}
\caption{\label{tab:radialorbital_splittings_final}Chirally extrapolated results for the radial and orbital energy splittings in GeV, calculated with the $v^4$ action.
The errors on the lattice results shown here are statistical/fitting/scale setting only; for a discussion of systematic errors, see Sec.~\ref{sec:radial_orbital_lattice_spacing_sys_err}.
The experimental values are from \cite{Aubert:2008vj, Aubert:2009pz, Bonvicini:2009hs} (for the $\eta_b(1S)$, which enters in $\overline{1S}$),
\cite{Bonvicini:2004yj, delAmoSanchez:2010kz} (for the $\Upsilon_2(1D)$) and from the Particle Data Group \cite{Amsler:2008zzb} (for all other states).}
\end{table*}

\begin{figure}[h!]
 \includegraphics[width=0.49\linewidth]{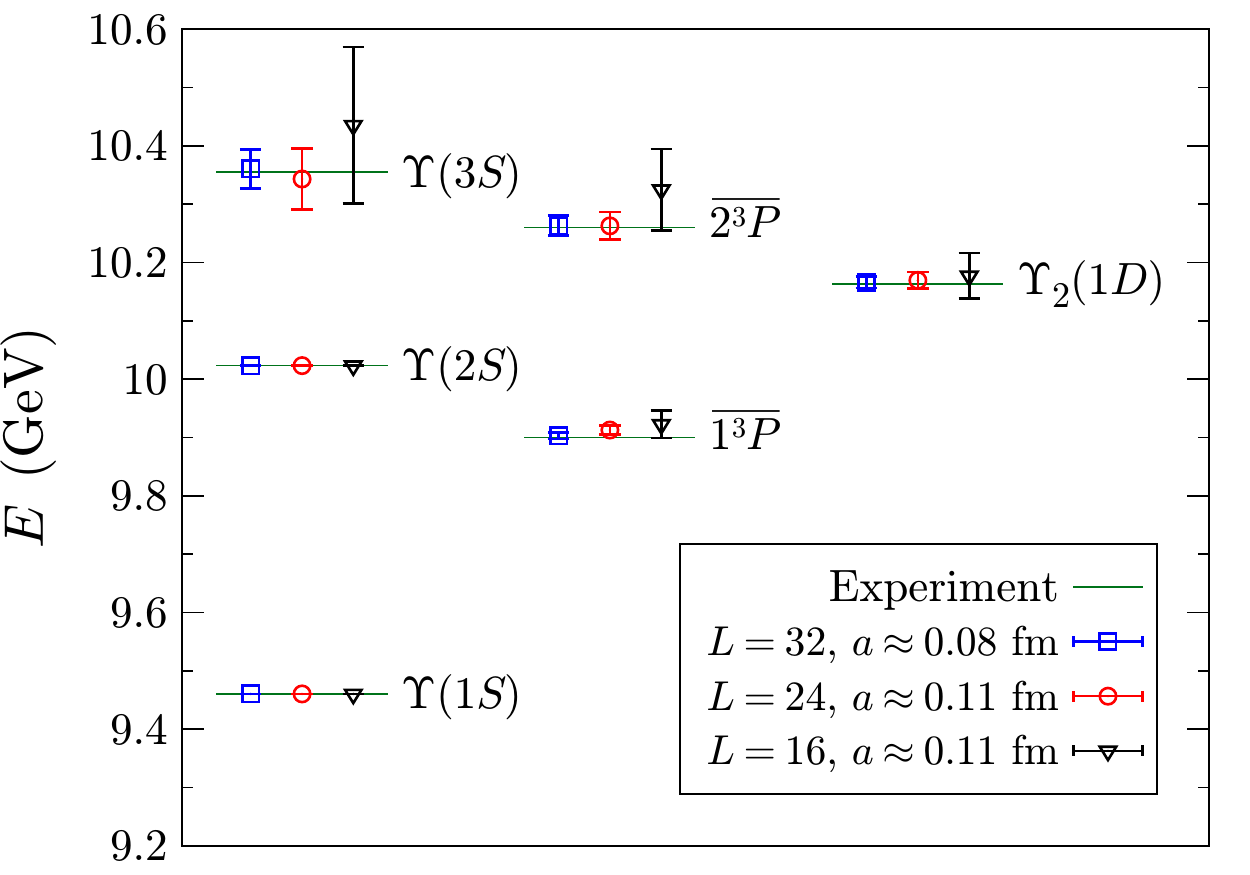}
\caption{\label{fig:radialorbital_splittings_final}Chirally extrapolated results for the radial and orbital energy splittings,
calculated with the $v^4$ action. The $\Upsilon(1S)$ and $\Upsilon(2S)$ masses are used as input in the lattice calculation.
The errors on the lattice results shown here are statistical/fitting/scale setting only.}
\end{figure}

The numerical results at the physical pion mass are given in Table \ref{tab:radialorbital_splittings_final}. In addition,
the energies of the radial and orbital excitations at the physical pion mass are plotted in Fig.~\ref{fig:radialorbital_splittings_final}.
The extrapolated data from the $L=16$ ensembles have significantly larger statistical errors than the other data. This is expected
for the following two main reasons: first, the four-dimensional volume of these ensembles is about 7 times smaller than that of the $L=24$
ensembles, providing less information.
Second, the lowest pion mass available at $L=16$ is larger than on the other two ensembles, requiring more extrapolation. Within the statistical errors, no
finite-volume effects are seen (also at the individual values of the quark masses, where the results are more precise; see Fig.~\ref{fig:radial_orbital_chiral_extrap}).
Finite-volume effects in bottomonium have been studied in detail using a lattice potential model in \cite{Bali:1998pi}. At
a box size of $1.8$ fm (corresponding to the $L=16$ lattices here) the most significant finite-size effects were found in the $3S$ energy, which was shifted by
about $-40$ MeV compared to the infinite-volume energy. At a size of $2.7$ fm, this shift was found to be negligible.

The results for the radial and orbital energy splittings obtained here from the $L=24$ and $L=32$ ensemble
show very little dependence on the lattice spacing. When going from $a\approx0.11$ fm to
$a\approx0.08$ fm, the $1P-1S$ splitting changes by about 2\%. However, this change is only 1 standard deviation and could also
be caused by a statistical fluctuation. All results at $a\approx0.08$ fm, where discretization
errors are expected to be the smallest, are in excellent agreement with experiment. The $\overline{1^3P}-\overline{1S}$ and $\Upsilon_2(1D)-\Upsilon(1S)$
splittings at $a\approx0.08$ fm have statistical errors of only about 1.3\%, and they agree fully with experiment. This indicates that the systematic errors are very small;
smaller than the tree-level estimates of gluon-discretization errors given in Table \ref{tab:gluonic_discr_errors} would suggest.

\subsection{\label{sec:spindep}Spin-dependent splittings}

\subsubsection{\label{sec:spindep_errors}Discussion of systematic errors}

Spin-dependent energy splittings first arise at order $v^4$ in the nonrelativistic expansion, via the terms with coefficients $c_3$ and $c_4$
in Eq.~(\ref{eq:dH_full}). For the order-$v^4$ action, it is expected that at lowest order the $S$-wave hyperfine splittings
\begin{equation}
 \Upsilon(nS)-\eta_b(nS)
\end{equation}

\vspace{-2ex}

\noindent and the $P$-wave tensor splittings

\vspace{-4ex}

\begin{equation}
-2\chi_{b0}(nP)+3\chi_{b1}(nP)-\chi_{b2}(nP)
\end{equation}
are proportional to $c_4^2$ and independent of $c_3$, while the $P$-wave spin-orbit splittings
\begin{equation}
-2\chi_{b0}(nP)-3\chi_{b1}(nP)+5\chi_{b2}(nP)
\end{equation}
are proportional to $c_3$ and independent of $c_4$ \cite{Gray:2005ur}. In this work, the coefficients $c_i$ are
set to their tree-level values, $c_i=1$. Therefore, spin-dependent energy splittings calculated directly will have
systematic errors of order $\alpha_s\approx0.2$. However, these unknown radiative corrections are expected to cancel in ratios of
quantities with equal dependence on the couplings $c_i$. Nonperturbative results for the dependence of various spin splittings
and ratios of spin splittings on $c_3$ and $c_4$ are given in Appendix \ref{sec:spindep_c3_c4} for both the $v^4$ and $v^6$ actions.
As can be seen there, ratios of hyperfine and tensor splittings are indeed independent of both $c_3$ and $c_4$ to a very good approximation.

Spin-splittings calculated with the $v^4$ action will also have relativistic errors of order $v^2\approx0.1$ due to the missing $v^6$ corrections.
Therefore, the spin-dependent $v^6$-corrections, given by the terms with coefficients $c_7$, $c_8$ and $c_9$ in Eq.~(\ref{eq:dH_full}),
are included in this work. Relativistic corrections for spin splittings calculated with the $v^6$ action are then expected to be of order $v^4\approx 0.01$
due to the missing order-$v^8$ terms. Missing radiative corrections to the order-$v^6$ terms lead to additional systematic errors of order
$\alpha_s v^2\approx 0.02$.

The terms with coefficients $c_3$ and $c_4$ in Eq.~(\ref{eq:dH_full}) include the tree-level discretization corrections via (\ref{eq:improved_D}) and (\ref{eq:improved_F}):
\begin{eqnarray}
\nonumber \bs{\sigma}\cdot \left(\bs{\widetilde{\nabla}}\times\bs{\widetilde{E}}-\bs{\widetilde{E}}\times\bs{\widetilde{\nabla}} \right)
&=& \epsilon_{jkl} \: \sigma_j \: \Big(\nabla_k F_{l0}-F_{k0} \nabla_l \Big) -\frac{a^2}{6} \:\epsilon_{jkl} \:\sigma_j \: \Big(\nabla_k^{(3)} F_{l0} -F_{k0} \nabla_l^{(3)} \Big)\\
&& -\frac{a^2}{6} \: \epsilon_{jkl} \: \sigma_j \: \Bigg(\nabla_k \Big[\nabla_l^{(2,{\rm ad})}F_{l0} + \nabla_0^{(2,{\rm ad})}F_{l0}\Big]  -\Big[\nabla_k^{(2,{\rm ad})}F_{k0} + \nabla_0^{(2,{\rm ad})}F_{k0}\Big] \nabla_l \Bigg),
\label{eq:spinorbit_discr}
\end{eqnarray}
\begin{eqnarray}
 \bs{\sigma}\cdot \bs{\widetilde{B}}&=&-\frac12 \: \sigma_j \:\epsilon_{jkl}\: F_{kl} + \frac{a^2}{12} \: \sigma_j \:\epsilon_{jkl}\: \Big[\nabla_k^{(2,{\rm ad})}F_{kl} + \nabla_l^{(2,{\rm ad})}F_{kl}\Big].
\label{eq:tensor_discr}
\end{eqnarray}
However, radiative corrections to the order-$a^2$ terms in (\ref{eq:spinorbit_discr}) and (\ref{eq:tensor_discr}) are missing. Their size can be estimated
using the NRQCD power-counting rules. We replace every spatial derivative by $m_b v$ and every temporal derivative by $\frac12 m_b v^2$ (by the leading-order
equations of motion, a temporal derivative is of the order of the nonrelativistic kinetic energy \cite{Lepage:1992tx}). The radiative discretization corrections
are of order $\alpha_s$ times the $a^2$ terms. For the spin-orbit splitting controlled by (\ref{eq:spinorbit_discr}), the terms with the spatial derivatives in
the $a^2$ terms are dominant, and the relative discretization errors become

\vspace{-3.5ex}

\begin{equation}
 \frac13 \alpha_s a^2 m_b^2 v^2.
\end{equation}

\noindent This is about $0.04$ at $a\approx0.11$ fm and about $0.02$ at $a\approx0.08$ fm. The relative discretization errors in
the hyperfine and tensor splittings are

\vspace{-3.5ex}

\begin{equation}
 \frac23 \alpha_s a^2 m_b^2 v^2. \label{eq:hfs_discr_err}
\end{equation}

\noindent Here, an additional factor of 2 was introduced to take into account the quadratic dependence of the hyperfine and tensor splittings on (\ref{eq:tensor_discr}).
Equation (\ref{eq:hfs_discr_err}) is equal to about $0.09$ at $a\approx0.11$ fm and about $0.05$ at $a\approx0.08$ fm.

In the spin-dependent order-$v^6$ corrections (the terms with coefficients $c_7$, $c_8$ and $c_9$ in the action), the lattice Laplacian
used here does not include discretization corrections. The tree-level corrected Laplacian would be
$\widetilde{\Delta}^{(2)}=\Delta^{(2)}-(a^2/12)\Delta^{(4)}$. The relative error in the hyperfine
and tensor splittings caused by the missing of this correction is then of order $a^2 m_b^2 v^4/6$, which is only 0.005 at $a\approx0.08$ fm.

Additional discretization errors in spin splittings may arise from the gluon action. These errors
are discussed in detail in Sec.~\ref{sec:gluon_errors_NP}. For the $S$-wave hyperfine splittings, the relative error caused by discretization
errors in the Iwasaki action is estimated to be about 5\% at $a\approx0.08$ fm; for the other spin splittings no significant gluon discretization
errors are found.

Finally, note that the lattice NRQCD calculation performed here does not include the effects of annihilation of the $b$ and $\overline{b}$.
This mainly affects the pseudoscalar states $\eta_b(nS)$, which can annihilate into two gluons (for the $\Upsilon(nS)$, at least three gluons are required).
The annihilation contribution to the $\eta_b$ mass can be related to the two-gluon decay width $\Gamma[\eta_b\rightarrow gg]$
as follows \cite{Bodwin:1994jh,Follana:2006rc}: \vspace{-1ex}
\begin{equation}
 \delta E_{\rm annihil.}[\eta_b] \approx \frac{\ln(2) -1}{\pi}\: \Gamma[\eta_b\rightarrow gg].
\end{equation}
For the $\eta_b(1S)$ and $\eta_b(2S)$, the widths were calculated in \cite{Kim:2004rz} to be 7 MeV and 3.5 MeV, respectively. This
gives $\delta E_{\rm annihil.}[\eta_b(1S)]=-0.7$ MeV and $\delta E_{\rm annihil.}[\eta_b(2S)]=-0.34$ MeV, which is only
about 1\% of the hyperfine splittings.

\subsubsection{\label{sec:spindep_results}Results}

The results for the spin-dependent energy splittings in lattice units at $a m_b=2.536$ (for $a\approx0.11$ fm, $L=24$) and
$a m_b=1.87$ (for $a\approx 0.08$ fm, $L=32$) for both the $v^4$ and $v^6$ NRQCD
actions are given in Appendix \ref{sec:spin_dep_lattice}. Here, all couplings
$c_i$ in the action were set to their tree-level values of 1.

In Appendix \ref{sec:spindep_c3_c4}, the dependence of the spin splittings on the couplings $c_3$ and $c_4$ in the
action (\ref{eq:dH_full}) is studied (on the $L=24$ ensemble with $a m_l=0.005$). Results are shown both for
the $v^4$ and $v^6$ actions. The naive expectations for the $c_3$- and $c_4$-dependence
of the order-$v^4$ spin splittings were discussed at the beginning of Sec.~\ref{sec:spindep_errors}. The dependence
of the $1P$ spin-orbit splitting on the coupling $c_3$ appears to be slightly weaker than expected: it changes
by only 13\% (for the $v^4$ action) or 15\% (for the $v^6$ action), when $c_3$ is changed by 20\%. Contrary
to the naive expectation, the $1P$ spin-orbit splitting also shows some dependence on $c_4$: about 6\%
(for the $v^4$ action) or 8\% (for the $v^6$ action), when $c_4$ is varied by 20\%.
On the other hand, the $1P$ tensor splitting behaves as expected: it shows no significant $c_3$-dependence,
and the dependence on $c_4$ is consistent with proportionality to $c_4^2$, both for the $v^4$ and $v^6$ actions.
The results for the $1S$ hyperfine splitting are also close to these expectations.
However, in the $1S$ hyperfine splitting the deviations from the naive expectations, while not large in
absolute terms, are statistically significant due to the very small statistical errors. The most important
result from Appendix \ref{sec:spindep_c3_c4} is that the ratio of the $2S$ and $1S$ hyperfine splitting
as well as the ratios of the $S$-wave hyperfine and the $1P$ tensor splitting show no significant dependence on either
$c_4$ or $c_3$, and this is true for both the $v^4$ and $v^6$ actions.

Next, in Appendix \ref{sec:spindep_splittings_mb_dep}, results for the spin-dependent
energy splittings for multiple values of $a m_b$, varying by about 15\%, are given, and visualized in
Fig.~\ref{fig:spin_dep_mb_dep}. As can be seen there, the results for most splittings
are compatible with a $1/m_b$-dependence, with the notable exception of the $1S$ hyperfine splitting. The $1/m_b$-dependence
of the spin splittings can be understood from the power-counting rules as follows: radial and orbital energy splittings,
which are of order $m_b v^2$, are nearly independent of $m_b$, as shown in Appendix \ref{sec:radial_orbital_mb_dep}. This
implies that
\begin{equation}
 v^2 \propto 1/m_b.
\end{equation}
Spin-dependent energy splittings are a factor of $v^2$ smaller than radial and orbital energy splittings. Since the latter are nearly
constant, spin-dependent energy splittings are expected to be proportional to $v^2$, and hence $1/m_b$.

The results for the $1S$ hyperfine splitting in lattice units have very small statistical errors, and are clearly incompatible
with a dependence proportional to $1/(a m_b)$. However, fits of the form $A/(a m_b)+B$ with a constant term $B$ describe the
date very well in the range considered here. The fit results $A$ and $B$ for the $1S$ hyperfine splittings
on all $L=24$ and all $L=32$ ensembles, for both the $v^4$ action and the $v^6$ action are given in Tables
\ref{tab:1S_hyperfine_L24_mb_dep}, \ref{tab:1S_hyperfine_L24_mb_dep_v6}, \ref{tab:1S_hyperfine_L32_mb_dep}, and \ref{tab:1S_hyperfine_L32_mb_dep_v6}.

To obtain physical results, all spin splittings were then interpolated to the physical $b$-quark masses given in Appendix \ref{sec:mb_tuning},
assuming a $1/(a m_b)$ dependence everywhere except for the $1S$ hyperfine splittings and the ratios. For the $1S$ hyperfine splittings
and the ratios involving them, the fit results $A$ and $B$ were used in the interpolation.

The interpolated spin splittings were then converted to physical units using the lattice spacings from the $2S-1S$ splittings
on the individual ensembles as obtained in Sec.~\ref{sec:radial_orbital_results}.
Note that the uncertainty in the lattice spacing enters with a factor of 2 here, due to the resulting uncertainty in the bare heavy quark mass
and the approximate $1/(a m_b)$ behavior of spin splittings.

\begin{figure*}[h!]
 \includegraphics[width=0.43\linewidth]{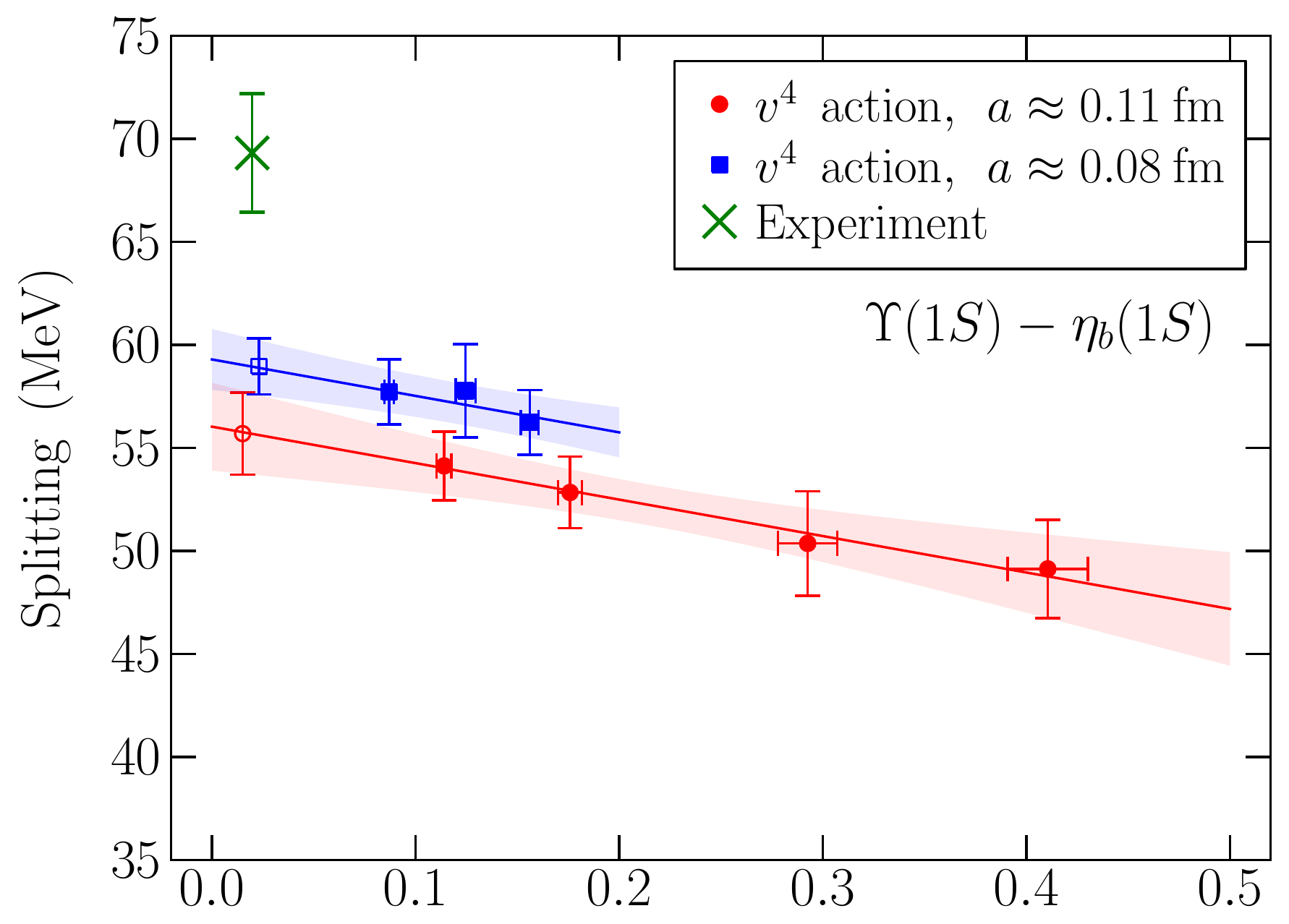} \hfill \includegraphics[width=0.43\linewidth]{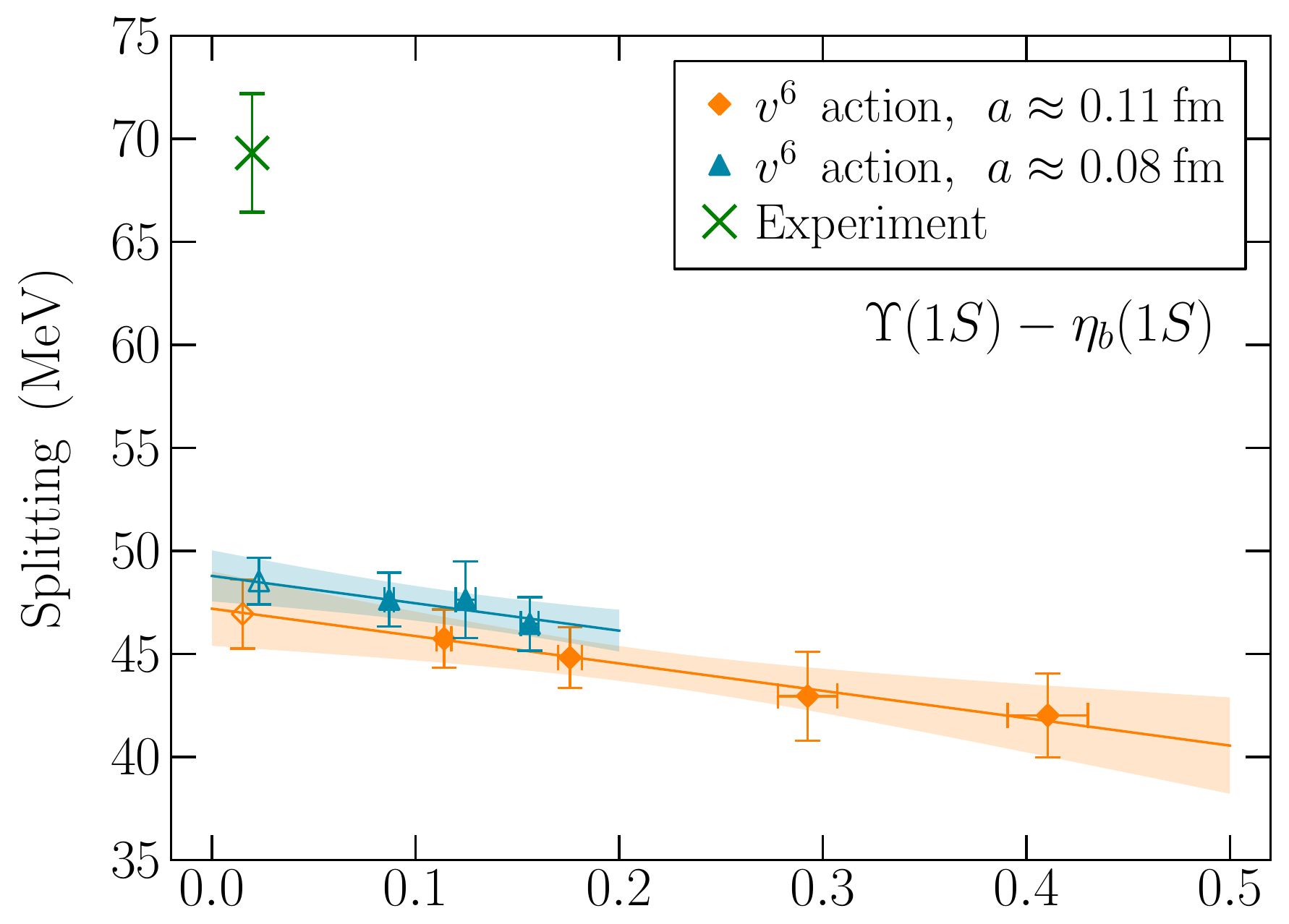}
 \includegraphics[width=0.43\linewidth]{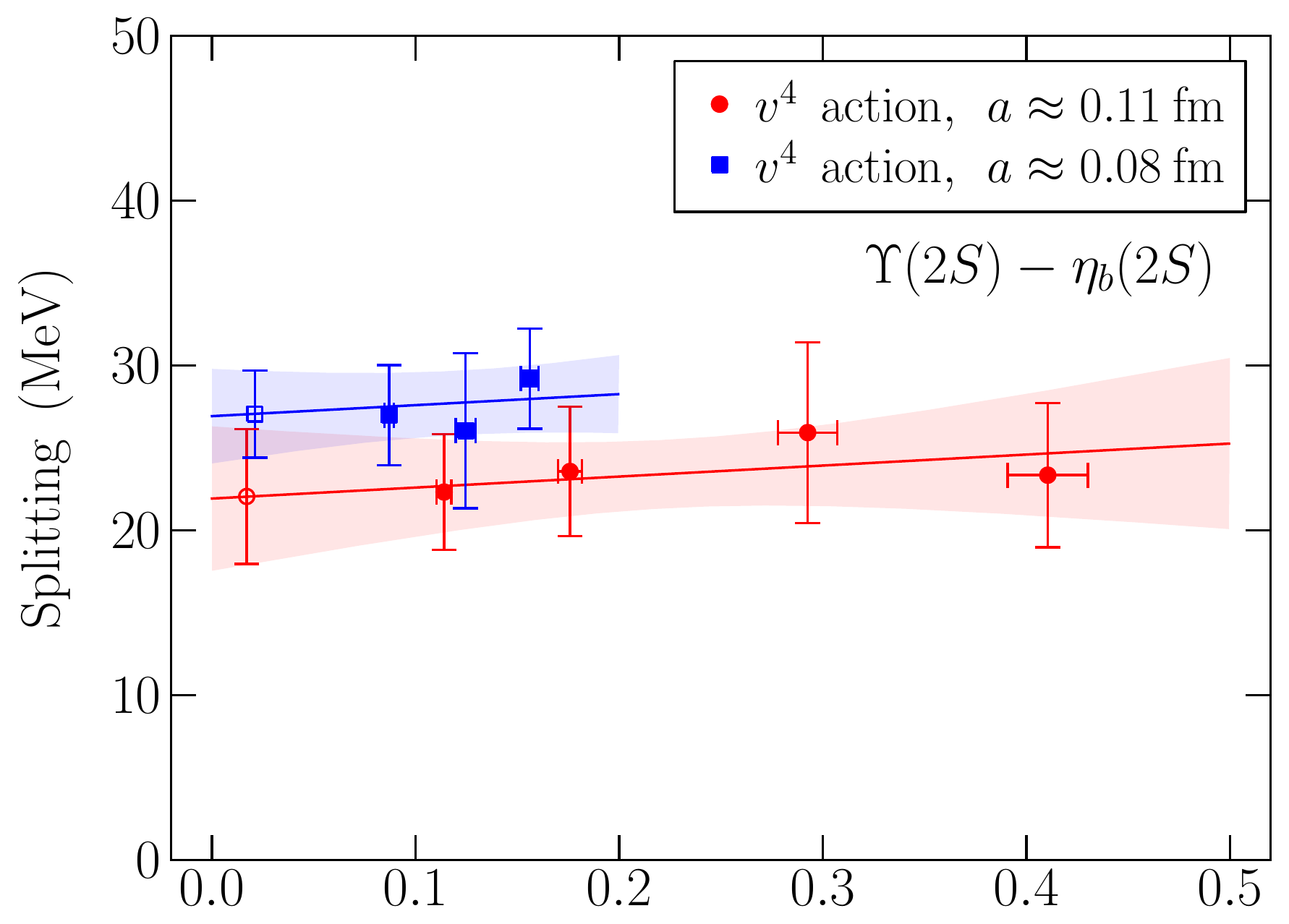} \hfill \includegraphics[width=0.43\linewidth]{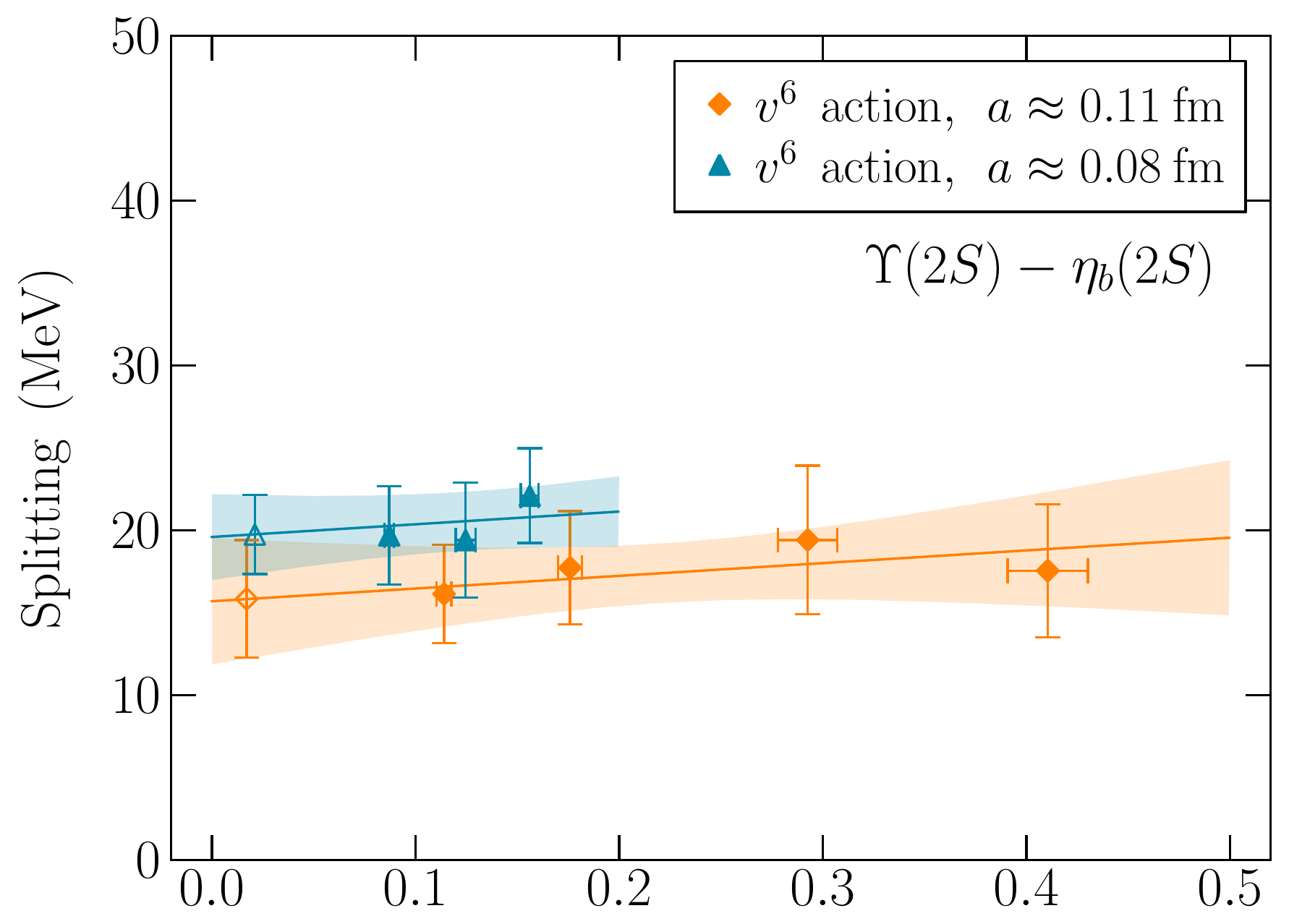}
 \includegraphics[width=0.43\linewidth]{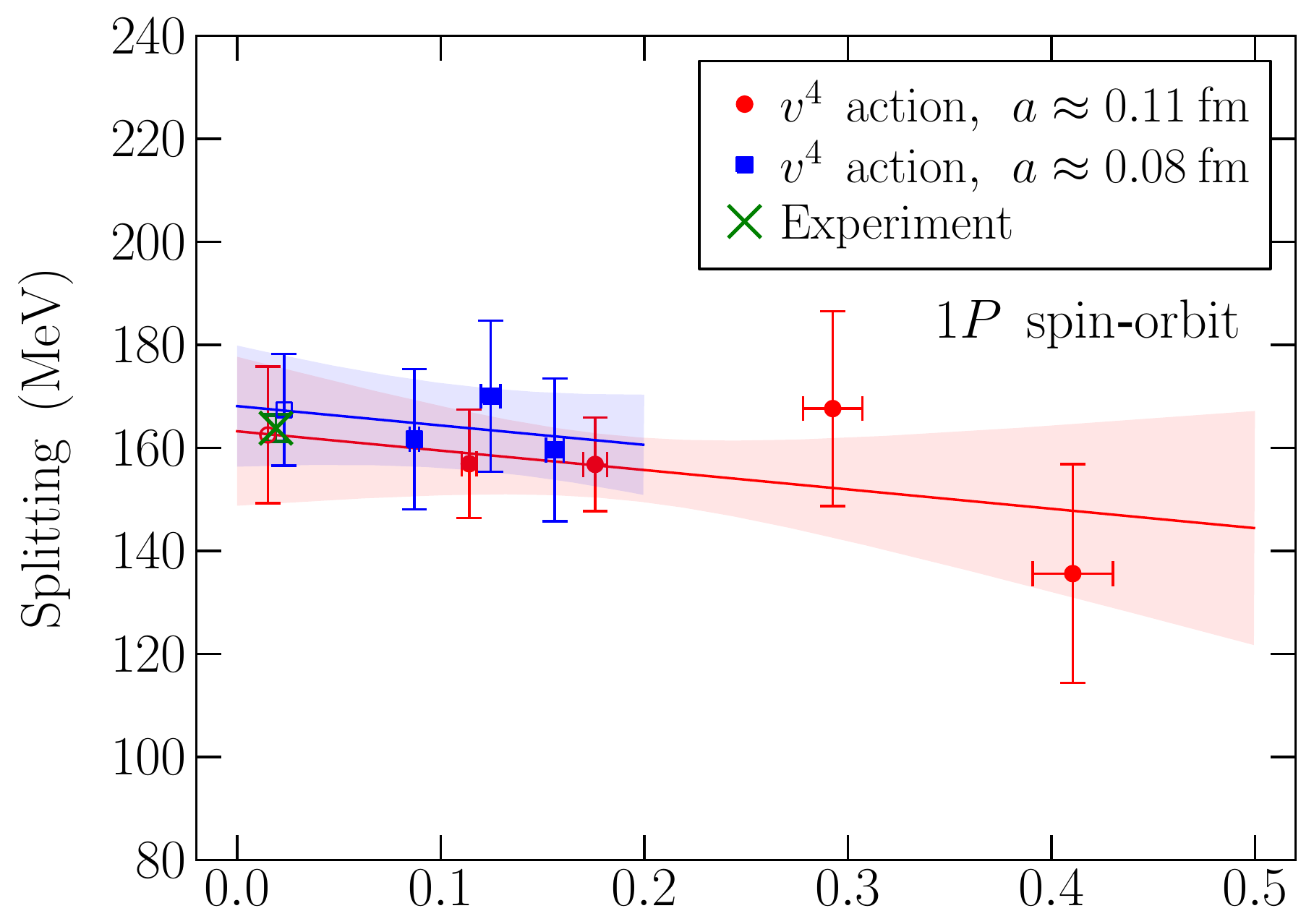}    \hfill \includegraphics[width=0.43\linewidth]{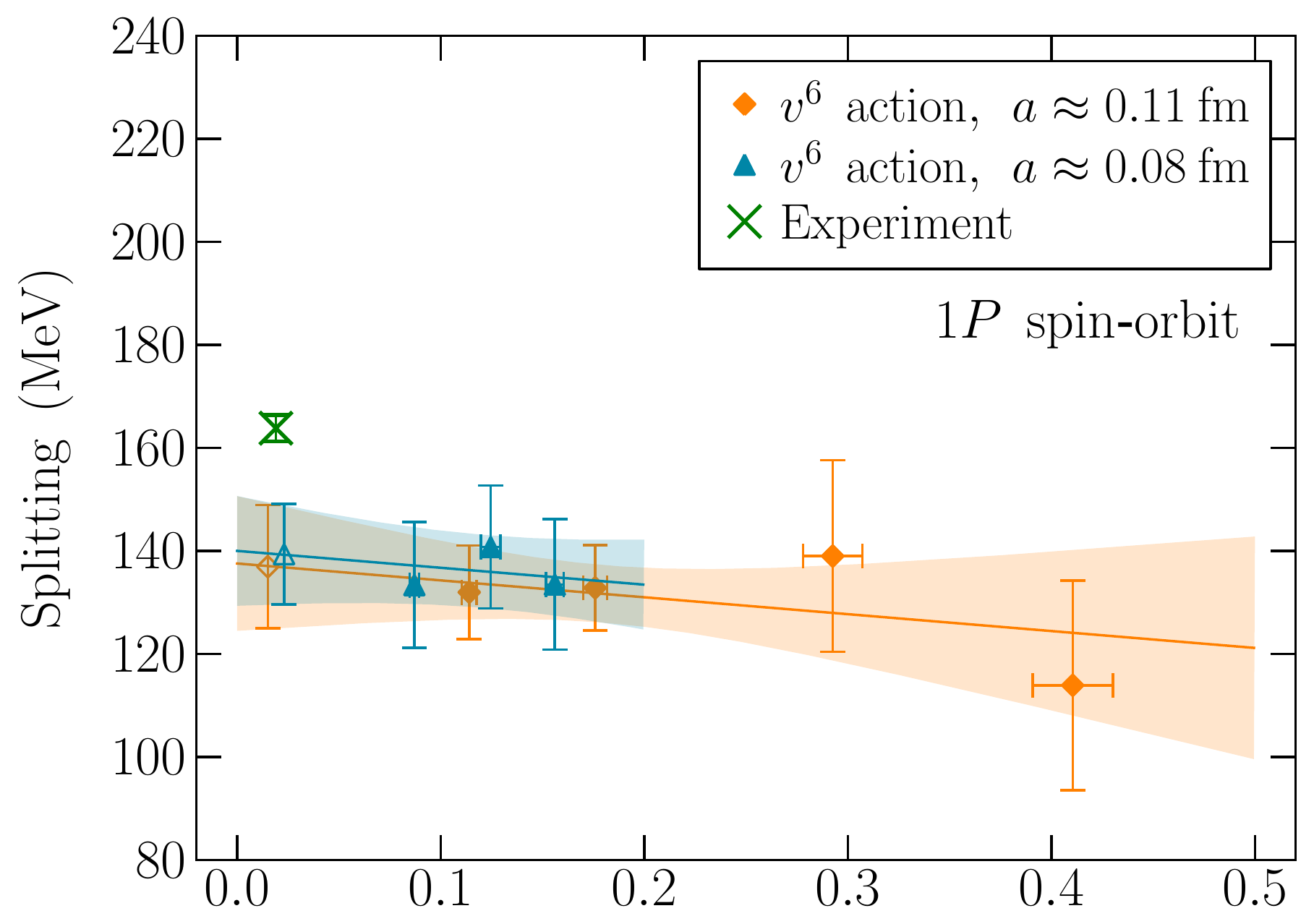}
 \includegraphics[width=0.43\linewidth]{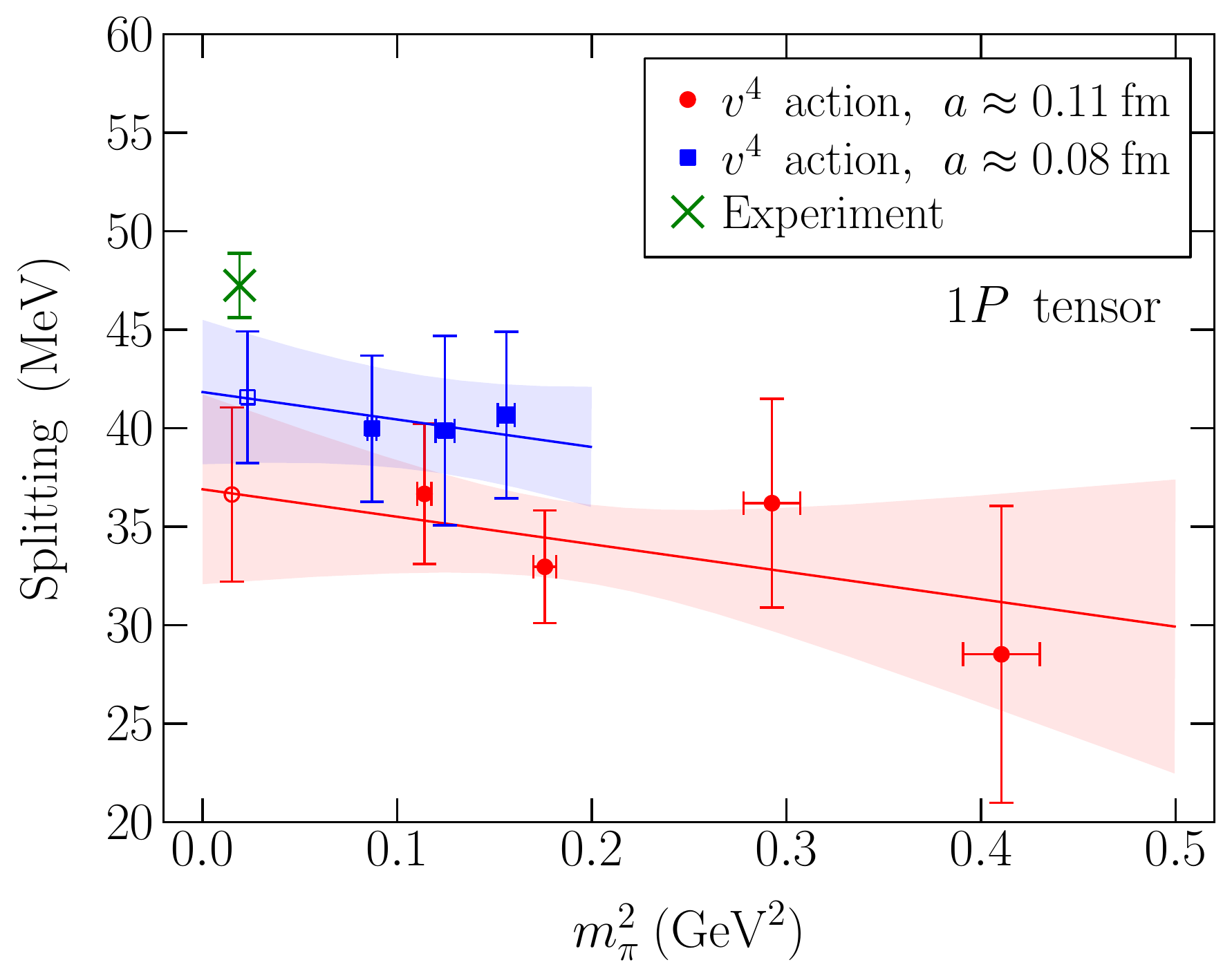}       \hfill \includegraphics[width=0.43\linewidth]{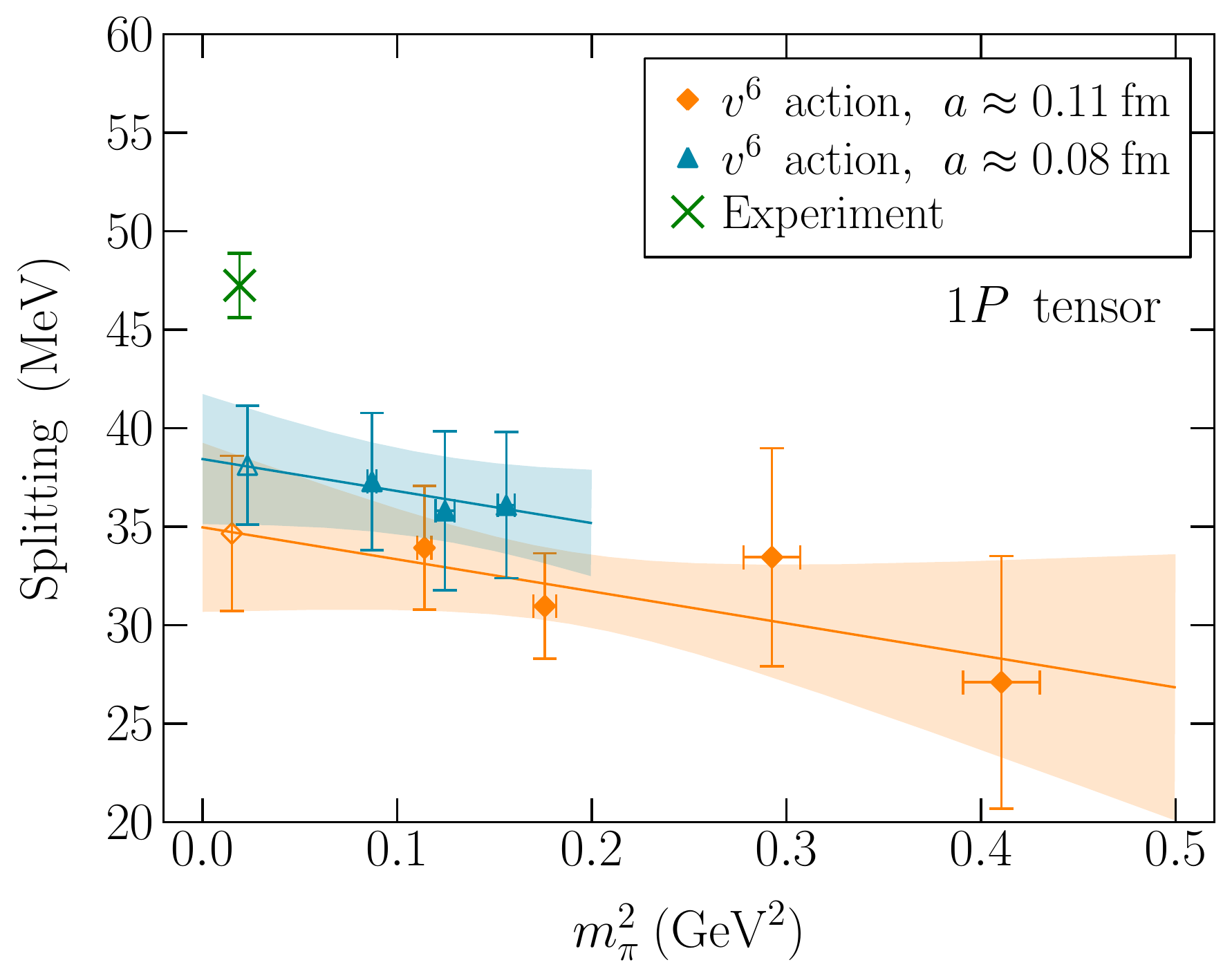}
\caption{\label{fig:spin_dep_chiral_extrap_pt1}Chiral extrapolation of spin-dependent splittings from the $L=24$ and $L=32$ ensembles, part I.
Extrapolated points are offset horizontally for legibility.}
\end{figure*}

\begin{figure*}[h!]
 \includegraphics[width=0.43\linewidth]{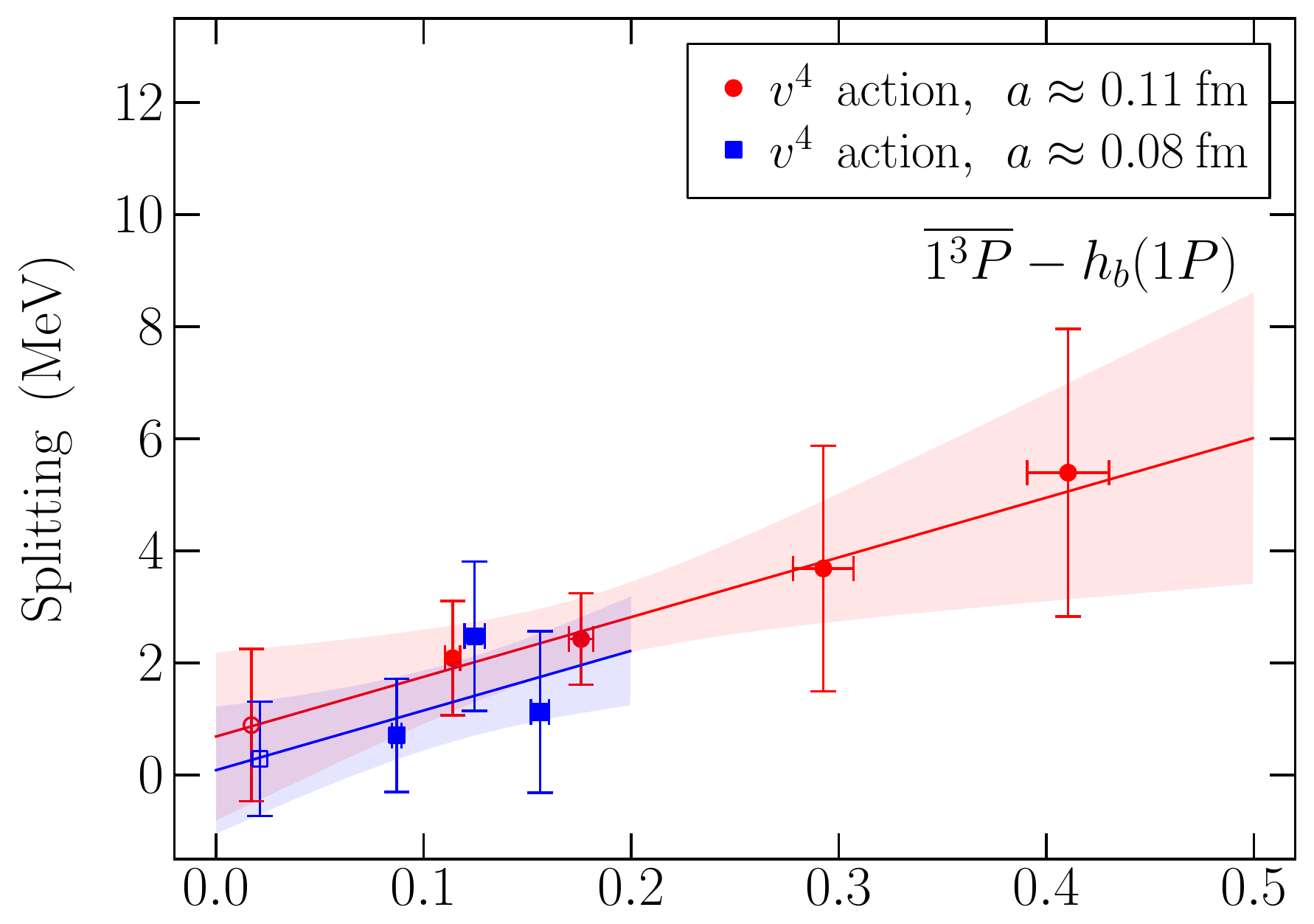}                  \hfill \includegraphics[width=0.43\linewidth]{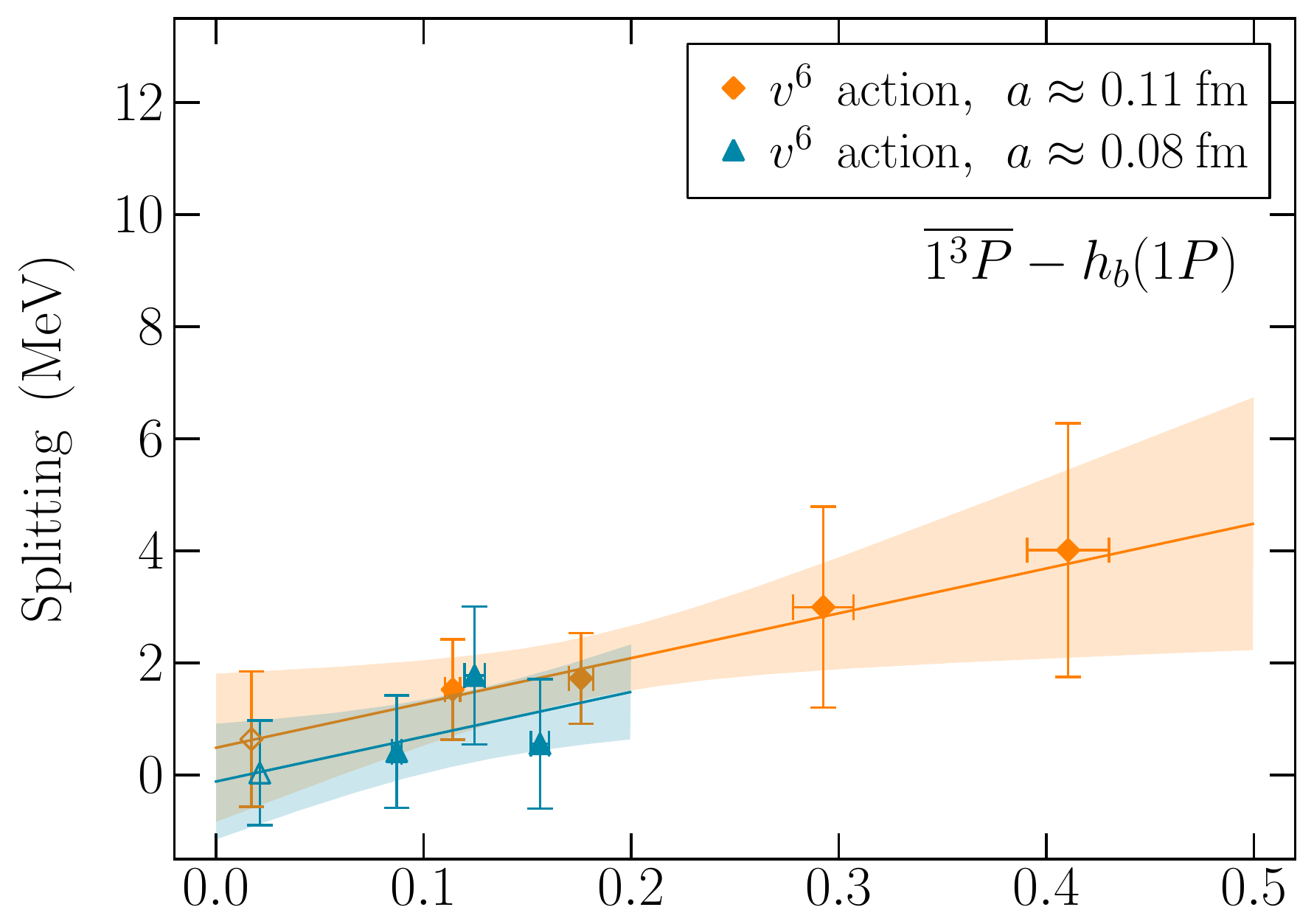}
 \includegraphics[width=0.43\linewidth]{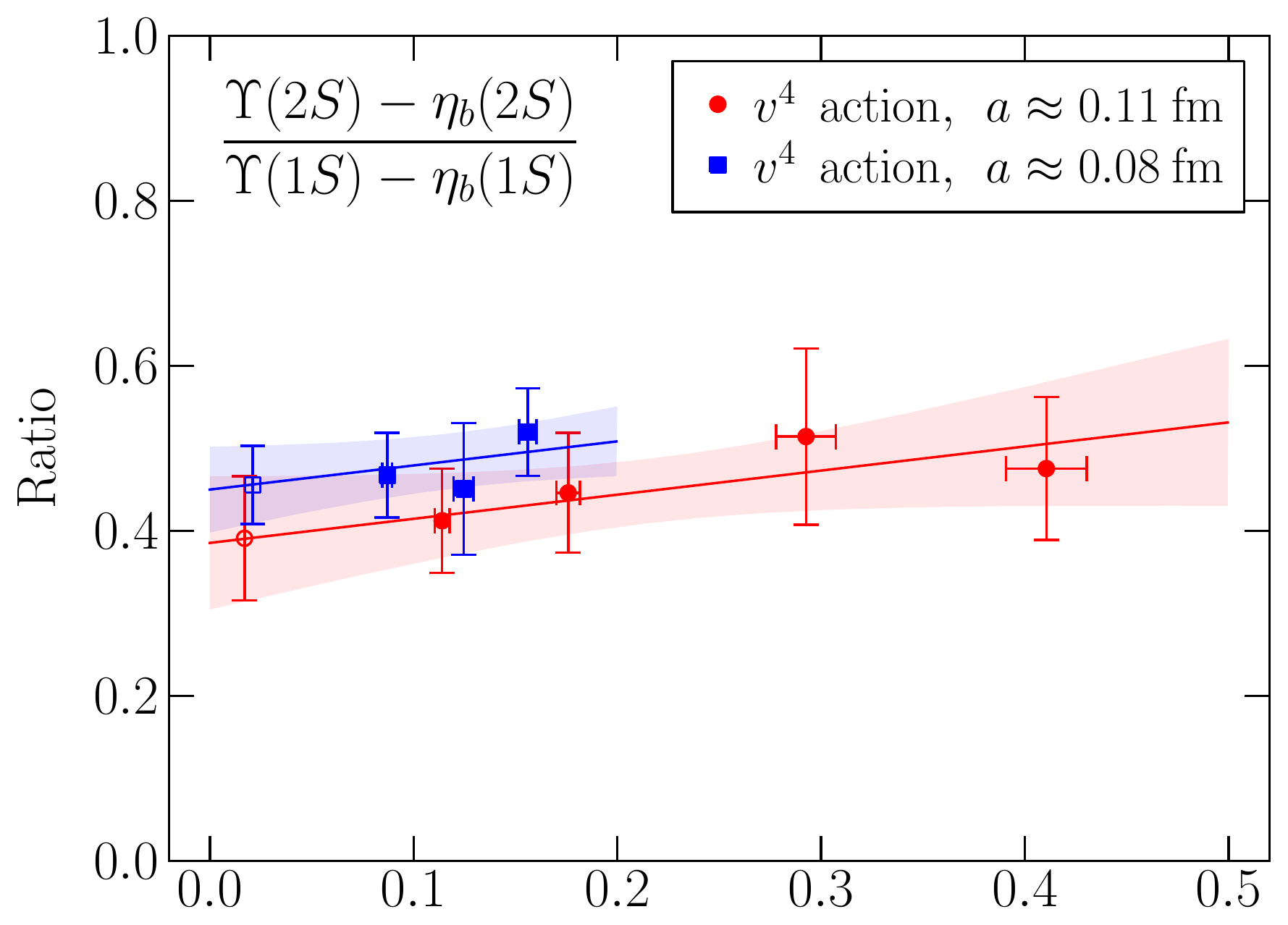}                    \hfill \includegraphics[width=0.43\linewidth]{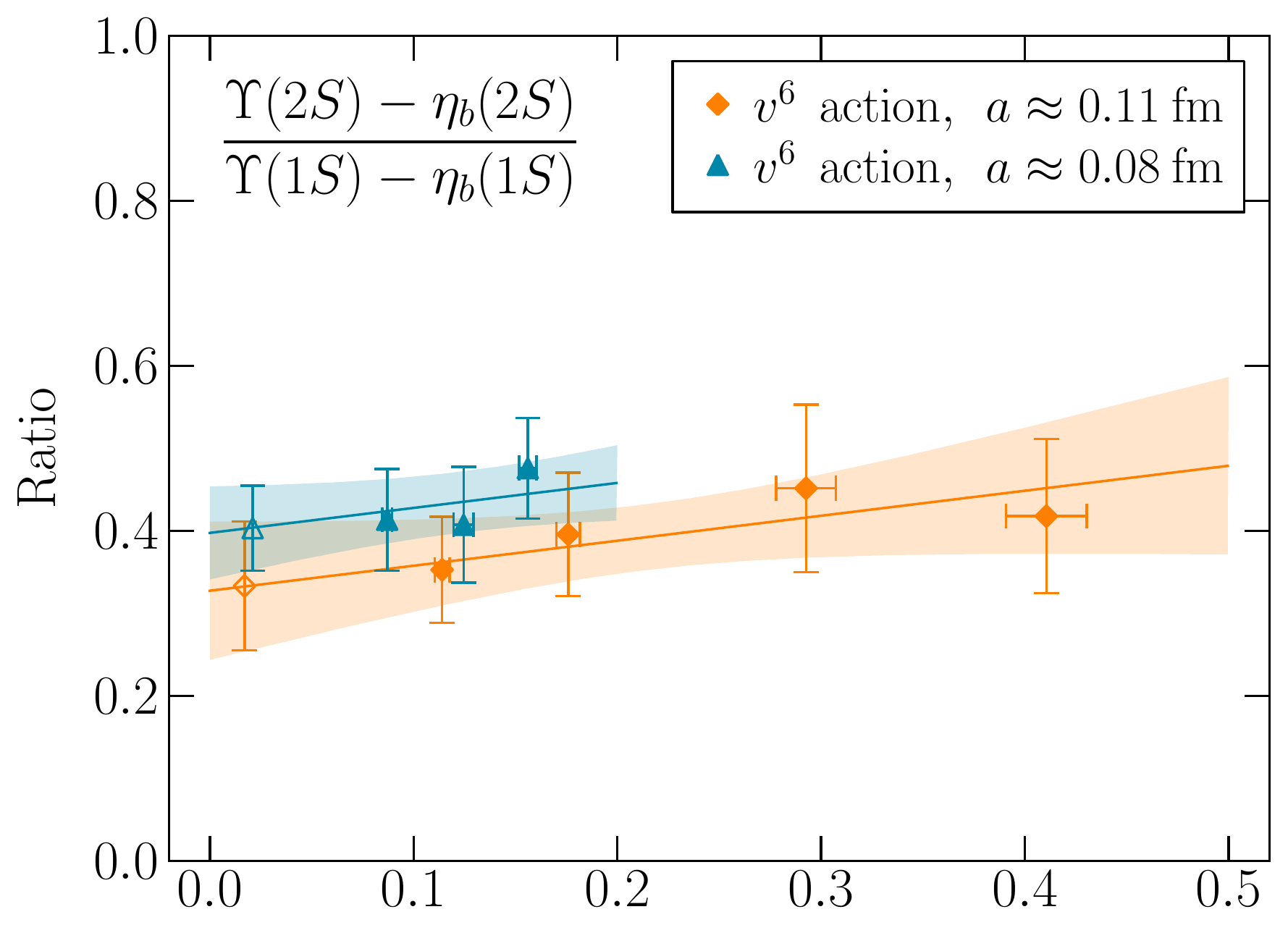}
 \includegraphics[width=0.43\linewidth]{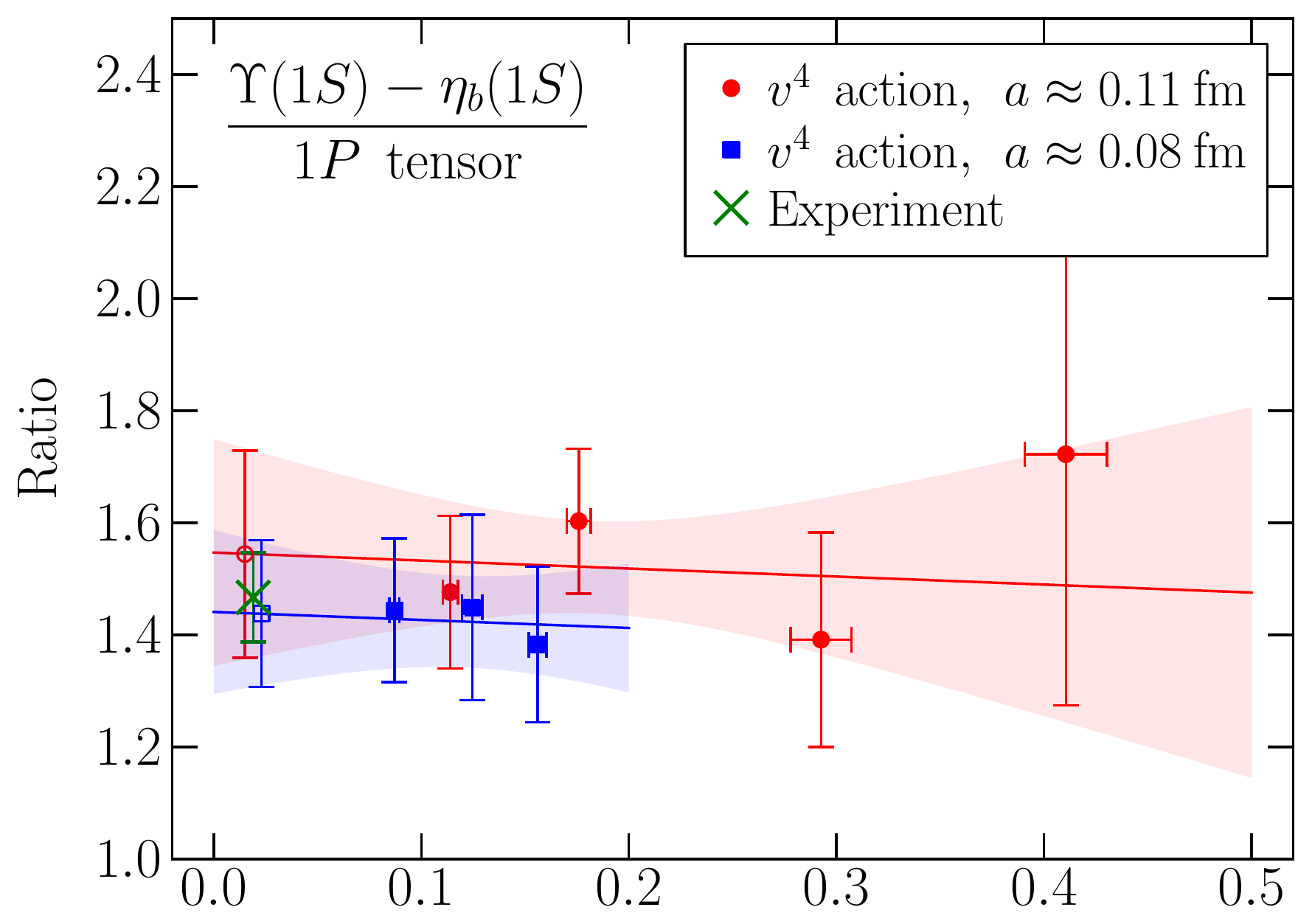} \hfill \includegraphics[width=0.43\linewidth]{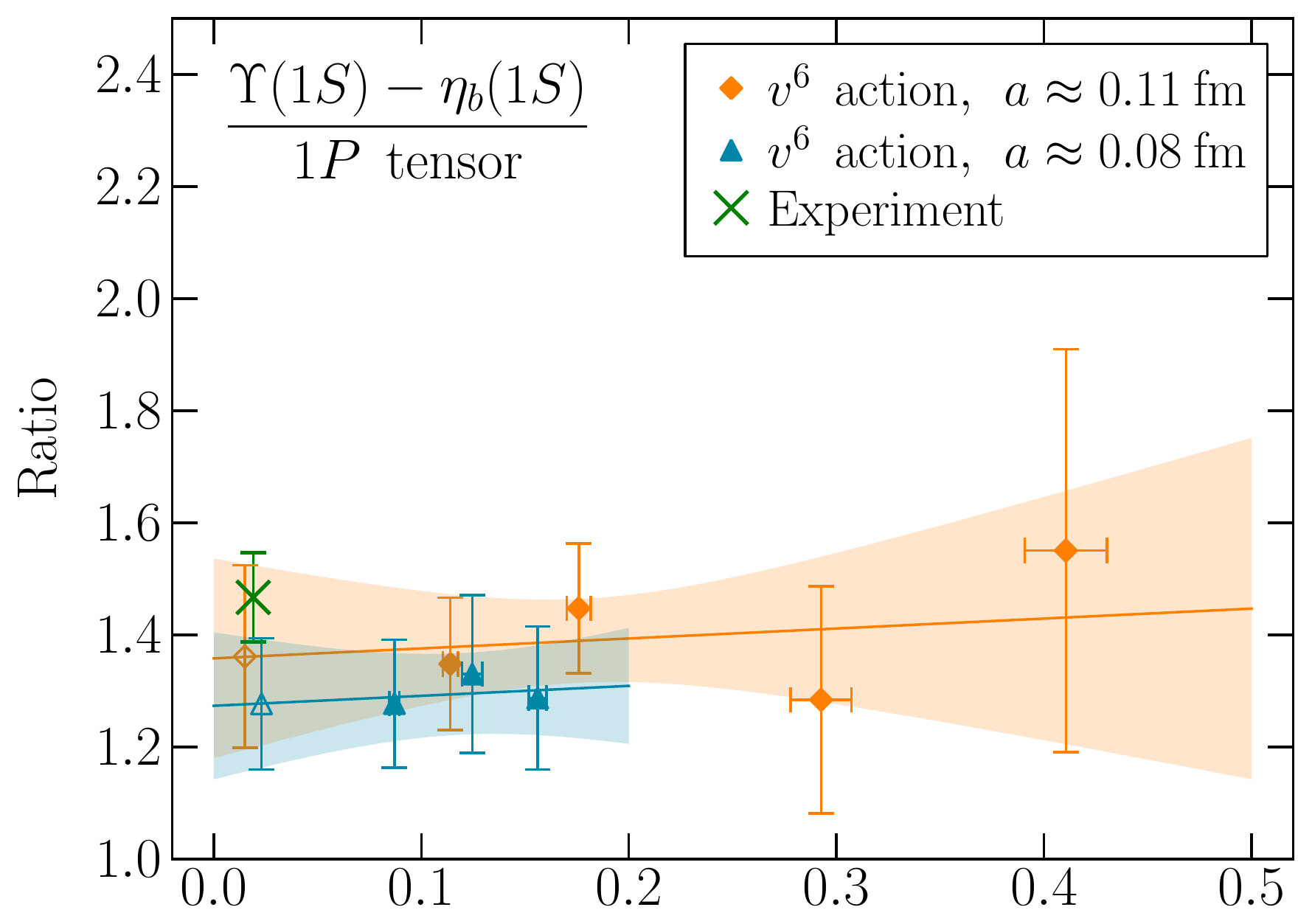}
 \includegraphics[width=0.43\linewidth]{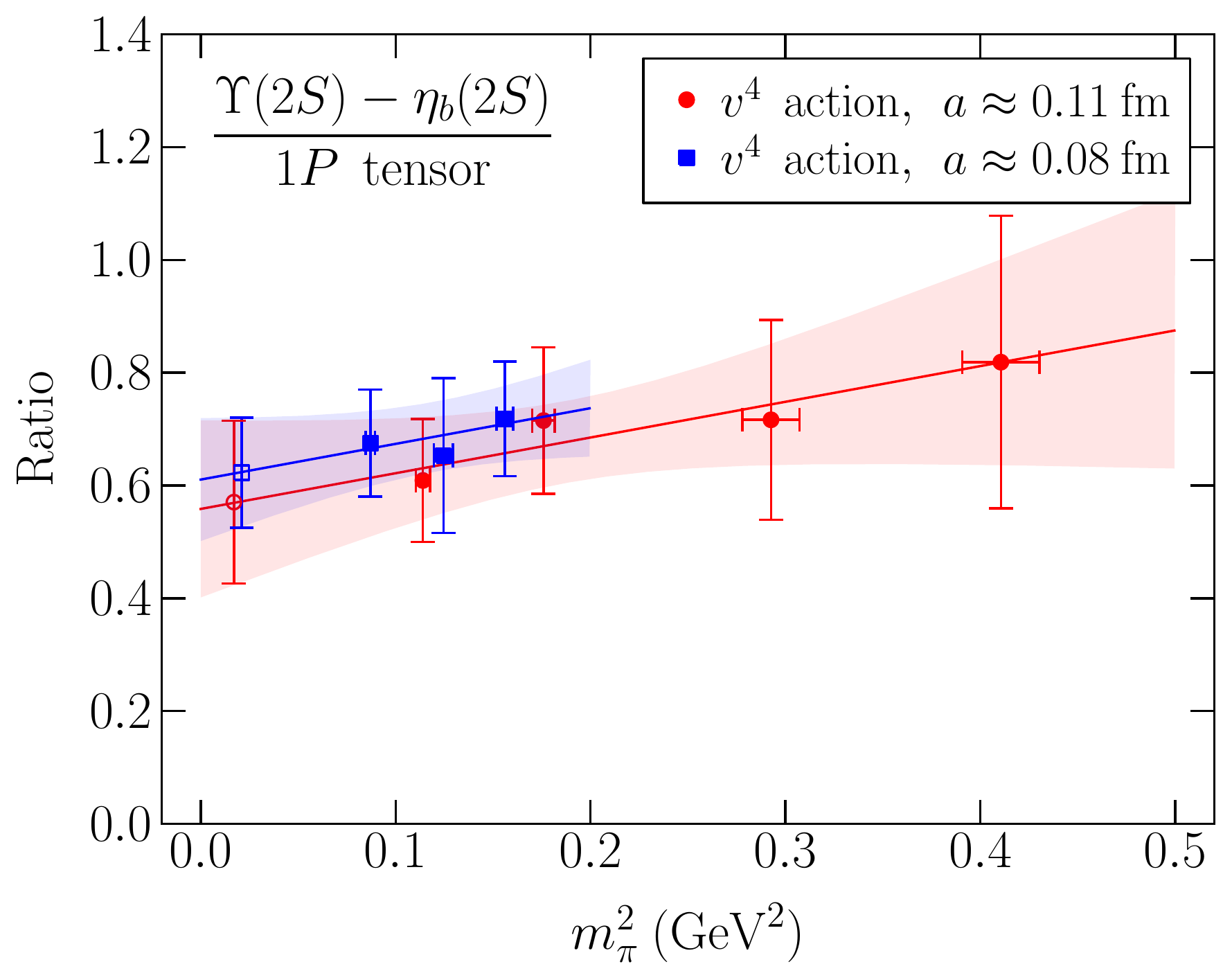} \hfill \includegraphics[width=0.43\linewidth]{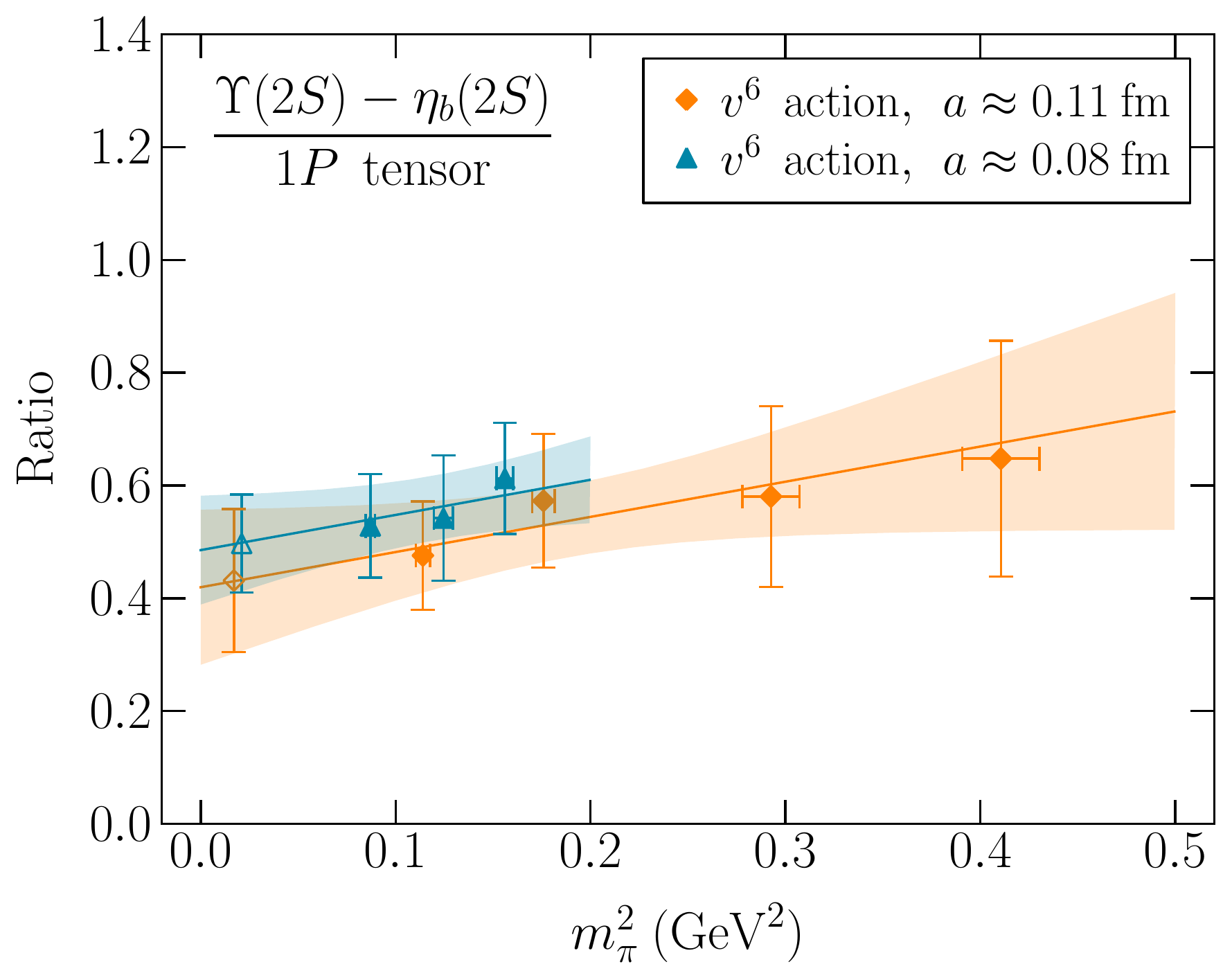}
\caption{\label{fig:spin_dep_chiral_extrap_pt2}Chiral extrapolation of spin-dependent splittings from the $L=24$ and $L=32$ ensembles, part II.
Extrapolated points are offset horizontally for legibility.}
\end{figure*}

Finally, simultaneous chiral extrapolations of the data at $a\approx0.11$ fm and $a\approx0.08$ fm to the physical pion mass,
using the functional form (\ref{eq:sim_chiral_extrap}), were performed (the data from the $v^4$ and $v^6$ actions
were treated independently).
These chiral extrapolations are visualized in Figs.~\ref{fig:spin_dep_chiral_extrap_pt1} and \ref{fig:spin_dep_chiral_extrap_pt2}.

\begin{figure*}[h!]
\vspace{4ex}
 \includegraphics[width=0.49\linewidth]{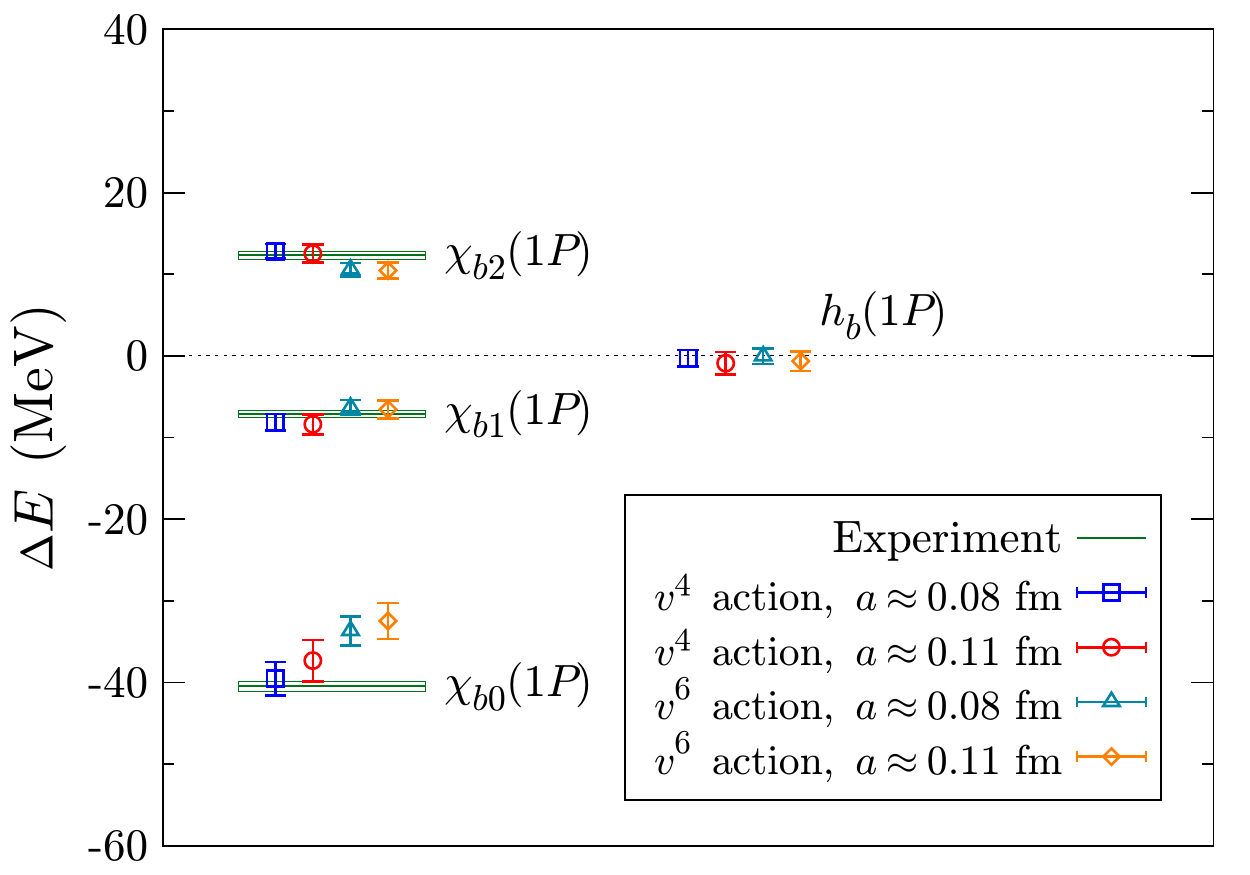}  \hfill \includegraphics[width=0.49\linewidth]{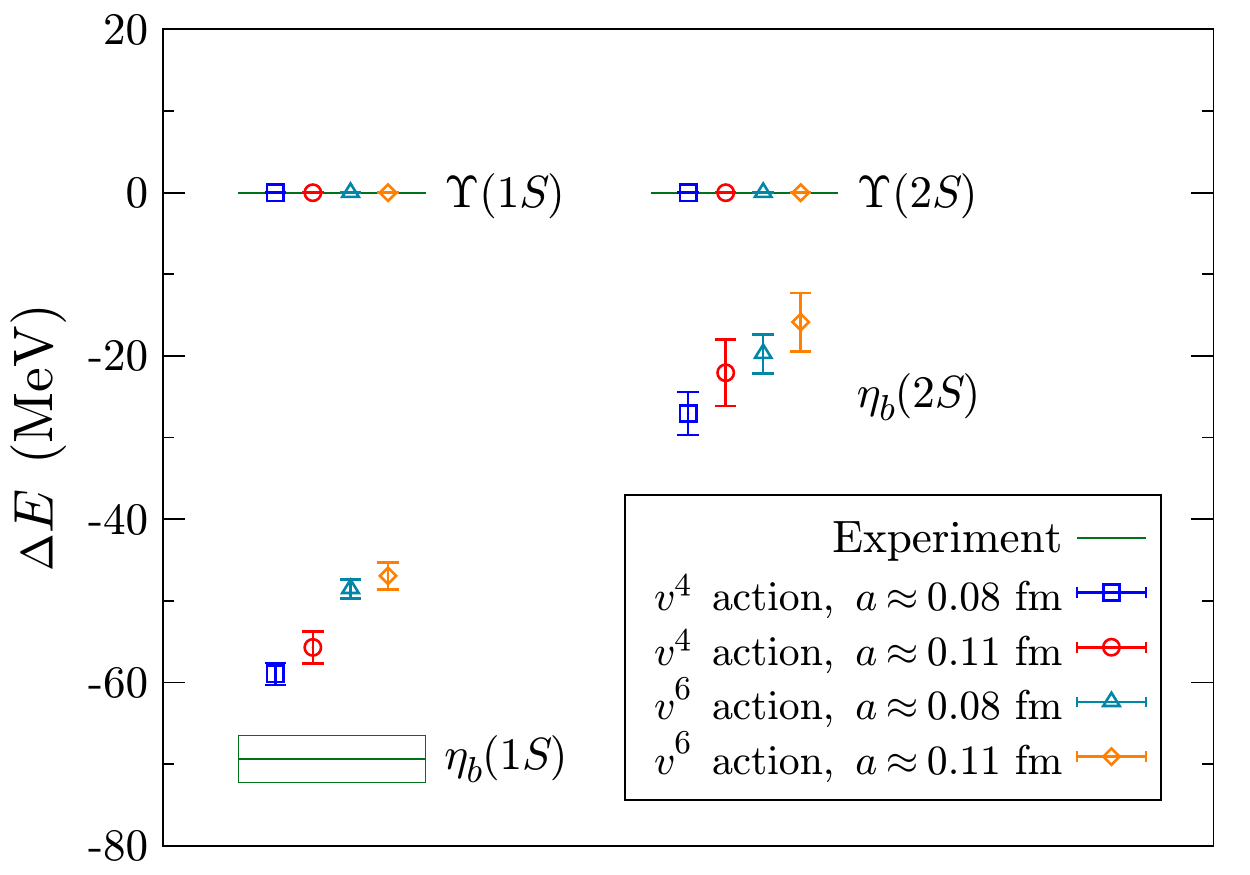}
\caption{\label{fig:spindep_splittings_final}Chirally extrapolated results for the spin-dependent energy splittings, from the $L=24$ and $L=32$ ensembles.
Left panel: $1P$ spin splittings (energies relative to $\overline{1^3P}$). Right panel: $S$-wave hyperfine splittings (energies relative to the $\Upsilon(1S)$ and $\Upsilon(2S)$ states,
respectively). The errors shown are statistical/fitting/scale setting only;  see Sec.~\ref{sec:spindep_errors} for a discussion of systematic errors and
Table \ref{tab:spindep_splittings_final_final}
for the final results that include estimates of the systematic errors.}
\end{figure*}

The numerical results for the spin-splittings at the physical pion mass are given in Table \ref{tab:spindep_splittings_final} and plotted
in Fig.~\ref{fig:spindep_splittings_final}. It can be seen that the results obtained with the $v^6$ and with the $v^4$ action differ
significantly. At $a\approx0.08$ fm, the $1S$ hyperfine and spin-orbit splitting are reduced by about 20\% by the $v^6$ terms, while
the $1P$ tensor splitting is reduced by 10\%. These changes are in line with the estimate of $v^2\approx 0.1$.

\clearpage

\begin{table*}[h!]
\begin{ruledtabular}
\begin{tabular}{llllll}
& \multicolumn{2}{c}{$v^4$ action} & \multicolumn{2}{c}{$v^6$ action} & \\
                                                                                               & $ a\approx 0.11$ fm &  $ a\approx 0.08$ fm  & $ a\approx 0.11$ fm  & $ a\approx 0.08$ fm  &  Experiment      \\
\hline
$\Upsilon(1S)-\eta_b(1S)$                                                                      &  $55.7(2.0)$  &  $59.0(1.4)$  &  $46.9(1.7)$  &  $48.5(1.1)$  &  $69.3(2.9)$    \\
$\Upsilon(2S)-\eta_b(2S)$                                                                      &  $22.1(4.1)$  &  $27.1(2.6)$  &  $15.8(3.6)$  &  $19.7(2.4)$  &  \hspace{4ex}-  \\
$\chi_{b2}(1P)-\chi_{b1}(1P)$                                                                  &  $20.9(2.3)$  &  $20.9(1.9)$  &  $17.0(2.0)$  &  $16.9(1.7)$  &  $19.43(57)$    \\
$\chi_{b1}(1P)-\chi_{b0}(1P)$                                                                  &  $28.9(2.1)$  &  $31.3(1.7)$  &  $26.0(1.8)$  &  $27.5(1.4)$  &  $33.34(66)$    \\
$\overline{1^3P}-h_b(1P)$                                                                      &  $0.9(1.4)$   &  $0.3(1.0)$   &  $0.6(1.2)$   &  $0.04(93)$   &  \hspace{4ex}-  \\
$-2\chi_{b0}(1P)-3\chi_{b1}(1P)+5\chi_{b2}(1P)$                                                &  $163(13)$    &  $167(11)$    &  $137(12)$    &  $139.4(9.8)$ &  $163.8(2.6)$   \\
$-2\chi_{b0}(1P)+3\chi_{b1}(1P)-\chi_{b2}(1P)$                                                 &  $36.6(4.4)$  &  $41.6(3.3)$  &  $34.7(3.9)$  &  $38.1(3.0)$  &  $47.3(1.6)$    \\
\\[-2ex]
$\displaystyle \frac{\Upsilon(2S)-\eta_b(2S)}{\Upsilon(1S)-\eta_b(1S)}$                        &  $0.391(75)$  &  $0.456(47)$  &  $0.333(78)$  &  $0.403(52)$  &  \hspace{4ex}-  \\
\\[-2ex]
$\displaystyle \frac{\Upsilon(1S)-\eta_b(1S)}{-2\chi_{b0}(1P)+3\chi_{b1}(1P)-\chi_{b2}(1P)}$   &  $1.54(18)$   &  $1.44(13)$   &  $1.36(16)$   &  $1.28(12)$   &  $ 1.467(79)$   \\
\\[-2ex]
$\displaystyle \frac{\Upsilon(2S)-\eta_b(2S)}{-2\chi_{b0}(1P)+3\chi_{b1}(1P)-\chi_{b2}(1P)}$   &  $0.57(14)$   &  $0.622(98)$  &  $0.43(13)$   &  $0.497(87)$  &  \hspace{4ex}-  \\
\end{tabular}
\end{ruledtabular}
\caption{\label{tab:spindep_splittings_final}
Spin-dependent energy splittings from the $L=24$ and $L=32$ ensembles, interpolated to the physical $b$ quark mass and extrapolated to the physical pion mass.
All results in MeV, except for the dimensionless ratios. For the lattice data, the errors shown here are statistical/fitting/scale setting only; see Sec.~\ref{sec:spindep_errors} for
a discussion of systematic errors and Table \ref{tab:spindep_splittings_final_final} for the final results that include estimates of the systematic errors.
The experimental value for $\Upsilon(1S)-\eta_b(1S)$ is the weighted average of the results from \cite{Aubert:2008vj}, \cite{Aubert:2009pz}, and \cite{Bonvicini:2009hs}.
All other experimental values are from the Particle Data Group \cite{Amsler:2008zzb}.}
\end{table*}

The dependence on the lattice spacing varies between the different quantities. For example, the $1S$ hyperfine splitting calculated with the $v^6$ action
increases by 3\% when going from $a\approx0.11$ fm to $a\approx0.08$ fm, but this change is only 0.8 standard deviations. With the $v^4$ action,
the lattice spacing dependence in the $1S$ hyperfine splitting appears to be stronger, about 6\% or 1.4 standard deviations.
This size of the $a$-dependence is in good agreement with the estimates of discretization errors obtained in
Sec.~\ref{sec:spindep_errors}. Note that the dependence on the lattice spacing in spin splittings calculated directly (as opposed to the ratios)
may be caused both by discretization errors and by the $a m_b$-dependence of the missing radiative corrections to $c_3$ and $c_4$.
When going from $a\approx0.11$ fm to $a\approx0.08$ fm, the $1P$ tensor splitting appears to change by about 10\%, but the effect
is less than 1 standard deviation. No $a$-dependence is seen in the $1P$ spin-orbit splitting. Recall that the estimates of discretization errors
obtained in Sec.~\ref{sec:spindep_errors} are indeed smaller by a factor of 2 in the spin-orbit splitting compared to the tensor and hyperfine splittings.
In addition, in potential models the spin-orbit splitting is not as sensitive to short distances as the hyperfine splitting.

The ratios of spin splittings calculated here also show no significant dependence on the lattice spacing: about 0.7 standard deviations in
the ratio of the $2S$ and $1S$ hyperfine splittings and about 0.5 standard deviations in the ratios of the $S$-wave hyperfine and $1P$ tensor splittings.

The most reliable results for the spin-dependent energy splittings obtained in this work, calculated with the $v^6$ action at $a\approx 0.08$ fm,
are summarized in Table \ref{tab:spindep_splittings_final_final}.
Here, estimates of the systematic errors based on the discussions in Sec.~\ref{sec:spindep_errors} and Sec.~\ref{sec:gluon_errors_NP} are given. The systematic
errors in the ratios of the hyperfine and tensor splittings, where the unknown radiative corrections to $c_3$ and $c_4$ cancel,
are dominated by discretization errors. The systematic error in the $\overline{1^3P}-h_b(1P)$ hyperfine splitting is dominated by the unknown radiative corrections of order $\alpha_s$.
However, the absolute systematic error in $\overline{1^3P}-h_b(1P)$ is only $0.2$ MeV (assuming $\alpha_s\approx0.2$), because the splitting
is found to be zero within the statistical error of about 1 MeV. The $\overline{1^3P}-h_b(1P)$ splitting vanishes in potential models because the wave function
at the origin is zero for $L\neq0$.

Using the lattice ratios of the $S$-wave hyperfine and the $1P$ tensor splitting, and the experimental result for the $1P$ tensor splitting \cite{Amsler:2008zzb},
the $1S$ and $2S$ hyperfine splittings can be calculated in MeV.
The $2S$ hyperfine splitting in MeV can be calculated alternatively from the lattice ratio of the $2S$ and $1S$ hyperfine splitting,
and the experimental value of the $1S$ hyperfine splitting
\cite{Aubert:2008vj, Aubert:2009pz, Bonvicini:2009hs}. The results obtained with both methods are shown in Table \ref{tab:spindep_splittings_final_final}.

\begin{table}[h!]
\begin{tabular}{lllll}
\hline\hline
                             & \hspace{4ex} &  This work                                                                   & \hspace{4ex}  &  Experiment                              \\
\hline
\\[-1.5ex]
$\displaystyle \frac{\Upsilon(2S)-\eta_b(2S)}{\Upsilon(1S)-\eta_b(1S)}$                        &&  $0.403(52)(27)$    && \null\hspace{4ex}- \\      
\\[-1.5ex]
$\displaystyle \frac{\Upsilon(1S)-\eta_b(1S)}{-2\chi_{b0}(1P)+3\chi_{b1}(1P)-\chi_{b2}(1P)}$   &&  $1.28(12)(10)$      && $ 1.467(79)$ \\
\\[-1.5ex]
$\displaystyle \frac{\Upsilon(2S)-\eta_b(2S)}{-2\chi_{b0}(1P)+3\chi_{b1}(1P)-\chi_{b2}(1P)}$   &&  $0.497(87)(44)$    && \null\hspace{4ex}- \\
\\[-1.5ex]
$\Upsilon(1S)-\eta_b(1S)$                   &&  $60.3(5.5)(5.0)(2.1)$ MeV \footnote{Using $1P$ tensor splitting from experiment}                &&  $69.3(2.9)$ MeV                       \\
\\[-1ex]
\multirow{2}{*}{$\Upsilon(2S)-\eta_b(2S)$}  &&  $23.5(4.1)(2.1)(0.8)$ MeV $^a$                                                                 &&  \multirow{2}{*}{\null\hspace{4ex}-}    \\
                                            &&  $28.0(3.6)(1.9)(1.2)$ MeV \footnote{Using $\Upsilon(1S)-\eta_b(1S)$ splitting from experiment}  &                                          \\
\\[-1ex]
$\overline{1^3P}-h_b(1P)$                                                                      &&  $0.04(93)(20)$ MeV    && \null\hspace{4ex}-  \\
\\[-1.5ex]
\hline\hline
\end{tabular}
\caption{\label{tab:spindep_splittings_final_final}
Final results for spin splittings, calculated with the $v^6$ action at $a\approx 0.08$ fm ($L=32$). For the lattice data,
the first error is statistical/fitting, the second error is an estimate of systematic uncertainties, and the third error (where given) is experimental.
The experimental value for the $1S$ hyperfine splitting is the weighted average of the results from \cite{Aubert:2008vj}, \cite{Aubert:2009pz}, and \cite{Bonvicini:2009hs};
the experimental value for the $1P$ tensor splitting is calculated using the $1^3P$ masses from the Particle Data Group \cite{Amsler:2008zzb}. }
\end{table}

Using the $\Upsilon(1S)$, $\Upsilon(2S)$ and $\overline{1^3P}$ masses from experiment \cite{Amsler:2008zzb}, the absolute masses of the $\eta_b(1S)$, $\eta_b(2S)$ and $h_b(1P)$
mesons can then be calculated. This gives
\begin{eqnarray}
\nonumber M[\eta_b(1S)]^{\phantom{a}}&=&9.4000(55)(50)(21)\:\:{\rm GeV},\\
\nonumber M[\eta_b(2S)]^a&=&9.9998(41)(21)(9)\:\:{\rm GeV},\\
\nonumber M[\eta_b(2S)]^b&=&9.9953(36)(19)(12)\:\:{\rm GeV},\\
 M[h_b(1S)]^{\phantom{a}} &=&9.89983(93)(20)(27)\:\:{\rm GeV}, \label{eq:predictions}
\end{eqnarray}
where the first error is statistical/fitting, the second error is systematic, and the third error is experimental.
For the $\eta_b(2S)$ mass, the results from both methods as discussed above are given.

\clearpage

\section{\label{sec:conclusions}Conclusions}

In this paper, a high-precision calculation of the bottomonium spectrum in lattice QCD with $2+1$ flavors
of dynamical light quarks was presented. One important improvement over \cite{Meinel:2009rd}
was the inclusion of a finer lattice spacing, giving better control of discretization errors.
The dependence of the results on the lattice spacing was seen to be weak, and consistent with the
estimates of NRQCD discretization errors based on power counting.
At $a\approx0.08$ fm, the radial and orbital energy splittings were found to be in excellent
agreement with experiment, within statistical errors as small as 1.3\% (see Fig.~\ref{fig:radialorbital_splittings_final}).
In addition, the square of the ``speed of light'', a quantity used on the lattice to measure deviations from
the relativistic continuum energy-momentum relationship, was found to be compatible with 1 within statistical errors smaller
than 0.4\% for bottomonium momenta up to 1.6 GeV (see Fig.~\ref{fig:speed_of_light}).
These results provide valuable tests of the lattice methods used here: NRQCD for the $b$ quarks,
the Iwasaki action for the gluons, and the domain wall action for the sea-quarks.
The discretization errors associated with the gluon action were studied further using a lattice potential
model based on tree-level perturbation theory, and by repeating the nonperturbative calculations on
MILC gauge field ensembles generated with the L\"uscher-Weisz action. These tests show that the Iwasaki action
works well at the lattice spacings considered here.

The focus of this work was the accurate calculation of spin splittings. To this end, ratios of
hyperfine and tensor splittings were calculated, in which the unknown radiative corrections to the leading
spin-dependent terms in the NRQCD action cancel. This cancellation was confirmed here directly through
numerical calculation of these ratios with different values of the spin-dependent couplings in the action.
Furthermore, systematic errors from relativistic corrections were reduced
by the inclusion of the spin-dependent order-$v^6$ terms in the NRQCD action. The results in
Table \ref{tab:spindep_splittings_final_final} are considerably more precise than those from
previous lattice computations. For example, the ratio of the $2S$ and $1S$ hyperfine splittings
is predicted here to be $0.403 \pm 0.052_{\rm\: stat } \pm 0.027_{\rm\: syst }$ (the result from \cite{Gray:2005ur}
is $0.5\pm0.3_{\rm\: stat }$). By the criterion of Ref.~\cite{Gregory:2009hq}, the results (\ref{eq:predictions})
for the $\eta_b(2S)$ mass obtained here are now the most accurate predictions of a gold-plated hadron mass from
lattice QCD to date. The prediction of the $h_b(1P)$ mass appears to be even
more accurate, but note that it is obtained by subtracting from the experimental result for the $\overline{1^3P}$ mass
a splitting that is zero within the statistical errors.

The result for the $1S$ hyperfine splitting obtained here is $60.3\pm5.5_{\rm\:stat } \pm5.0_{\rm\: syst } \pm2.1_{\rm\: exp }$ MeV.
This is consistent with the value of $54 \pm 12$ MeV calculated with the Fermilab method in \cite{Burch:2009az}. It
is also in excellent agreement with the prediction of 60 MeV obtained in \cite{Ebert:2002pp}
using a relativistic quark model, and with the result of $58\pm1$ MeV from \cite{Badalian:2009cx} for $n_f=3$.

The $1S$ hyperfine splitting calculated here is only about 1 standard deviation below the weighted average
of the experimental results from \cite{Aubert:2008vj, Aubert:2009pz, Bonvicini:2009hs}.
The splitting obtained here is larger than many results from perturbative QCD \cite{Recksiegel:2003fm, Kniehl:2003ap, Penin:2009wf}.
Penin argues in \cite{Penin:2009wf} using continuum perturbation theory (where in fact lattice perturbation theory should be used)
that the inclusion of radiative corrections in the NRQCD action could reduce the
lattice value for the hyperfine splitting by about 20 MeV and bring it in agreement with perturbative QCD.
This statement does not apply to the result obtained here, where the hyperfine splitting is calculated
from the ratio to the $P$-wave tensor splitting so that the radiative corrections cancel.
Interestingly, it is noted in \cite{Radford:2009qi} that the perturbative prediction for the $1S$ hyperfine splitting increases
significantly and becomes consistent with experiment when the delta function terms in the potential are not softened.

\vspace{10ex}

\begin{acknowledgments}
I would like to thank Abdou Abdel-Rehim, Christopher Aubin, Christine Davies,
Will Detmold, Ron Horgan, Peter Lepage, Kostas Orginos, Andr\'e Walker-Loud and Matt Wingate for useful discussions. I thank the RBC/UKQCD
and MILC collaborations for making their gauge field ensembles available. This work was supported by the U.S.~Department of Energy under
grant number {D}{E}-{S}{C00}01{784}. The computations were performed using resources at NERSC and Teragrid resources at NCSA (grant number TG-PHY080014N).
The Chroma software \cite{Edwards:2004sx} was used for the gauge transformations.
\end{acknowledgments}

\clearpage

\appendix

\section{\label{sec:autocorr}Analysis of autocorrelations}

In this work, bottomonium two-point functions were calculated for 32 different source locations spread evenly across
the lattice on each gauge field configuration in order to increase statistics.
The question is whether data from different source locations are statistically independent,
and also whether the data from successive (in molecular dynamics time) gauge field configurations are statistically independent.
Possible autocorrelations in the data can be reduced by binning, that is, by averaging the data
within blocks of some size $B$ prior to the further statistical analysis.
Increases in the statistical errors in an observable under binning of the data would indicate the presence of autocorrelations.

\begin{figure*}[h!]
 \includegraphics[width=0.45\linewidth]{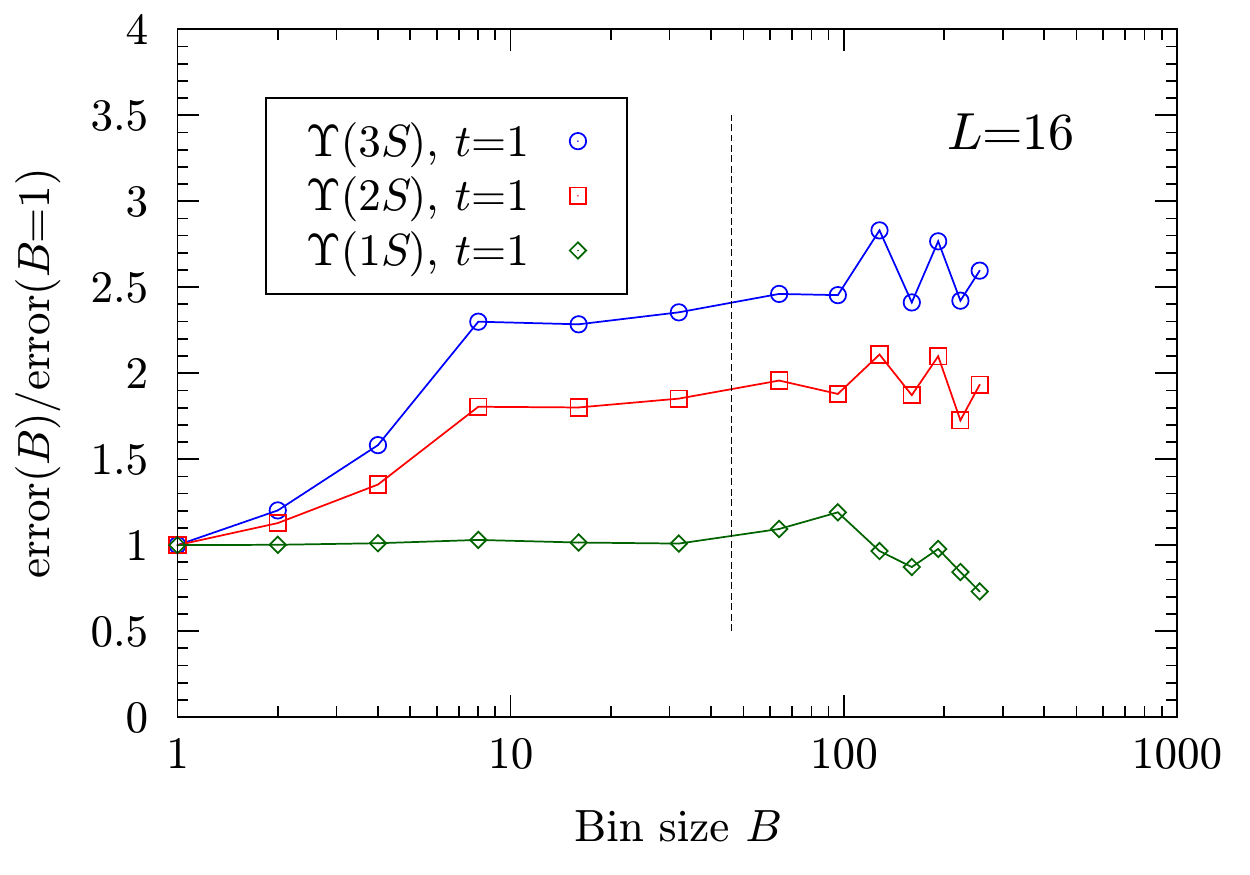} \hfill \includegraphics[width=0.45\linewidth]{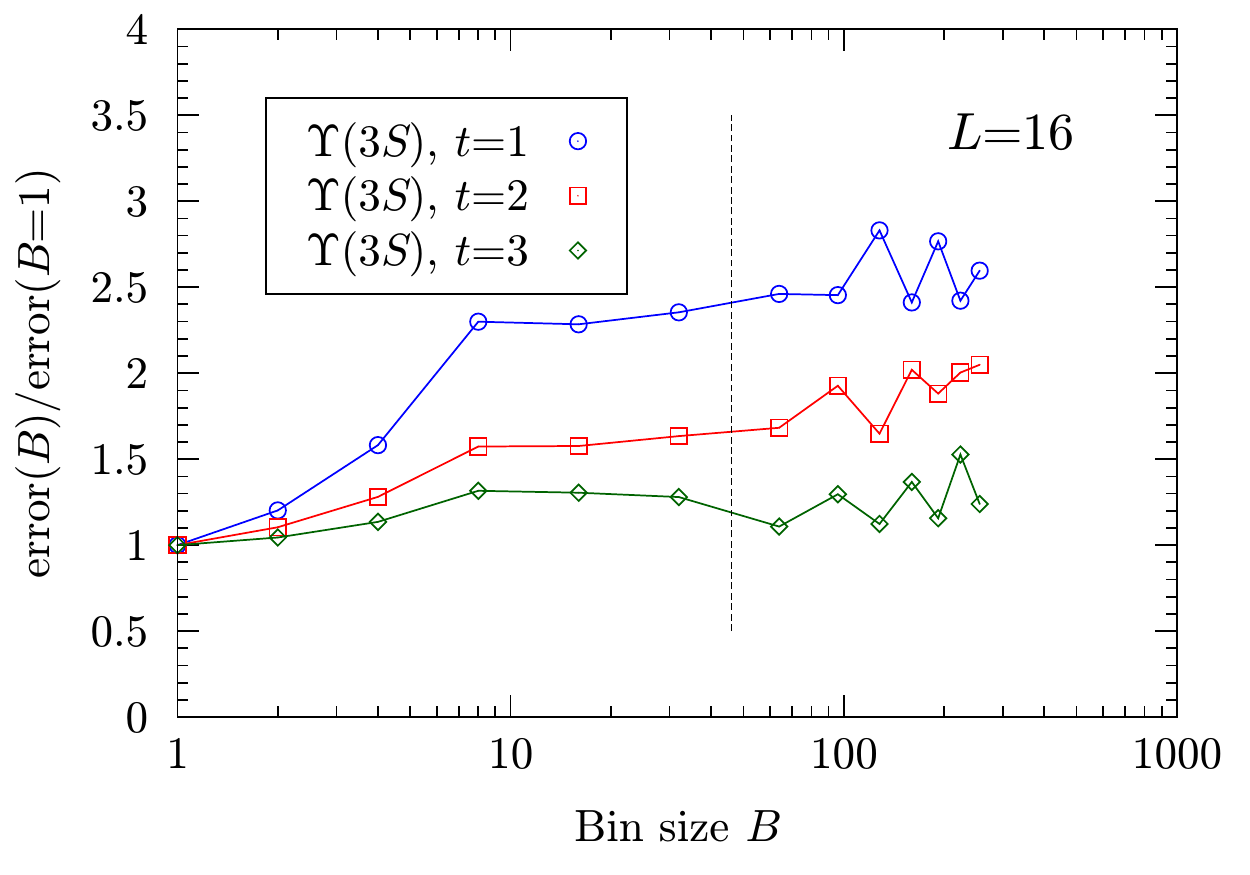}
 \includegraphics[width=0.45\linewidth]{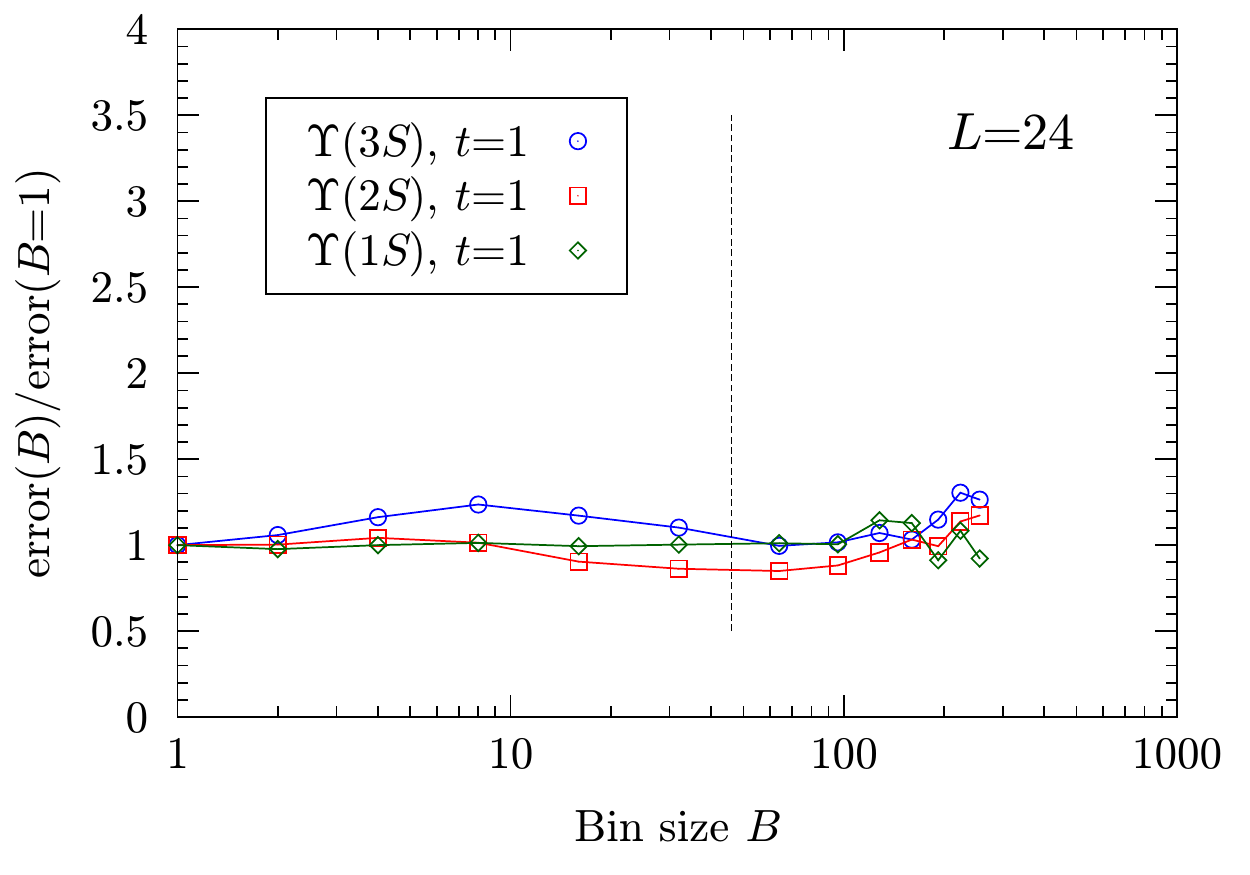} \hfill \includegraphics[width=0.45\linewidth]{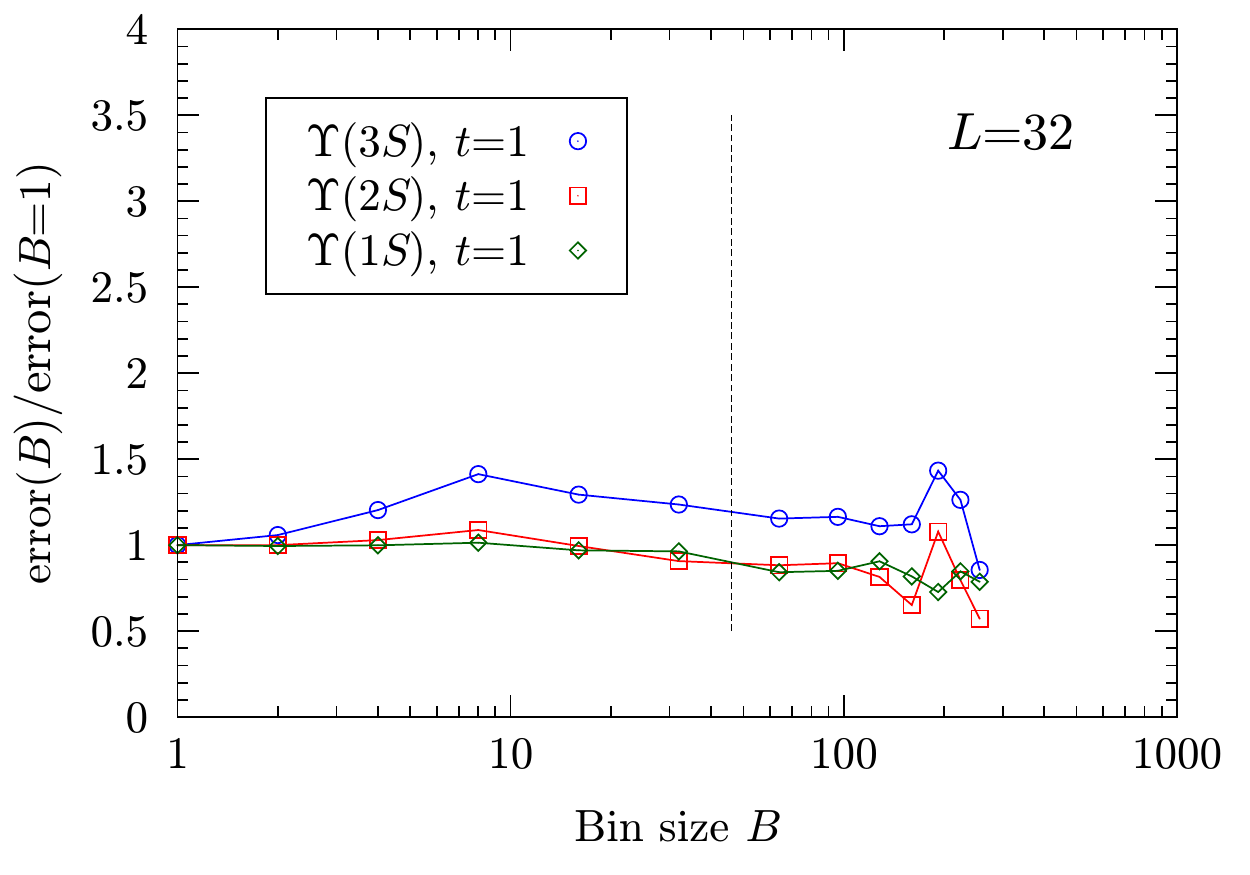}
\caption{\label{fig:autocorr}Analysis of autocorrelations in the two-point functions of the $\Upsilon(1S)$, $\Upsilon(2S)$ and $\Upsilon(3S)$ interpolating fields.
The data are from the $L=16$ ensemble with $a m_l=0.01$, the $L=24$ ensemble with $ a m_l=0.005$, and the $L=32$ ensemble with $a m_l=0.004$.}
\end{figure*}

Performing the binning analysis for the energies obtained from the matrix fits
used here with their large number of degrees of freedom is problematic due to
spurious finite-sample-size effects for the estimates of the data correlation matrix
when the bin size becomes too large. Therefore, in the following analysis, the statistical errors in the two-point functions
themselves are considered instead. Figure \ref{fig:autocorr} shows the errors in the diagonal
$\Upsilon(1S)$, $\Upsilon(2S)$ and $\Upsilon(3S)$ two-point functions for a given source-sink separation $t$,
versus the bin size $B$ (all errors relative to the corresponding error at $B=1$).
To the left of the vertical dashed lines in the graphs, binning is performed over neighboring source locations
only ($B=1, 2, 4,..., 32$). To the right of the vertical dashed lines, binning is performed over neighboring
gauge configurations ($B=64, 96, 128, ..., 256$). Here, ``neighboring'' gauge configurations are separated by the
step sizes given in Table \ref{tab:lattices}. As can be seen in Fig.~\ref{fig:autocorr}, for the $L=16$ ensemble,
which has a box size of about $1.8$ fm, significant autocorrelations between the data from
the different source locations can be seen at short time separations for the interpolating fields optimized for the
excited states ($\Upsilon(2S)$ and $\Upsilon(3S)$). The stronger autocorrelations
for excited states compared to ground states can be explained by the larger physical size of the excited states.
No significant autocorrelations are seen in molecular dynamics time. Note that the source locations were always shifted randomly from
configuration to configuration in this work. For the $L=24$ and $L=32$ ensembles, which have a box size of about
$2.7$ fm, no significant autocorrelations are seen either between source locations or in molecular dynamics time.
The same qualitative behavior was found for other bottomonium two-point functions.

\clearpage

\section{\label{sec:mb_tuning}Tuning of the bare $b$ quark mass}

\begin{figure*}[h!]
 \includegraphics[width=0.46\linewidth]{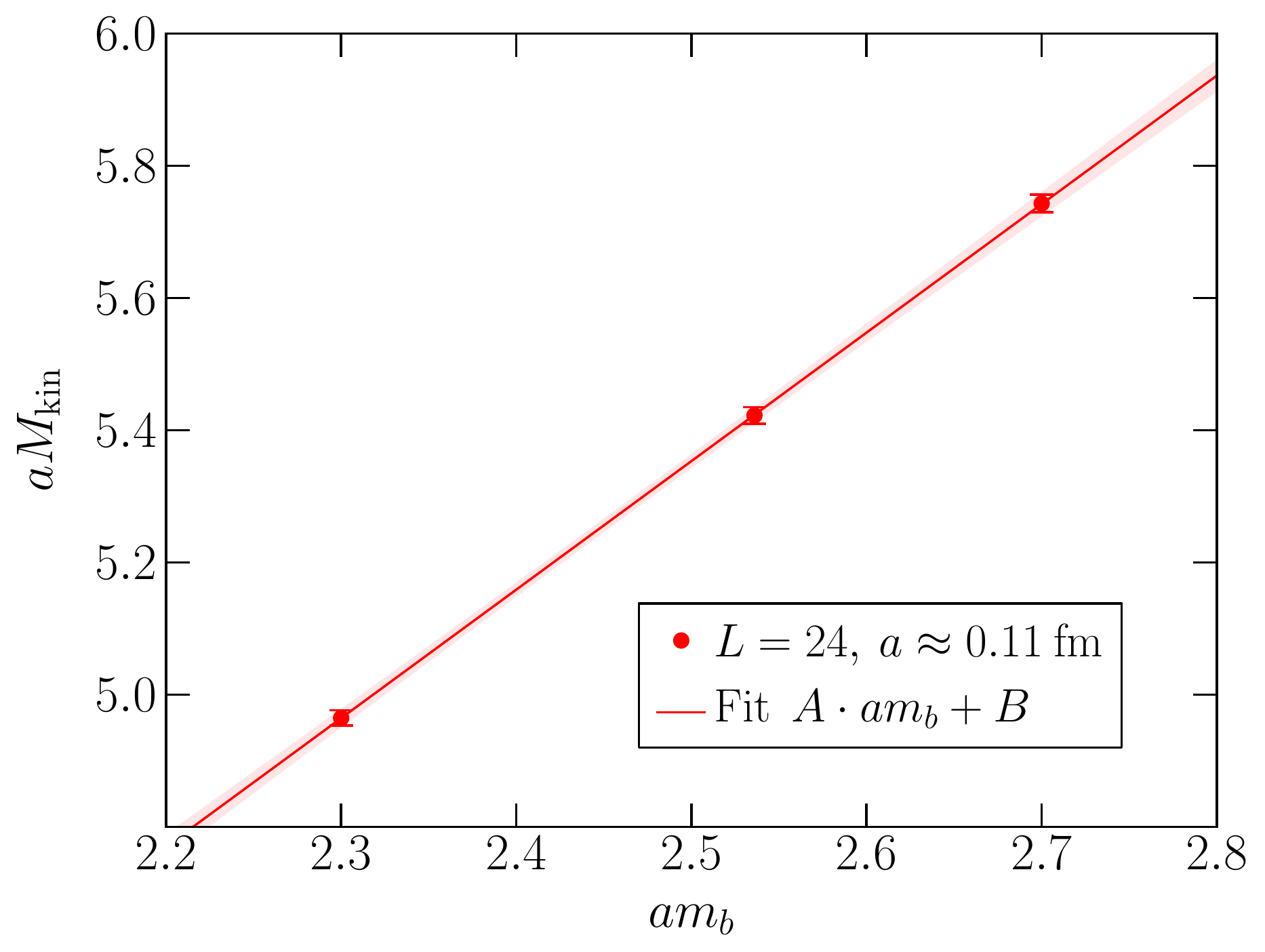}  \hfill \includegraphics[width=0.46\linewidth]{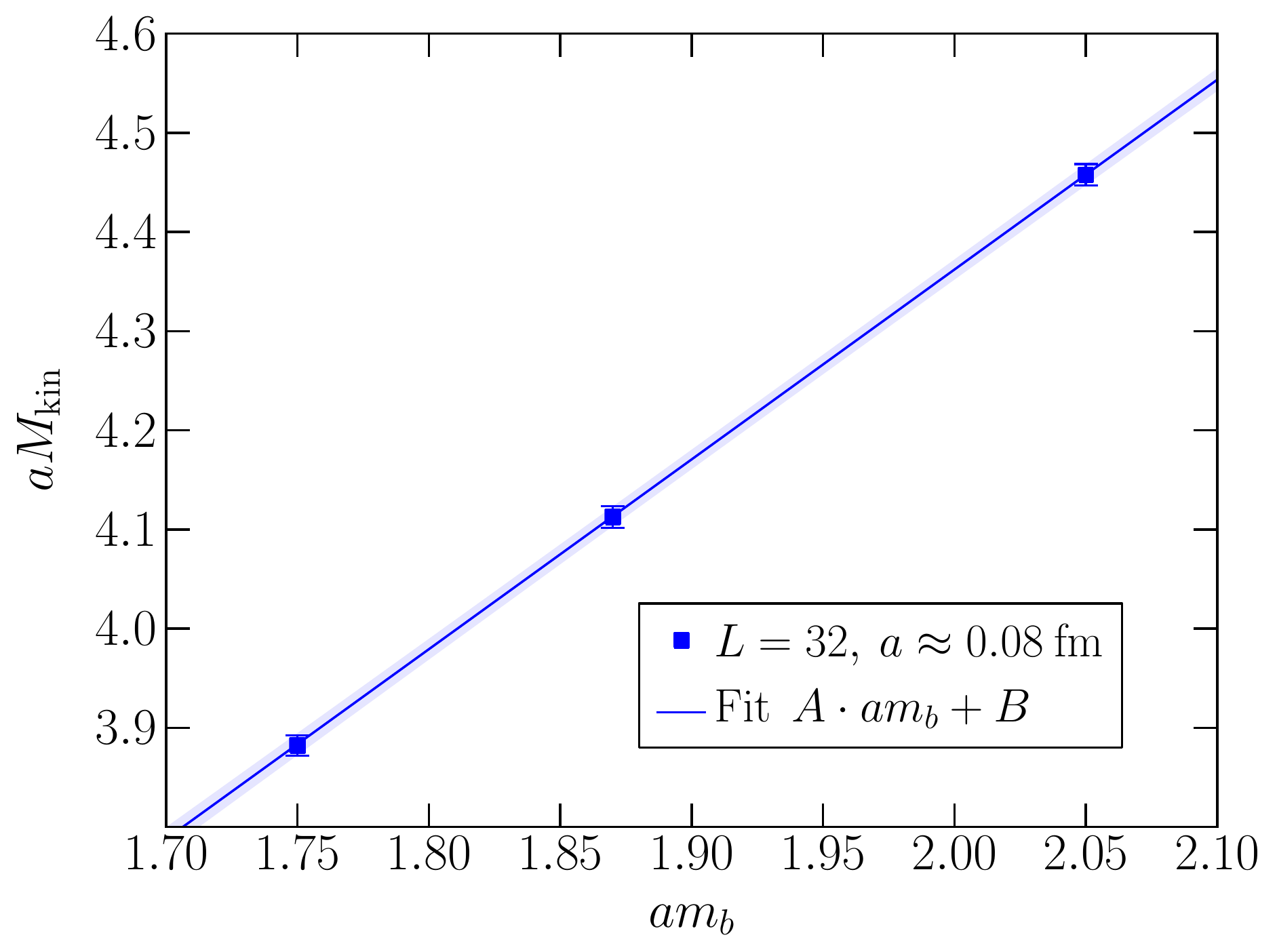}
\caption{\label{fig:Mkin_vs_mb}Kinetic mass of the $\eta_b(1S)$ meson plotted as a function of the bare heavy quark mass (both in lattice units).
The lines and error bands are from correlated fits using the functional form $aM_{\rm kin} = A \cdot a m_b + B$.
Left panel: $L=24$, $a m_l=0.005$, right panel:
$L=32$, $a m_l=0.004$.}
\end{figure*}

Results for the kinetic mass of the $\eta_b(1S)$ meson in lattice units, calculated from Eq.~(\ref{eq:mkin}) with
the smallest possible magnitude of the lattice momentum $|\bm{p}|=1\cdot 2\pi /(aL)$, are given in Tables \ref{tab:Mkin_L24_mb_dep}
and \ref{tab:Mkin_L32_mb_dep} for the $L=24$ and $L=32$ ensembles, respectively. Note that $aM_{\rm kin}$ is nearly independent
of $\bm{p}$, as demonstrated by the calculation of the ``speed of light'' in Sec.~\ref{sec:speed_of_light}.
As can be seen in Fig.~\ref{fig:Mkin_vs_mb}, the dependence of $a M_{\rm kin}$ on $a m_b$
is consistent with the linear function $aM_{\rm kin} = A \cdot a m_b + B$ in the ranges considered.
The fit results for $A$ and $B$ are also shown in the tables. Using these fit results,
and the lattice spacings from the $\Upsilon(2S)-\Upsilon(1S)$ splitting given in
Tables \ref{tab:lattice_spacing_L24} and \ref{tab:lattice_spacing_L32},
the ``physical'' values of the bare $b$ quark mass $a m_b^{\rm (phys.)}$ were then determined
such that $M_{\rm kin}$ agrees with the experimental value of 9.3910(29) GeV
\cite{Aubert:2008vj, Aubert:2009pz, Bonvicini:2009hs}.

Note that using the spin-averaged $\overline{1S}$ kinetic mass instead of the $\eta_b(1S)$ kinetic mass for the
tuning gives values of $a m_b^{\rm (phys.)}$ that are about 1\% larger.
The resulting shifts in the spin splittings (in physical units) would be smaller than the statistical
errors obtained here. For the ratios of spin splittings the shifts would be only about 0.05 standard
deviations, which is negligible.

The tuning of the $b$-quark mass was performed here only for the $v^4$ action, and the values of
$a m_b^{\rm (phys.)}$ obtained with this action were then used also for the calculations with the $v^6$ action.
The $v^6$ action employed here does not include the spin-independent order-$v^6$ terms, and therefore a calculation of
$M_{\rm kin}$ with this action would not be complete to this order. Ignoring this issue and doing the tuning for the $v^6$ action
was found to increase $a m_b^{\rm (phys.)}$ by about 2\% relative to the values obtained for the $v^4$ action using the $\overline{1S}$ kinetic mass.
Again, the effect of this shift on the ratios of spin splittings would be negligible.
For the $v^6$ action the values of $a m_b^{\rm (phys.)}$ calculated using the kinetic masses of the
$\eta_b(1S)$, the spin average $\overline{1S}$, and the $\Upsilon(1S)$ were found to be in agreement.

\begin{table*}[h!]
\begin{ruledtabular}
\begin{tabular}{llllcl}
               &  $am_b=2.3$   & $am_b=2.536$  & $am_b=2.7$ & Fit result & $a m_b^{\rm (phys.)}$\\
\hline
$a m_l=0.005$  & $4.965(11)$   &  $5.414(22)$  &  $5.743(13)$  &   $A=1.944(12)$, $B=0.494(27)$     & $2.487(39)$ \\
$a m_l=0.01$   & $4.986(13)$   &  $5.447(15)$  &  $5.768(17)$  &   $A=1.958(18)$, $B=0.483(44)$     & $2.522(42)$ \\
$a m_l=0.02$   & $4.979(45)$   &  $5.443(50)$  &  $5.763(54)$  &   $A=1.967(56)$, $B=0.46(12)\nb$   & $2.622(70)$ \\
$a m_l=0.03$   & $4.933(29)$   &  $5.391(31)$  &  $5.712(32)$  &   $A=1.947(38)$, $B=0.454(95)$     & $2.691(66)$ \\
\end{tabular}
\end{ruledtabular}
\caption{\label{tab:Mkin_L24_mb_dep}Heavy-quark mass dependence of the kinetic mass of the $\eta_b(1S)$ meson in lattice units,
calculated with the $v^4$ action on the $L=24$ ensembles. Also shown are the results of correlated fits using the functional form $aM_{\rm kin} = A \cdot a m_b + B$, and the
value of $a m_b$ that would yield agreement of the $\eta_b(1S)$ kinetic mass with experiment.}
\end{table*}

\begin{table*}[h!]
\begin{ruledtabular}
\begin{tabular}{llllcl}
                    &  $am_b=1.75$   & $am_b=1.87$  & $am_b=2.05$ & Fit result & $a m_b^{\rm (phys.)}$\\
\hline
$a m_l=0.004$       & $3.882(10)$    &  $4.112(11)$   &  $4.458(11)$   &   $A=1.914(16)$, $B=0.534(29)$  & $1.831(25)$ \\
$a m_l=0.006$       & $3.8823(85)$   &  $4.1118(93)$  &  $4.4560(98)$  &   $A=1.912(10)$, $B=0.536(19)$  & $1.829(36)$ \\
$a m_l=0.008$       & $3.889(13)$    &  $4.122(14)$   &  $4.468(15)$   &   $A=1.927(22)$, $B=0.518(40)$  & $1.864(27)$ \\
\end{tabular}
\end{ruledtabular}
\caption{\label{tab:Mkin_L32_mb_dep}Heavy-quark mass dependence of the kinetic mass of the $\eta_b(1S)$ meson in lattice units,
calculated with the $v^4$ action on the $L=32$ ensembles. Also shown are the results of correlated fits using the functional form $aM_{\rm kin} = A \cdot a m_b + B$, and the
value of $a m_b$ that would yield agreement of the $\eta_b(1S)$ kinetic mass with experiment.}
\end{table*}

\clearpage

\section{\label{sec:results_lattice_units}Results in lattice units}

\subsection{\label{sec:radial_orbital_lattice}Radial and orbital energy splittings}

\begin{table*}[h!]
\begin{ruledtabular}
\begin{tabular}{lllll}
                                  & $am_l=0.01$     & $am_l=0.02$     & $am_l=0.03$  \\
\hline
$\Upsilon(2S)-\Upsilon(1S)$       &  $0.3187(93)$   &  $0.3337(91)$   &  $0.3409(68)$  \\
$\overline{2S}-\overline{1S}$     &  $0.3233(84)$   &  $0.3372(84)$   &  $0.3448(70)$  \\
$\Upsilon(3S)-\Upsilon(1S)$       &  $0.536(42)$    &  $0.545(37)$    &  $0.549(26)$   \\
$\overline{1^3P}-\Upsilon(1S)$    &  $0.2559(24)$   &  $0.2620(20)$   &  $0.2646(15)$  \\
$\overline{1^3P}-\overline{1S}$   &  $0.2634(24)$   &  $0.2697(20)$   &  $0.2725(15)$  \\
$\overline{2^3P}-\overline{1^3P}$ &  $0.226(21)$    &  $0.244(16)$    &  $0.236(16)$  \\
$\overline{2^3P}-\Upsilon(1S)$    &  $0.478(23)$    &  $0.506(17)$    &  $0.501(16)$  \\
$\overline{2^3P}-\overline{1S}$   &  $0.485(23)$    &  $0.513(17)$    &  $0.509(16)$  \\
$\Upsilon_2(1D)-\Upsilon(1S)$     &  $0.4052(84)$   &  $0.4254(60)$   &  $0.4318(42)$  \\
\end{tabular}
\end{ruledtabular}
\caption{\label{tab:radial_orbital_splittings_L16}Radial and orbital energy splittings in lattice units,
calculated with the $v^4$ action on the $L=16$ ensembles, for $a m_b=2.536$.}
\end{table*}

\begin{table*}[h!]
\begin{ruledtabular}
\begin{tabular}{llllll}
                                  &  $am_l=0.005$   & $am_l=0.01$  & $am_l=0.02$  & $am_l=0.03$  \\
\hline
$\Upsilon(2S)-\Upsilon(1S)$       &   $0.3193(49)$  &  $0.3250(53)$  &  $0.3359(84)$  &  $0.3413(82)$  \\
$\overline{2S}-\overline{1S}$     &   $0.3236(46)$  &  $0.3291(51)$  &  $0.3395(87)$  &  $0.3453(81)$  \\
$\Upsilon(3S)-\Upsilon(1S)$       &   $0.497(22)$   &  $0.542(23)$   &  $0.540(53)$   &  $0.573(51)$  \\
$\overline{1^3P}-\Upsilon(1S)$    &   $0.2523(20)$  &  $0.2580(19)$  &  $0.2617(42)$  &  $0.2632(51)$  \\
$\overline{1^3P}-\overline{1S}$   &   $0.2598(20)$  &  $0.2656(19)$  &  $0.2695(42)$  &  $0.2710(51)$  \\
$\overline{2^3P}-\overline{1^3P}$ &   $0.207(12)$   &  $0.2240(79)$  &  $0.246(13)$   &  $0.263(19)$  \\
$\overline{2^3P}-\Upsilon(1S)$    &   $0.460(14)$   &  $0.4820(84)$  &  $0.508(15)$   &  $0.526(21)$  \\
$\overline{2^3P}-\overline{1S}$   &   $0.467(14)$   &  $0.4896(84)$  &  $0.516(15)$   &  $0.534(21)$  \\
$\Upsilon_2(1D)-\Upsilon(1S)$     &   $0.3998(54)$  &  $0.4140(49)$  &  $0.4330(82)$  &  $0.430(11)$  \\
\end{tabular}
\end{ruledtabular}
\caption{\label{tab:radial_orbital_splittings_L24}Radial and orbital energy splittings in lattice units,
calculated with the $v^4$ action on the $L=24$ ensembles, for $a m_b=2.536$.}
\end{table*}

\begin{table*}[h!]
\begin{ruledtabular}
\begin{tabular}{lllll}
                                  &  $am_l=0.004$  & $am_l=0.006$  & $am_l=0.008$  \\
\hline
$\Upsilon(2S)-\Upsilon(1S)$       &   $0.2421(33)$  &   $0.2418(47)$  &  $0.2464(34)$  \\
$\overline{2S}-\overline{1S}$     &   $0.2454(33)$  &   $0.2452(45)$  &  $0.2493(34)$  \\
$\Upsilon(3S)-\Upsilon(1S)$       &   $0.394(15)$   &   $0.395(16)$   &  $0.401(16)$  \\
$\overline{1^3P}-\Upsilon(1S)$    &   $0.1907(20)$  &   $0.1888(19)$  &  $0.1888(19)$  \\
$\overline{1^3P}-\overline{1S}$   &   $0.1969(20)$  &   $0.1949(19)$  &  $0.1950(19)$  \\
$\overline{2^3P}-\overline{1^3P}$ &   $0.1629(94)$  &   $0.1649(85)$  &  $0.170(11)$  \\
$\overline{2^3P}-\Upsilon(1S)$    &   $0.3519(94)$  &   $0.3533(94)$  &  $0.359(11)$  \\
$\overline{2^3P}-\overline{1S}$   &   $0.3580(94)$  &   $0.3594(94)$  &  $0.365(11)$  \\
$\Upsilon_2(1D)-\Upsilon(1S)$     &   $0.3051(40)$  &   $0.3045(56)$  &  $0.3088(54)$  \\
\end{tabular}
\end{ruledtabular}
\caption{\label{tab:radial_orbital_splittings_L32}Radial and orbital energy splittings in lattice units,
calculated with the $v^4$ action on the $L=32$ ensembles, for $a m_b=1.87$.}
\end{table*}

\clearpage

\subsection{\label{sec:radial_orbital_mb_dep}Heavy-quark mass dependence of radial and orbital energy splittings}

\begin{table*}[h!]
\begin{ruledtabular}
\begin{tabular}{lllll}
                                  & $am_b=2.3$   & $am_b=2.536$  & $am_b=2.7$  \\
\hline
$\Upsilon(2S)-\Upsilon(1S)$       &  $0.3211(50)$  &  $0.3193(49)$  &   $0.3184(49)$     \\
$\overline{2S}-\overline{1S}$     &  $0.3257(47)$  &  $0.3236(46)$  &   $0.3225(48)$     \\
$\overline{1^3P}-\Upsilon(1S)$    &  $0.2509(21)$  &  $0.2523(20)$  &   $0.2536(19)$     \\
$\overline{1^3P}-\overline{1S}$   &  $0.2590(21)$  &  $0.2598(20)$  &   $0.2608(19)$     \\
$\overline{2^3P}-\overline{1^3P}$ &  $0.213(15)$   &  $0.207(12)$   &   $0.204(11)$      \\
$\overline{2^3P}-\Upsilon(1S)$    &  $0.464(16)$   &  $0.460(14)$   &   $0.458(12)$     \\
$\overline{2^3P}-\overline{1S}$   &  $0.472(16)$   &  $0.467(14)$   &   $0.465(12)$     \\
$\Upsilon_2(1D)-\Upsilon(1S)$     &  $0.4002(52)$  &  $0.3998(54)$  &   $0.4001(57)$     \\
\end{tabular}
\end{ruledtabular}
\caption{\label{tab:radial_orbital_splittings_L24_mb_dep}Heavy-quark mass dependence of radial and orbital energy splittings in lattice units,
calculated with the $v^4$ action on the $L=24$ ensemble with $a m_l=0.005$.}
\end{table*}

\begin{table*}[h!]
\begin{ruledtabular}
\begin{tabular}{lllll}
                                  & $am_b=1.75$   & $am_b=1.87$  & $am_b=2.05$  \\
\hline
$\Upsilon(2S)-\Upsilon(1S)$       &  $0.2422(31)$  &  $0.2421(33)$  &   $0.2418(31)$  \\
$\overline{2S}-\overline{1S}$     &  $0.2456(32)$  &  $0.2454(33)$  &   $0.2448(31)$  \\
$\overline{1^3P}-\Upsilon(1S)$    &  $0.1901(22)$  &  $0.1907(20)$  &   $0.1918(19)$    \\
$\overline{1^3P}-\overline{1S}$   &  $0.1965(22)$  &  $0.1969(20)$  &   $0.1975(19)$    \\
$\overline{2^3P}-\overline{1^3P}$ &  $0.1645(99)$  &  $0.1629(94)$  &   $0.1592(80)$    \\
$\overline{2^3P}-\Upsilon(1S)$    &  $0.353(10)$   &  $0.3519(94)$  &   $0.3494(82)$    \\
$\overline{2^3P}-\overline{1S}$   &  $0.359(10)$   &  $0.3580(94)$  &   $0.3552(82)$    \\
$\Upsilon_2(1D)-\Upsilon(1S)$     &  $0.3048(39)$  &  $0.3051(40)$  &   $0.3059(42)$    \\
\end{tabular}
\end{ruledtabular}
\caption{\label{tab:radial_orbital_splittings_L32_mb_dep}Heavy-quark mass dependence of radial and orbital energy splittings in lattice units,
calculated with the $v^4$ action on the $L=32$ ensemble with $a m_l=0.004$.}
\end{table*}

\subsection{\label{sec:spin_dep_lattice}Spin-dependent energy splittings}

\begin{table*}[h!]
\begin{ruledtabular}
\begin{tabular}{llllll}
                                                            & $ am_l=0.005$   & $ am_l=0.01$  & $ am_l=0.02$  & $ am_l=0.03$  \\
\hline
$\Upsilon(1S)-\eta_b(1S)$                                   &  $0.030216(78)$  & $0.03037(10)$   &  $0.03087(23)$  &  $0.03132(24)$  \\
$\Upsilon(2S)-\eta_b(2S)$                                   &  $0.0124(19)$    & $0.0135(22)$    &  $0.0160(33)$   &  $0.0150(27)$   \\
$\chi_{b2}(1P)-\chi_{b1}(1P)$                               &  $0.01116(97)$   & $0.01185(76)$   &  $0.0135(19)$   &  $0.0115(24)$   \\
$\chi_{b1}(1P)-\chi_{b0}(1P)$                               &  $0.01573(79)$   & $0.01540(73)$   &  $0.0179(15)$   &  $0.0150(20)$   \\
$\overline{1^3P}-h_b(1P)$                                   &  $0.00116(56)$     & $0.00139(46)$     &  $0.0023(13)$     &  $0.0035(16)$     \\
$-2\chi_{b0}(1P)-3\chi_{b1}(1P)+5\chi_{b2}(1P)$             &  $0.0873(52)$    &  $0.0900(43)$   & $0.103(10)$     &  $0.087(13)$    \\
$-2\chi_{b0}(1P)+3\chi_{b1}(1P)-\chi_{b2}(1P)$              &  $0.0204(19)$    &  $0.0189(15)$   & $0.0223(31)$    &  $0.0183(48)$   \\
\\[-2ex]
$\displaystyle \frac{\Upsilon(2S)-\eta_b(2S)}{\Upsilon(1S)-\eta_b(1S)}$                        &  $0.411(63)$     & $0.446(73)$     &  $0.52(11)$     &  $0.480(87)$    \\
\\[-2ex]
$\displaystyle \frac{\Upsilon(1S)-\eta_b(1S)}{-2\chi_{b0}(1P)+3\chi_{b1}(1P)-\chi_{b2}(1P)}$   &  $1.48(14)$      & $1.60(13)$      &  $1.38(19)$     &  $1.71(44)$    \\
\\[-2ex]
$\displaystyle \frac{\Upsilon(2S)-\eta_b(2S)}{-2\chi_{b0}(1P)+3\chi_{b1}(1P)-\chi_{b2}(1P)}$   &  $0.61(11)$      & $0.72(13)$      &  $0.72(18)$     &  $0.82(26)$    \\
\end{tabular}
\end{ruledtabular}
\caption{\label{tab:spindep_splittings_L24}Spin-dependent energy splittings in lattice units, from the $v^4$ action on the $L=24$ lattices, for $a m_b=2.536$.}
\end{table*}

\begin{table*}[h!]
\begin{ruledtabular}
\begin{tabular}{llllll}
                                                            & $ am_l=0.005$   & $ am_l=0.01$  & $ am_l=0.02$  & $ am_l=0.03$  \\
\hline
$\Upsilon(1S)-\eta_b(1S)$                                   &  $0.025607(67)$  &  $0.025781(84)$  &  $0.02620(19)$  &  $0.02657(19)$  \\
$\Upsilon(2S)-\eta_b(2S)$                                   &  $0.0090(16)$    &  $0.0102(19)$    &  $0.0120(27)$   &  $0.0113(25)$   \\
$\chi_{b2}(1P)-\chi_{b1}(1P)$                               &  $0.00910(80)$   &  $0.00974(69)$   &  $0.0108(19)$   &  $0.0093(22)$  \\
$\chi_{b1}(1P)-\chi_{b0}(1P)$                               &  $0.01396(71)$   &  $0.01380(64)$   &  $0.0158(14)$   &  $0.0134(17)$  \\
$\overline{1^3P}-h_b(1P)$                                   &  $0.00085(50)$     &  $0.00099(46)$  &  $0.0018(11)$  &  $0.0026(15)$  \\
$-2\chi_{b0}(1P)-3\chi_{b1}(1P)+5\chi_{b2}(1P)$             &  $0.0734(45)$    &  $0.0763(40)$    &  $0.086(11)$    &  $0.073(13)$  \\
$-2\chi_{b0}(1P)+3\chi_{b1}(1P)-\chi_{b2}(1P)$              &  $0.0189(16)$    &  $0.0178(14)$    &  $0.0206(33)$   &  $0.0174(40)$  \\
\\[-2ex]
$\displaystyle \frac{\Upsilon(2S)-\eta_b(2S)}{\Upsilon(1S)-\eta_b(1S)}$                        &  $0.350(64)$  &  $0.395(75)$  & $0.46(10)$   &  $0.425(95)$  \\
\\[-2ex]
$\displaystyle \frac{\Upsilon(1S)-\eta_b(1S)}{-2\chi_{b0}(1P)+3\chi_{b1}(1P)-\chi_{b2}(1P)}$   &  $1.36(12)$  &  $1.45(12)$  &  $1.27(20)$  & $1.52(35)$   \\
\\[-2ex]
$\displaystyle \frac{\Upsilon(2S)-\eta_b(2S)}{-2\chi_{b0}(1P)+3\chi_{b1}(1P)-\chi_{b2}(1P)}$   &  $0.476(96)$      & $0.57(12)$      &  $0.58(16)$     &  $0.65(21)$    \\
\end{tabular}
\end{ruledtabular}
\caption{\label{tab:spindep_splittings_L24_v6}Spin-dependent energy splittings in lattice units, from the $v^6$ action on the $L=24$ lattices, for $a m_b=2.536$.}
\end{table*}

\begin{table*}[h!]
\begin{ruledtabular}
\begin{tabular}{lllll}
                                                            & $ am_l=0.004$   & $ am_l=0.006$   & $ am_l=0.008$  \\
\hline
$\Upsilon(1S)-\eta_b(1S)$                                   & $0.024441(74)$  & $0.024408(74)$  & $0.02455(11)$   \\
$\Upsilon(2S)-\eta_b(2S)$                                   & $0.0114(12)$    & $0.0109(19)$    & $0.0127(13)$   \\
$\chi_{b2}(1P)-\chi_{b1}(1P)$                               & $0.00856(93)$   & $0.00910(98)$   & $0.0087(11)$   \\
$\chi_{b1}(1P)-\chi_{b0}(1P)$                               & $0.01268(78)$   & $0.01286(85)$   & $0.01321(78)$   \\
$\overline{1^3P}-h_b(1P)$                                   & $0.00030(43)$     & $0.00104(56)$     & $0.00049(63)$   \\
$-2\chi_{b0}(1P)-3\chi_{b1}(1P)+5\chi_{b2}(1P)$             & $0.0681(54)$    & $0.0714(55)$    & $0.0696(57)$  \\
$-2\chi_{b0}(1P)+3\chi_{b1}(1P)-\chi_{b2}(1P)$              & $0.0168(15)$    & $0.0168(19)$    & $0.0177(18)$  \\
\\[-2ex]
$\displaystyle \frac{\Upsilon(2S)-\eta_b(2S)}{\Upsilon(1S)-\eta_b(1S)}$                        & $0.465(51)$     & $0.448(79)$     & $0.520(53)$   \\
\\[-2ex]
$\displaystyle \frac{\Upsilon(1S)-\eta_b(1S)}{-2\chi_{b0}(1P)+3\chi_{b1}(1P)-\chi_{b2}(1P)}$   & $1.45(13)$      & $1.46(17)$      & $1.38(14)$    \\
\\[-2ex]
$\displaystyle \frac{\Upsilon(2S)-\eta_b(2S)}{-2\chi_{b0}(1P)+3\chi_{b1}(1P)-\chi_{b2}(1P)}$   &  $0.675(95)$      & $0.65(14)$      &  $0.72(10)$   \\
\end{tabular}
\end{ruledtabular}
\caption{\label{tab:spindep_splittings_L32}Spin-dependent energy splittings in lattice units, from the $v^4$ action on the $L=32$ lattices, for $a m_b=1.87$.}
\end{table*}

\begin{table*}[h!]
\begin{ruledtabular}
\begin{tabular}{lllll}
                                                            & $ am_l=0.004$   & $ am_l=0.006$   & $ am_l=0.008$  \\
\hline
$\Upsilon(1S)-\eta_b(1S)$                                   &  $0.020215(69)$ & $0.020170(59)$  &  $0.020292(96)$  \\
$\Upsilon(2S)-\eta_b(2S)$                                   &  $0.0083(12)$   & $0.0082(14)$    &  $0.0096(12)$    \\
$\chi_{b2}(1P)-\chi_{b1}(1P)$                               &  $0.00677(85)$  & $0.00735(79)$   &  $0.00710(95)$   \\
$\chi_{b1}(1P)-\chi_{b0}(1P)$                               &  $0.01122(69)$  & $0.01116(70)$   &  $0.01141(68)$  \\
$\overline{1^3P}-h_b(1P)$                                   &  $0.00018(42)$ & $0.00075(52)$  &  $0.00024(50)$  \\
$-2\chi_{b0}(1P)-3\chi_{b1}(1P)+5\chi_{b2}(1P)$             &  $0.0562(49)$   & $0.0591(44)$    &  $0.0582(53)$  \\
$-2\chi_{b0}(1P)+3\chi_{b1}(1P)-\chi_{b2}(1P)$              &  $0.0157(14)$   & $0.0150(16)$    & $0.0157(16)$   \\
\\[-2ex]
$\displaystyle \frac{\Upsilon(2S)-\eta_b(2S)}{\Upsilon(1S)-\eta_b(1S)}$                        & $0.410(61)$  & $0.404(70)$  &  $0.476(61)$  \\
\\[-2ex]
$\displaystyle \frac{\Upsilon(1S)-\eta_b(1S)}{-2\chi_{b0}(1P)+3\chi_{b1}(1P)-\chi_{b2}(1P)}$   & $1.29(11)$   & $1.34(14)$   &  $1.29(13)$  \\
\\[-2ex]
$\displaystyle \frac{\Upsilon(2S)-\eta_b(2S)}{-2\chi_{b0}(1P)+3\chi_{b1}(1P)-\chi_{b2}(1P)}$   &  $0.528(92)$      & $0.54(11)$      &  $0.613(99)$    \\
\end{tabular}
\end{ruledtabular}
\caption{\label{tab:spindep_splittings_L32_v6}Spin-dependent energy splittings in lattice units, from the $v^6$ action on the $L=32$ lattices, for $a m_b=1.87$.}
\end{table*}

\clearpage

\subsection{\label{sec:spindep_c3_c4}Dependence of spin splittings on the couplings $c_3$ and $c_4$}

\begin{table*}[h!]
\begin{ruledtabular}
\begin{tabular}{lllllll}
                                                            & $c_3=0.8$ & $c_3=1.2$  & $c_4=0.8$  & $c_4=1.2$  \\
\hline
$\Upsilon(1S)-\eta_b(1S)$                                   &  $0.98016(18)$  &  $1.02148(19)$  &  $0.67151(53)$  &  $1.3808(12)$  &   \\
$\Upsilon(2S)-\eta_b(2S)$                                   &  $0.983(87)$    &  $1.025(91)$    &  $0.68(10)$     &  $1.35(14)$    &   \\
$\chi_{b2}(1P)-\chi_{b1}(1P)$                               &  $0.832(49)$    &  $1.162(54)$    &  $1.020(62)$    &  $0.954(61)$   &   \\
$\chi_{b1}(1P)-\chi_{b0}(1P)$                               &  $0.935(38)$    &  $1.064(35)$    &  $0.786(34)$    &  $1.239(48)$   &   \\
$\overline{1^3P}-h_b(1P)$                                   &  $0.95(37)$     &  $1.01(43)$     &  $0.67(36)$     &  $1.32(63)$    &   \\
$-2\chi_{b0}(1P)-3\chi_{b1}(1P)+5\chi_{b2}(1P)$             &  $0.871(29)$    &  $1.129(31)$    &  $0.936(32)$    &  $1.059(39)$   &   \\
$-2\chi_{b0}(1P)+3\chi_{b1}(1P)-\chi_{b2}(1P)$              &  $0.991(84)$    &  $1.008(76)$    &  $0.658(67)$    &  $1.40(11)$    &   \\
\\[-2ex]
$\displaystyle \frac{\Upsilon(2S)-\eta_b(2S)}{\Upsilon(1S)-\eta_b(1S)}$                        &  $1.003(89)$  &  $1.003(89)$  &  $1.02(15)$  &  $0.98(10)$  &   \\
\\[-2ex]
$\displaystyle \frac{\Upsilon(1S)-\eta_b(1S)}{-2\chi_{b0}(1P)+3\chi_{b1}(1P)-\chi_{b2}(1P)}$   &  $0.989(83)$  &  $1.013(78)$  &  $1.02(10)$  &  $0.989(77)$  &   \\
\\[-2ex]
$\displaystyle \frac{\Upsilon(2S)-\eta_b(2S)}{-2\chi_{b0}(1P)+3\chi_{b1}(1P)-\chi_{b2}(1P)}$   &  $0.99(16)$  &  $1.02(15)$  &  $1.05(23)$  &  $0.98(16)$  &   \\
\end{tabular}
\end{ruledtabular}
\caption{\label{tab:spindep_splittings_L24_c_dep}Dependence of the spin splittings, calculated with the $v^4$ action, on the couplings $c_3$ and $c_4$.
Shown is the ratio of the splitting with either $c_3\neq 1$ or $c_4\neq 1$ to the splitting with all $c_i=1$, calculated using bootstrap. The data are for the $L=24$ ensemble with
$a m_l=0.005$ and $a m_b=2.536$.}
\end{table*}

\begin{table*}[h!]
\begin{ruledtabular}
\begin{tabular}{lllllll}
                                                            & $c_3=0.8$ & $c_3=1.2$  & $c_4=0.8$  & $c_4=1.2$  \\
\hline
$\Upsilon(1S)-\eta_b(1S)$                                   &  $0.97788(17)$  &  $1.02411(20)$  & $0.64656(47)$   &  $1.4180(11)$  &   \\
$\Upsilon(2S)-\eta_b(2S)$                                   &  $0.98(13)$     &  $1.03(13)$     & $0.63(12)$      &  $1.44(19)$    &   \\
$\chi_{b2}(1P)-\chi_{b1}(1P)$                               &  $0.795(53)$    &  $1.205(58)$    & $1.016(55)$     &  $0.961(68)$   &   \\
$\chi_{b1}(1P)-\chi_{b0}(1P)$                               &  $0.924(34)$    &  $1.071(30)$    & $0.765(34)$     &  $1.265(53)$   &   \\
$\overline{1^3P}-h_b(1P)$                                   &  $0.92(42)$     &  $1.02(49)$     & $0.66(41)$      &  $1.36(66)$    &   \\
$-2\chi_{b0}(1P)-3\chi_{b1}(1P)+5\chi_{b2}(1P)$             &  $0.845(28)$    &  $1.154(32)$    & $0.920(29)$     &  $1.077(40)$   &   \\
$-2\chi_{b0}(1P)+3\chi_{b1}(1P)-\chi_{b2}(1P)$              &  $0.987(71)$    &  $1.006(62)$    & $0.641(59)$     &  $1.41(11)$    &   \\
\\[-2ex]
$\displaystyle \frac{\Upsilon(2S)-\eta_b(2S)}{\Upsilon(1S)-\eta_b(1S)}$                        &  $1.00(13)$   &  $1.00(13)$   &  $0.97(19)$   &  $1.01(14)$  &   \\
\\[-2ex]
$\displaystyle \frac{\Upsilon(1S)-\eta_b(1S)}{-2\chi_{b0}(1P)+3\chi_{b1}(1P)-\chi_{b2}(1P)}$   &  $0.991(75)$  &  $1.018(62)$  &  $1.008(95)$  & $1.002(74)$   &   \\
\\[-2ex]
$\displaystyle \frac{\Upsilon(2S)-\eta_b(2S)}{-2\chi_{b0}(1P)+3\chi_{b1}(1P)-\chi_{b2}(1P)}$   &  $1.00(21)$   &  $1.04(19)$   &  $0.96(26)$   &  $1.02(20)$  &   \\
\end{tabular}
\end{ruledtabular}
\caption{\label{tab:spindep_splittings_L24_c_dep_v6}Dependence of the spin splittings, calculated with $v^6$ action, on the couplings $c_3$ and $c_4$.
Shown is the ratio of the splitting with either $c_3\neq 1$ or $c_4\neq 1$ to the splitting with all $c_i=1$, calculated using bootstrap. The data are for the $L=24$ ensemble with
$a m_l=0.005$ and $a m_b=2.536$.}
\end{table*}

\clearpage

\subsection{\label{sec:spindep_splittings_mb_dep}Heavy-quark mass dependence of spin splittings}

\begin{table*}[h!]
\begin{ruledtabular}
\begin{tabular}{lllll}
                                                            &  $ am_b=2.3$   & $ am_b=2.536$  & $ am_b=2.7$  \\
\hline
$\Upsilon(2S)-\eta_b(2S)$                                   &  $0.0138(21)$   &  $0.0124(19)$   &  $0.0116(18)$      \\
$\chi_{b2}(1P)-\chi_{b1}(1P)$                               &  $0.0122(11)$   &  $0.01116(97)$  &  $0.01060(86)$     \\
$\chi_{b1}(1P)-\chi_{b0}(1P)$                               &  $0.01734(86)$  &  $0.01573(79)$  &  $0.01469(71)$     \\
$\overline{1^3P}-h_b(1P)$                                   &  $0.00120(62)$    &  $0.00116(56)$    &  $0.00120(54)$       \\
$-2\chi_{b0}(1P)-3\chi_{b1}(1P)+5\chi_{b2}(1P)$             &  $0.0956(62)$   &  $0.0873(52)$   &  $0.0824(46)$      \\
$-2\chi_{b0}(1P)+3\chi_{b1}(1P)-\chi_{b2}(1P)$              &  $0.0225(22)$   &  $0.0204(19)$   &  $0.0189(18)$      \\
\end{tabular}
\end{ruledtabular}
\caption{\label{tab:spindep_splittings_L24_mb_dep}Heavy-quark mass dependence of spin splittings in lattice units, from the $v^4$ action on the $L=24$ ensemble with $a m_l=0.005$
(The results for the $\Upsilon(1S)-\eta_b(1S)$ splitting are given in Table \ref{tab:1S_hyperfine_L24_mb_dep}).}
\end{table*}

\begin{table*}[h!]
\begin{ruledtabular}
\begin{tabular}{llllc}
                    &  $am_b=2.3$   & $am_b=2.536$     & $am_b=2.7$      & Fit result\\
\hline
$a m_l=0.005$      & $0.032685(90)$ &  $0.030216(78)$  & $0.028727(75)$  &  $A=0.0615(10)$,  $\nb B=0.00597(41)$  \\
$a m_l=0.01$       & $0.03291(11)$  &  $0.03037(10)$   & $0.02885(10)$   &  $A=0.06295(90)$, $    B=0.00554(34)$  \\
$a m_l=0.02$       & $0.03344(25)$  &  $0.03087(23)$   & $0.02933(24)$   &  $A=0.0636(28)$,  $\nb B=0.0058(11)\nb$  \\
$a m_l=0.03$       & $0.03403(26)$  &  $0.03132(24)$   & $0.02968(23)$   &  $A=0.0675(20)$,  $\nb B=0.00470(79)$  \\
\end{tabular}
\end{ruledtabular}
\caption{\label{tab:1S_hyperfine_L24_mb_dep}Heavy-quark mass dependence of the $1S$ hyperfine splitting in lattice units, from the $v^4$ action on the $L=24$ ensembles. 
In the last column of the table, the results of correlated fits using the functional form $aE_{\Upsilon(1S)}-aE_{\eta_b(1S)} = A/(a m_b) + B$ are shown.}
\end{table*}

\begin{table*}[h!]
\begin{ruledtabular}
\begin{tabular}{lllll}
                                                            &  $ am_b=2.3$   & $ am_b=2.536$  & $ am_b=2.7$  \\
\hline
$\Upsilon(2S)-\eta_b(2S)$                                   &  $0.0097(18)$    &  $0.0090(16)$    &  $0.0085(15)$    \\
$\chi_{b2}(1P)-\chi_{b1}(1P)$                               &  $0.00964(87)$   &  $0.00910(80)$   &  $0.00877(83)$    \\
$\chi_{b1}(1P)-\chi_{b0}(1P)$                               &  $0.01517(80)$   &  $0.01396(71)$   &  $0.01320(64)$    \\
$\overline{1^3P}-h_b(1P)$                                   &  $0.00083(54)$     &  $0.00085(50)$     &  $0.00093(52)$    \\
$-2\chi_{b0}(1P)-3\chi_{b1}(1P)+5\chi_{b2}(1P)$             &   $0.0786(47)$   &  $0.0734(45)$    &  $0.0703(47)$     \\
$-2\chi_{b0}(1P)+3\chi_{b1}(1P)-\chi_{b2}(1P)$              &   $0.0208(19)$   &  $0.0189(16)$    &  $0.0177(15)$     \\
\end{tabular}
\end{ruledtabular}
\caption{\label{tab:spindep_splittings_L24_mb_dep_v6}Heavy-quark mass dependence of spin splittings in lattice units, from the $v^6$ action on the $L=24$ ensemble with $a m_l=0.005$
(The results for the $\Upsilon(1S)-\eta_b(1S)$ splitting are given in Table \ref{tab:1S_hyperfine_L24_mb_dep_v6}).}
\end{table*}

\begin{table*}[h!]
\begin{ruledtabular}
\begin{tabular}{llllc}
                    &  $am_b=2.3$   & $am_b=2.536$     & $am_b=2.7$      & Fit result\\
\hline
$a m_l=0.005$      &  0.027319(70)  &  0.025607(67)  &  0.024567(66)  &  $A=0.0427(10)$,  $\nb B=0.00876(40)$  \\
$a m_l=0.01$       &  0.027548(91)  &  0.025781(84)  &  0.024712(85)  &  $A=0.04398(75)$, $B=0.00843(29)$  \\
$a m_l=0.02$       &  0.02800(19)   &  0.02620(19)   &  0.02513(19)   &  $A=0.0447(25)$,  $\nb B=0.0086(10)\nb$  \\
$a m_l=0.03$       &  0.02851(21)   &  0.02657(19)   &  0.02540(19)   &  $A=0.0482(16)$,  $\nb B=0.00753(65)$  \\
\end{tabular}
\end{ruledtabular}
\caption{\label{tab:1S_hyperfine_L24_mb_dep_v6}Heavy-quark mass dependence of the $1S$ hyperfine splitting in lattice units, from the $v^6$ action on the $L=24$ ensembles. 
In the last column of the table, the results of correlated fits using the functional form $aE_{\Upsilon(1S)}-aE_{\eta_b(1S)} = A/(a m_b) + B$ are shown.}
\end{table*}

\begin{table*}[h!]
\begin{ruledtabular}
\begin{tabular}{lllll}
                                                            &  $ am_b=1.75$   & $ am_b=1.87$  & $ am_b=2.05$  \\
\hline
$\Upsilon(2S)-\eta_b(2S)$                                   & $0.0122(13)$    & $0.0114(12)$   & $0.0108(12)$      \\
$\chi_{b2}(1P)-\chi_{b1}(1P)$                               & $0.0091(10)$    & $0.00856(93)$  & $0.00804(76)$     \\
$\chi_{b1}(1P)-\chi_{b0}(1P)$                               & $0.01352(84)$   & $0.01268(78)$  & $0.01172(70)$     \\
$\overline{1^3P}-h_b(1P)$                                   & $0.00028(49)$     & $0.00030(43)$    & $0.00033(39)$       \\
$-2\chi_{b0}(1P)-3\chi_{b1}(1P)+5\chi_{b2}(1P)$             & $0.0725(60)$    & $0.0681(54)$   & $0.0636(44)$      \\
$-2\chi_{b0}(1P)+3\chi_{b1}(1P)-\chi_{b2}(1P)$              & $0.0179(18)$    & $0.0168(15)$   & $0.0154(13)$      \\
\end{tabular}
\end{ruledtabular}
\caption{\label{tab:spindep_splittings_L32_mb_dep}Heavy-quark mass dependence of spin splittings in lattice units, from the $v^4$ action on the $L=32$ ensemble with $a m_l=0.004$
(The results for the $\Upsilon(1S)-\eta_b(1S)$ splitting are given in Table \ref{tab:1S_hyperfine_L32_mb_dep}).}
\end{table*}

\begin{table*}[h!]
\begin{ruledtabular}
\begin{tabular}{llllc}
                    &  $am_b=1.75$   & $am_b=1.87$    & $am_b=2.05$    & Fit result\\
\hline
$a m_l=0.004$       &  $0.025679(82)$  &  $0.024441(74)$  &  $0.022850(76)$  &   $A=0.03380(58)$, $B=0.00636(28)$ \\
$a m_l=0.006$       &  $0.025669(78)$  &  $0.024408(74)$  &  $0.022787(70)$  &   $A=0.03443(54)$, $B=0.00598(27)$ \\
$a m_l=0.008$       &  $0.02582(12)$   &  $0.02455(11)$   &  $0.02291(11)$   &   $A=0.0348(11)$,  $\nb B=0.00592(57)$ \\
\end{tabular}
\end{ruledtabular}
\caption{\label{tab:1S_hyperfine_L32_mb_dep}Heavy-quark mass dependence of the $1S$ hyperfine splitting in lattice units, from the $v^4$ action on the $L=32$ ensembles. 
In the last column of the table, the results of correlated fits using the functional form $aE_{\Upsilon(1S)}-aE_{\eta_b(1S)} = A/(a m_b) + B$ are shown.}
\end{table*}

\begin{table*}[h!]
\begin{ruledtabular}
\begin{tabular}{lllll}
                                                            &  $ am_b=1.75$   & $ am_b=1.87$  & $ am_b=2.05$  \\
\hline
$\Upsilon(2S)-\eta_b(2S)$                                   &  $0.0088(13)$    &  $0.0083(12)$    &   $0.0077(13)$     \\
$\chi_{b2}(1P)-\chi_{b1}(1P)$                               &  $0.00700(89)$   &  $0.00677(85)$   &   $0.00643(72)$     \\
$\chi_{b1}(1P)-\chi_{b0}(1P)$                               &  $0.01187(73)$   &  $0.01122(69)$   &   $0.01040(61)$     \\
$\overline{1^3P}-h_b(1P)$                                   &  $0.00020(42)$     &  $0.00018(42)$     &   $0.00020(32)$    \\
$-2\chi_{b0}(1P)-3\chi_{b1}(1P)+5\chi_{b2}(1P)$             &  $0.0587(51)$    &  $0.0562(49)$    &   $0.0530(41)$     \\
$-2\chi_{b0}(1P)+3\chi_{b1}(1P)-\chi_{b2}(1P)$              &  $0.0168(15)$    &  $0.0157(14)$    &   $0.0144(12)$     \\
\end{tabular}
\end{ruledtabular}
\caption{\label{tab:spindep_splittings_L32_mb_dep_v6}Heavy-quark mass dependence of spin splittings in lattice units, from the $v^6$ action on the $L=32$ ensemble with $a m_l=0.004$
(The results for the $\Upsilon(1S)-\eta_b(1S)$ splitting are given in Table \ref{tab:1S_hyperfine_L32_mb_dep_v6}).}
\end{table*}

\begin{table*}[h!]
\begin{ruledtabular}
\begin{tabular}{llllc}
                    &  $am_b=1.75$   & $am_b=1.87$    & $am_b=2.05$    & Fit result\\
\hline
$a m_l=0.004$       &  $0.021117(74)$  &  $0.020215(69)$  &  $0.019088(62)$  &  $A=0.02424(44)$,  $B=0.00726(22)$  \\
$a m_l=0.006$       &  $0.021083(63)$  &  $0.020170(59)$  &  $0.019029(56)$  &  $A=0.02450(56)$,  $B=0.00707(27)$  \\
$a m_l=0.008$       &  $0.021222(92)$  &  $0.020292(96)$  &  $0.019126(91)$  &  $A=0.0251(11)$,   $\nb B=0.00685(58)$  \\
\end{tabular}
\end{ruledtabular}
\caption{\label{tab:1S_hyperfine_L32_mb_dep_v6}Heavy-quark mass dependence of the $1S$ hyperfine splitting in lattice units, from the $v^6$ action on the $L=32$ ensembles. 
In the last column of the table, the results of correlated fits using the functional form $aE_{\Upsilon(1S)}-aE_{\eta_b(1S)} = A/(a m_b) + B$ are shown.}
\end{table*}

\begin{figure*}[h!]
 \includegraphics[width=0.415\linewidth]{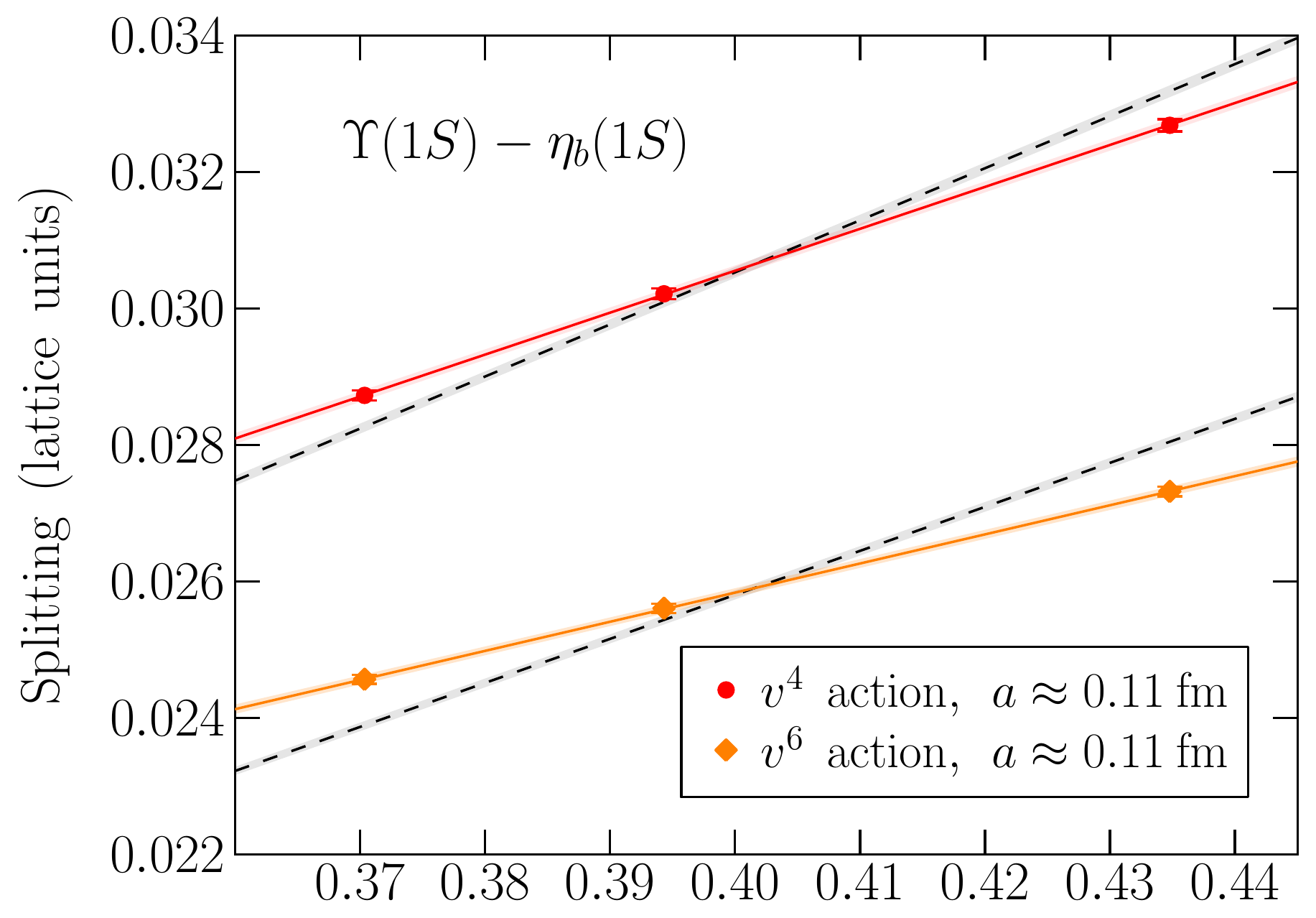}  \hfill \includegraphics[width=0.415\linewidth]{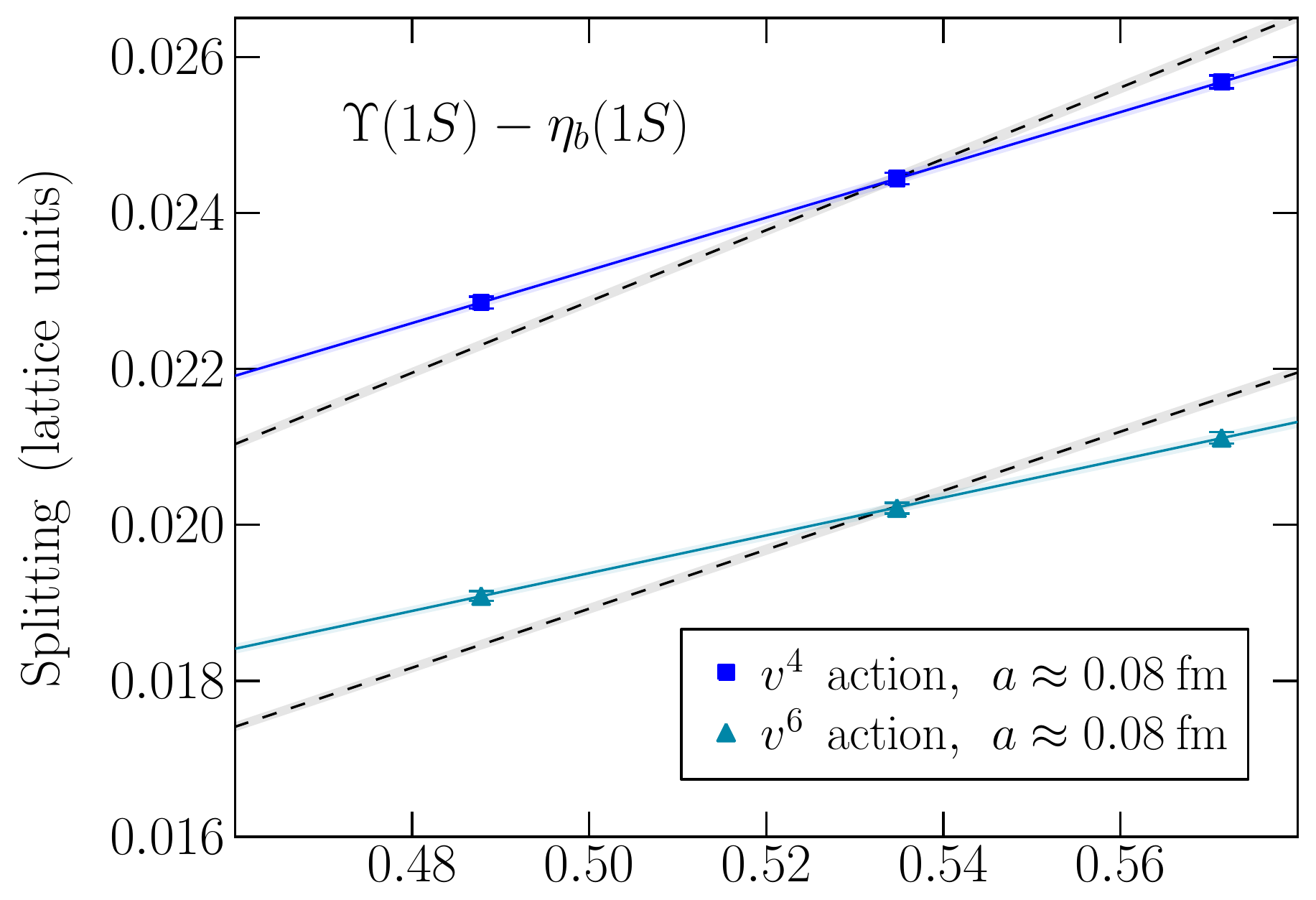}
 \includegraphics[width=0.415\linewidth]{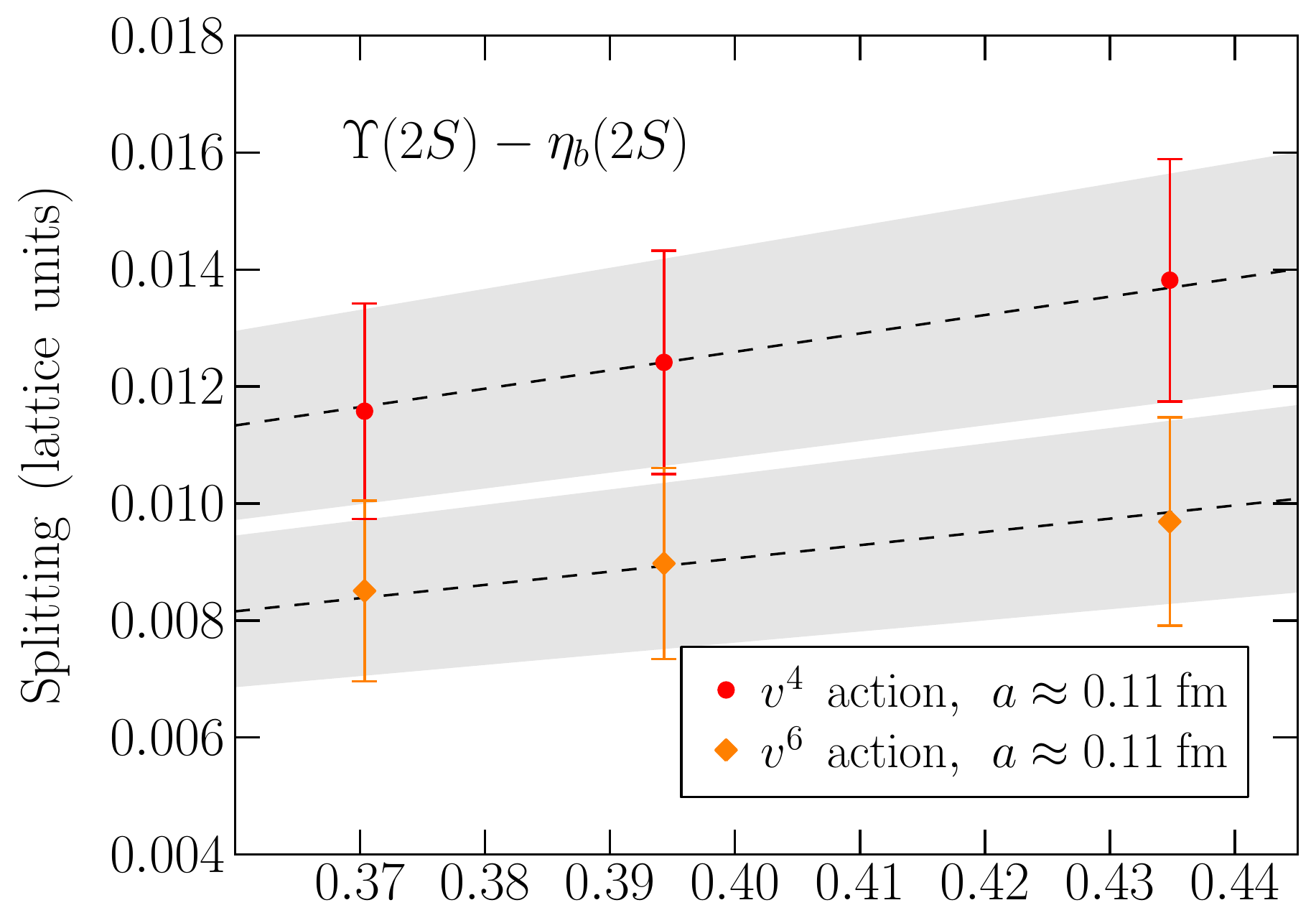}  \hfill \includegraphics[width=0.415\linewidth]{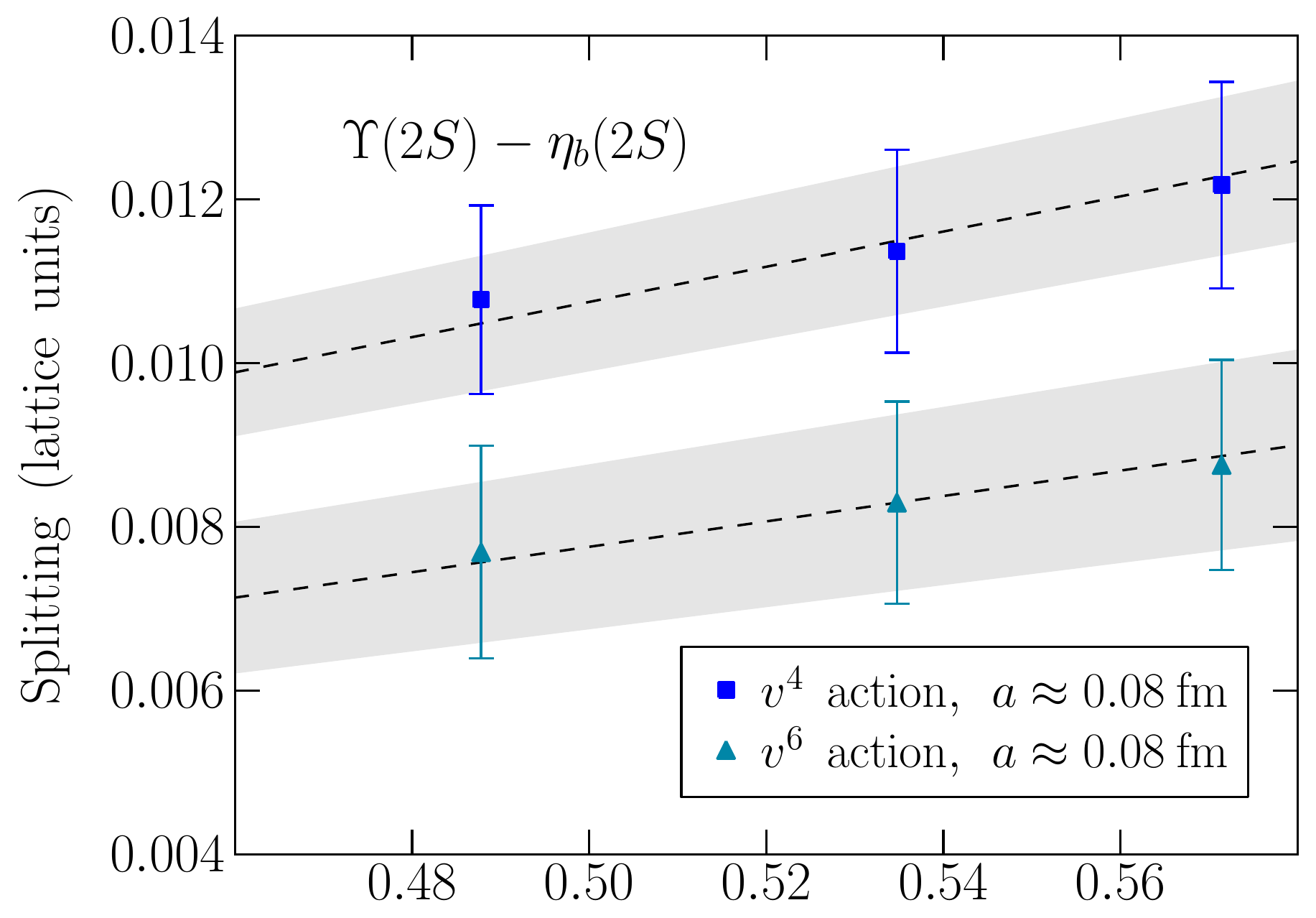}
 \includegraphics[width=0.415\linewidth]{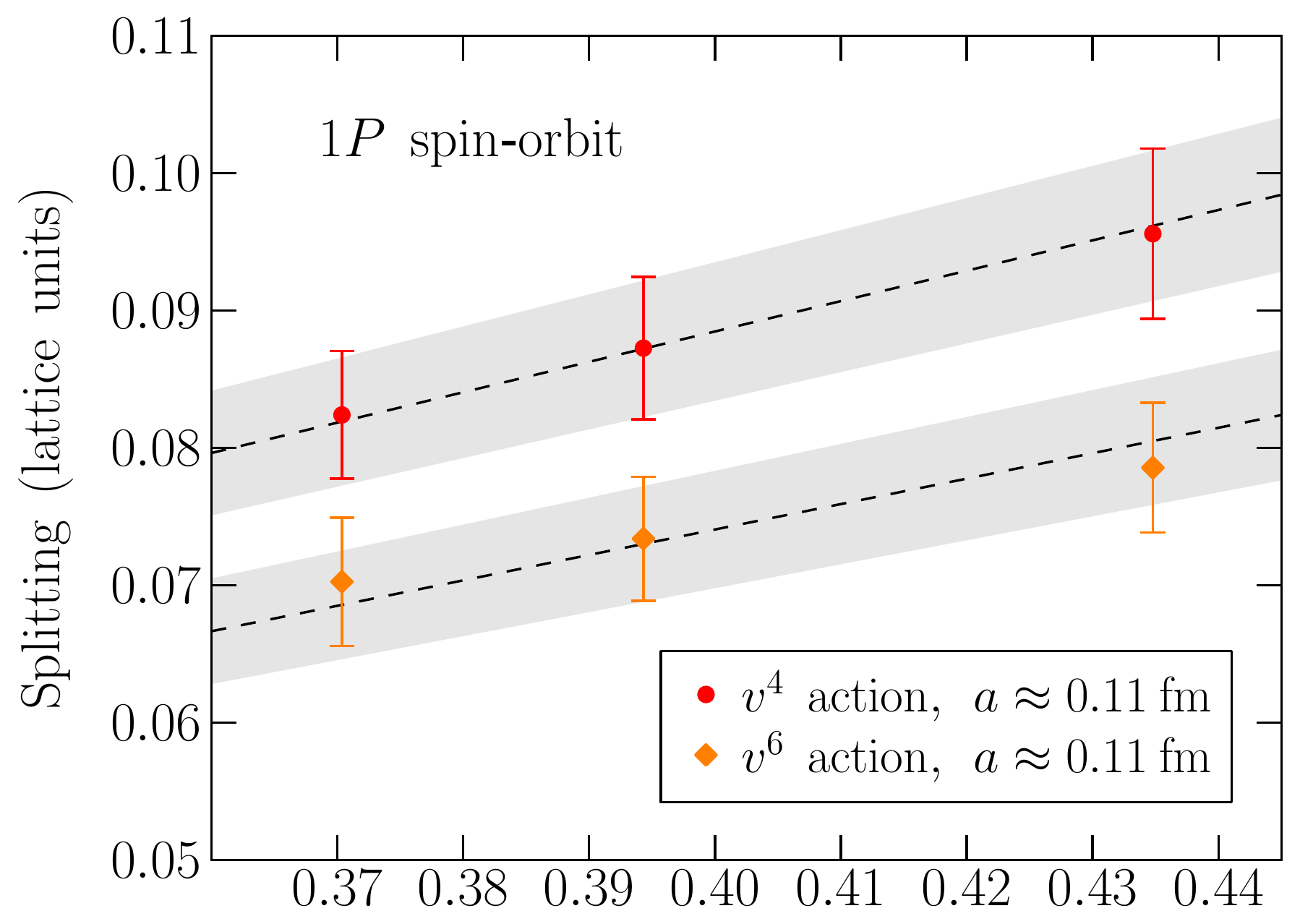}    \hfill \includegraphics[width=0.415\linewidth]{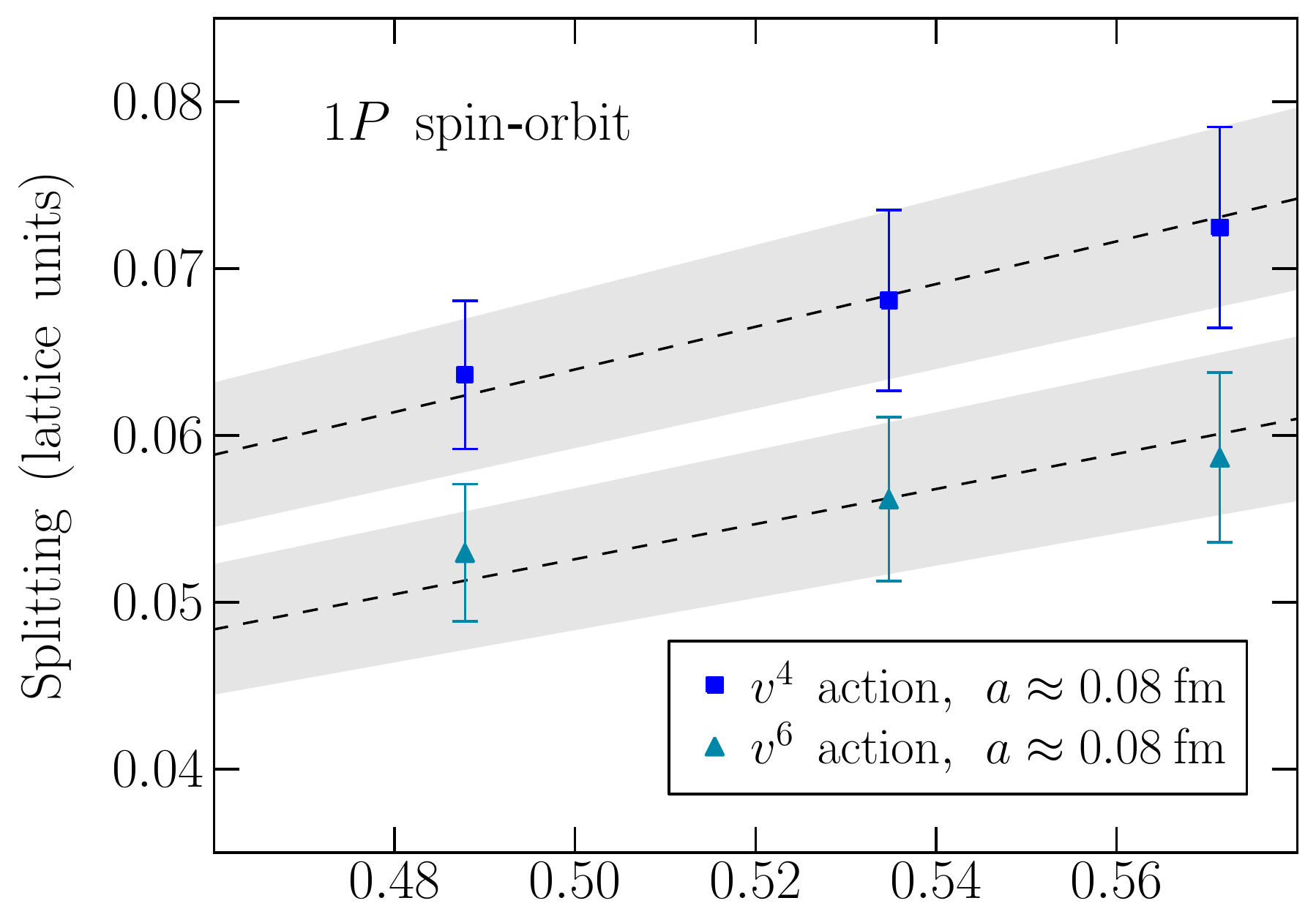}
 \includegraphics[width=0.415\linewidth]{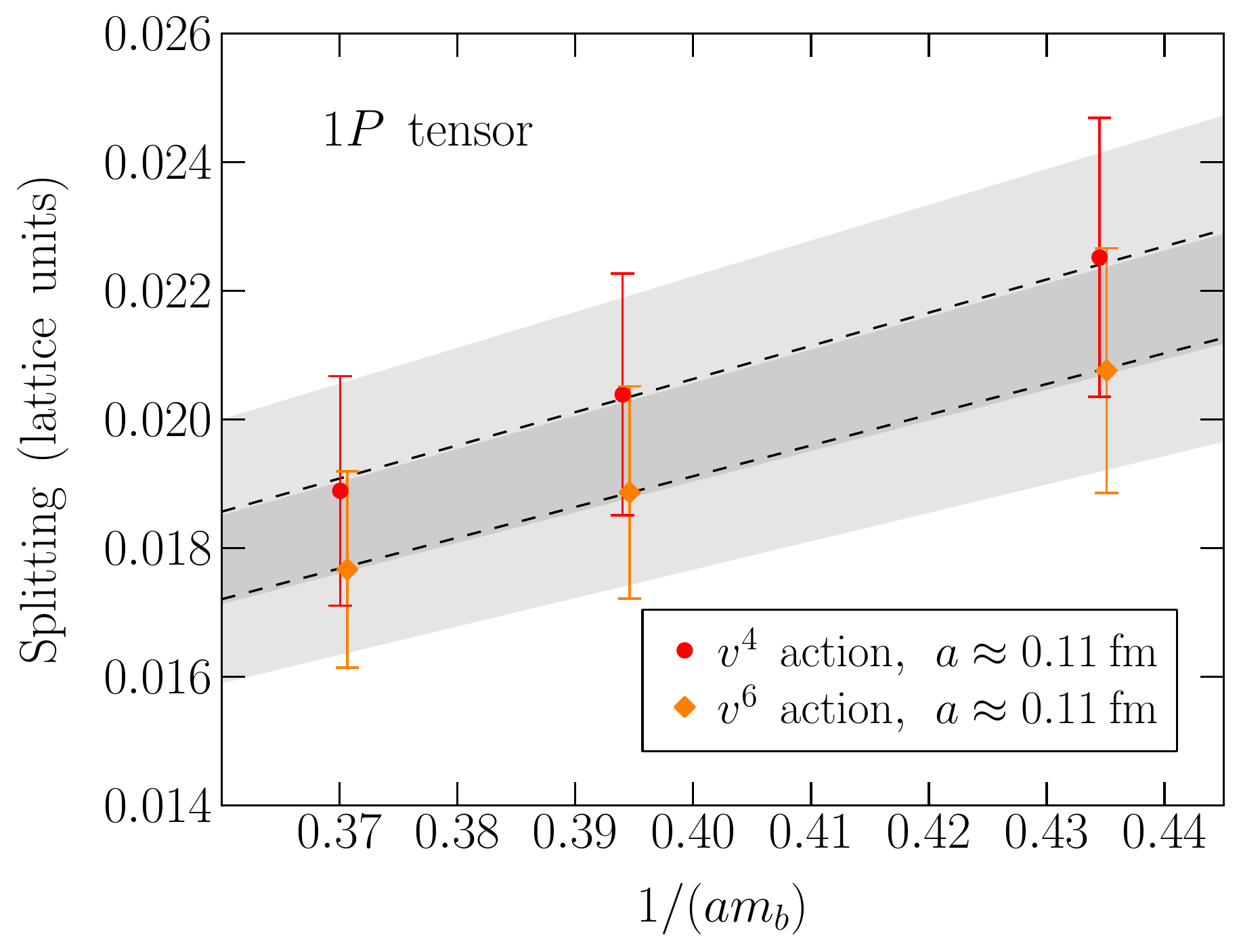}        \hfill \includegraphics[width=0.415\linewidth]{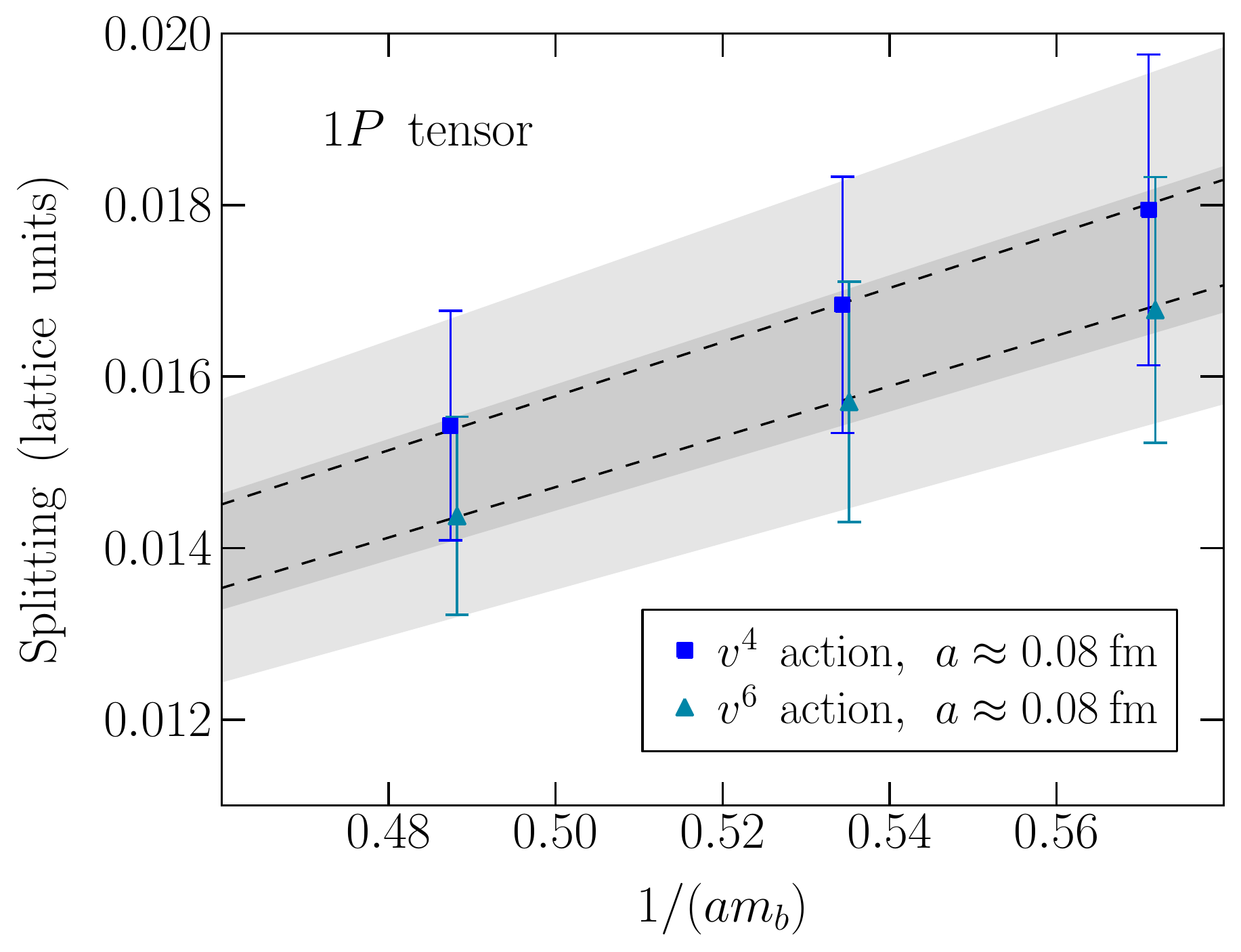}
\caption{\label{fig:spin_dep_mb_dep} Heavy-quark mass dependence of the spin splittings on the $L=24,\: a m_l=0.005$ and $L=32,\: a m_l=0.004$ ensembles.
Results are shown for both the $v^4$ and the $v^6$ actions. The dashed lines and gray error bands are correlated fits using the function $A/(a m_b)$.
The data for the $1S$ hyperfine splittings are incompatible with this form, and additional fits using the function $A/(a m_b)+B$ are shown, which describe the data very well.}
\end{figure*}

\clearpage

\section{\label{sec:gluon_errors}Gluon discretization errors}

The Iwasaki gluon action used in this work belongs to a class of actions with the form
\begin{equation}
 S_G[U]=-\frac{\beta}{3}\sum_x\left[(1-8c_1)\sum_{\mu<\nu}P[U]_{x,\mu\nu} + c_1 \sum_{\mu\neq\nu}R[U]_{x,\mu\nu} \right], \label{eq:gauge_action}
\end{equation}
where $P[U]_{x,\mu\nu}$ and $R[U]_{x,\mu\nu}$ are the real part of the trace of the $1\times1$ plaquette and
$1\times2$ rectangle terms, respectively (the coefficient $c_1$ in (\ref{eq:gauge_action}) should not be confused with the one in the NRQCD action (\ref{eq:dH_full})).
The Iwasaki action uses $c_1=-0.331$, derived from a renormalization-group transformation \cite{Iwasaki:1983ck,Iwasaki:1984cj, Iwasaki:1996sn}.
Note that tree-level order-$a^2$ improvement would require $c_1=-1/12$, corresponding to the tree-level L\"uscher-Weisz action \cite{Luscher:1984xn, Luscher:1985zq}).
However, nonperturbatively and at coarse lattice spacings, the Iwasaki action has been shown to yield reduced lattice artifacts compared to the tree-level L\"uscher-Weisz action
\cite{deForcrand:1999bi, Necco:2003vh}.

In this appendix, gluonic discretization errors in bottomonium energy splittings will be investigated. In Sec.~\ref{sec:gluon_errors_tree},
the shifts in radial and orbital energy splittings are studied using tree-level perturbation theory for four different choices of $c_1$.
The tree-level energy shift for the simple plaquette action ($c_1=0$) has previously been estimated at order $a^2$ in \cite{Davies:1994ei, Davies:1998im}.
In the following, a new analysis based on a lattice potential model is presented. This analysis does not make use of an expansion in powers of $a$, which
would not be appropriate for the Iwasaki action.

Then, to go beyond tree-level, in Sec.~\ref{sec:gluon_errors_NP} nonperturbative bottomonium results obtained from the RBC/UKQCD ensembles (using the Iwasaki gluon action)
and from the MILC ensembles \cite{Bazavov:2009bb} (using the tadpole-improved one-loop L\"uscher-Weisz action \cite{Alford:1995hw}) are compared. This comparison
also includes all the bottomonium spin splittings considered in this paper, and leads to nonperturbative estimates of gluonic discretization errors for them.

\subsection{\label{sec:gluon_errors_tree}Lattice potential model using tree-level perturbation theory}

\subsubsection{The model}

The discretization errors caused by the gluon action in radial and orbital bottomonium energy splittings
can be estimated using a potential model on a three-dimensional cubic lattice with Hamiltonian
\begin{equation}
 H = -\frac{\Delta}{m_b}+V, \label{eq:H_pot}
\end{equation}
where $\Delta$ is a lattice Laplace operator and $V(\bs{r})$ is the static quark-antiquark potential derived from the lattice
gluon action in use. In the model employed here, $V(\bs{r})$ is taken to be of the form
\begin{equation}
 V(\bs{r})=V^{\rm lat,0}(\bs{r})+\kappa|\bs{r}|, \label{eq:Vpot}
\end{equation}
where $V^{\rm lat,0}(\bs{r})$ is the tree-level lattice potential that is obtained from the tree-level lattice 
gluon propagator $G^{\rm lat}_{\mu\nu}(q)$ as follows:
\begin{equation}
V^{\rm lat,0}(\bs{r})= -\frac43\: g^2 \int_{|q_j|\leq\frac{\pi}{a}}\frac{\mathrm{d}^3q}{(2\pi)^3} \: e^{i\bs{q}\cdot\bs{r}}\: G^{\rm lat}_{00}(\bs{q},0).
\label{eq:Vlat}
\end{equation}
For $a\rightarrow0$, the potential $V^{\rm lat,0}(\bs{r})$ approaches the continuum Coulomb potential $V^{0}(\bs{r})=-(4/3)\alpha_s/|\bs{r}|$ (with $\alpha_s=g^2/(4\pi)$).
The linear term $\kappa|\bs{r}|$ in (\ref{eq:Vpot}) is added to describe the nonperturbative long-distance behavior of the quark-antiquark potential.
In Ref.~\cite{Bali:1998pi} a similar model on a cubic lattice was considered, with a discrete Laplacian but with the continuum form of the potential. In the following, the same parameters as in \cite{Bali:1998pi} are used:
$\sqrt{\kappa}=468$ MeV, $\alpha_s=0.24$, and $m_b=4.676$ GeV. Note that in \cite{Bali:1998pi} the coordinate system was chosen such that the origin $\bs{r}=0$
was at the center of an elementary cube, in order to avoid the singularity of the continuum Coulomb potential. In contrast, here the point $\bs{r}=0$
is a regular lattice point, and the lattice potential (\ref{eq:Vlat}) is finite at that point (one has $V^{\rm lat,0}(0)\propto 1/a$).
In fact, the dominant gluon discretization errors arise at and near the point $\bs{r}=0$.

The propagator $G^{\rm lat}_{\mu\nu}(q)$ for the action (\ref{eq:gauge_action}) can be found in \cite{Weisz:1983bn}. The 0-0-component at $q_0=0$ has the simple form
\begin{equation}
G^{\rm lat}_{00}(\bs{q},0)=\frac{1}{\left(\frac{2}{a}\right)^2\sum_{j=1}^3 \sin^2\left(\frac{a q_j}{2}\right) - c_1 a^2 \left(\frac{2}{a}\right)^4 \sum_{j=1}^3 \sin^4\left(\frac{a q_j}{2}\right) }.
\end{equation}
The Iwasaki gluon action \cite{Iwasaki:1983ck,Iwasaki:1984cj, Iwasaki:1996sn} has $c_1=-0.331$. Results will also be given
for $c_1=0$ (the simple plaquette action), $c_1=-1/12$ (the tree-level L\"uscher-Weisz action \cite{Luscher:1984xn, Luscher:1985zq}),
and $c_1=-1.40686$ (the DBW2 action \cite{Takaishi:1996xj, deForcrand:1999bi}). For these choices of $c_1$ and for all points $\bs{r}$ with $|r_i/a|\leq60$
the integral (\ref{eq:Vlat}) was computed numerically.

For the Laplace operator $\Delta$ in (\ref{eq:H_pot}), three different discretizations are considered:
\begin{equation}
 \Delta=\left\{\begin{array}{ll} \sum_{j=1}^3\nabla^+_j\nabla^-_j, & \hspace{4ex}{\rm unimproved}, \\ 
\sum_{j=1}^3\nabla^+_j\nabla^-_j-(a^2/12)\sum_{j=1}^3\left[\nabla^+_j\nabla^-_j\right]^2, & \hspace{4ex}\mathcal{O}(a^2){\rm\mbox{-}improved}, \\
\sum_{j=1}^3\nabla^+_j\nabla^-_j-(a^2/12)\sum_{j=1}^3\left[\nabla^+_j\nabla^-_j\right]^2+(a^4/90)\sum_{j=1}^3\left[\nabla^+_j\nabla^-_j\right]^3, & \hspace{4ex}\mathcal{O}(a^4){\rm\mbox{-}improved}, \end{array} \right. \label{eq:Laplace}
\end{equation}
with $\nabla^+_j\psi(\bs{r})=\left[\psi(\bs{r}+a\bs{e}_j)-\psi(\bs{r})\right]/a$ and $\nabla^-_j\psi(\bs{r})=\left[\psi(\bs{r})-\psi(\bs{r}-a\bs{e}_j)\right]/a$.
As shown by the results in the next section, when the $\mathcal{O}(a^4)$-improved Laplacian is used, the discretization errors associated with $\Delta$ are in most cases much
smaller than the gluonic discretization errors associated with (\ref{eq:Vlat}).

The low-lying eigenvalues and eigenfunctions of the Hamiltonian (\ref{eq:H_pot}) were computed numerically
for lattices with a physical side length of $2.7$ fm and lattice spacings in the range from $0.0223$ fm to $0.208$ fm.
As in \cite{Bali:1998pi}, only one octant of the lattice was simulated. $S$-wave states ($A_1$ representation) and $D$-wave states ($E$ representation)
were obtained by using periodic boundary conditions in all three lattice directions;
$P$-wave states ($T_1$ representation) were obtained by using antiperiodic boundary conditions in the $r_3$-direction and
periodic boundary conditions in the $r_1$- and $r_2$-directions.

\subsubsection{\label{sec:gluon_errors_tree_results}Results}

\begin{figure*}[h!]
 \includegraphics[width=0.46\linewidth]{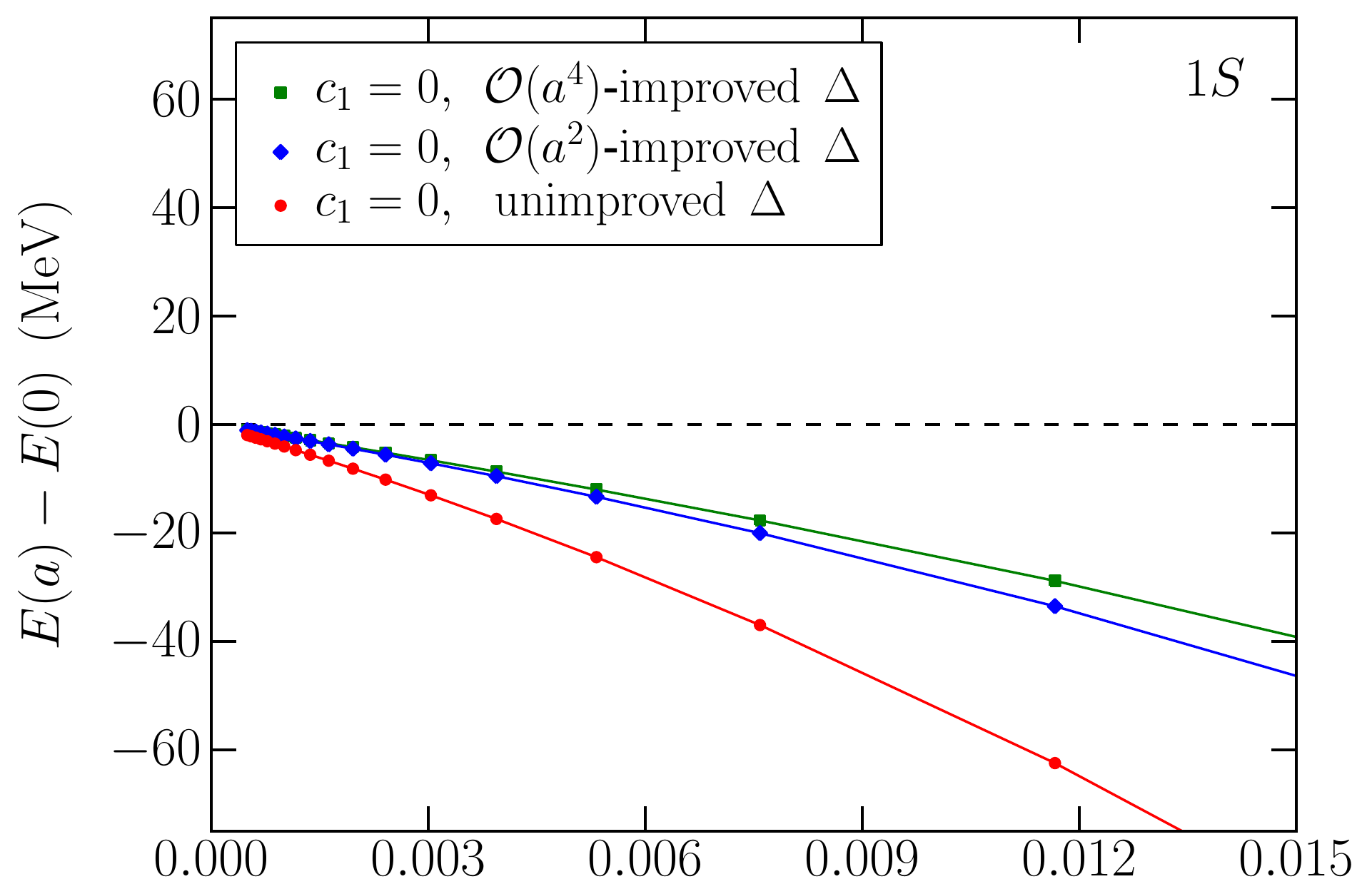}   \hfill \includegraphics[width=0.46\linewidth]{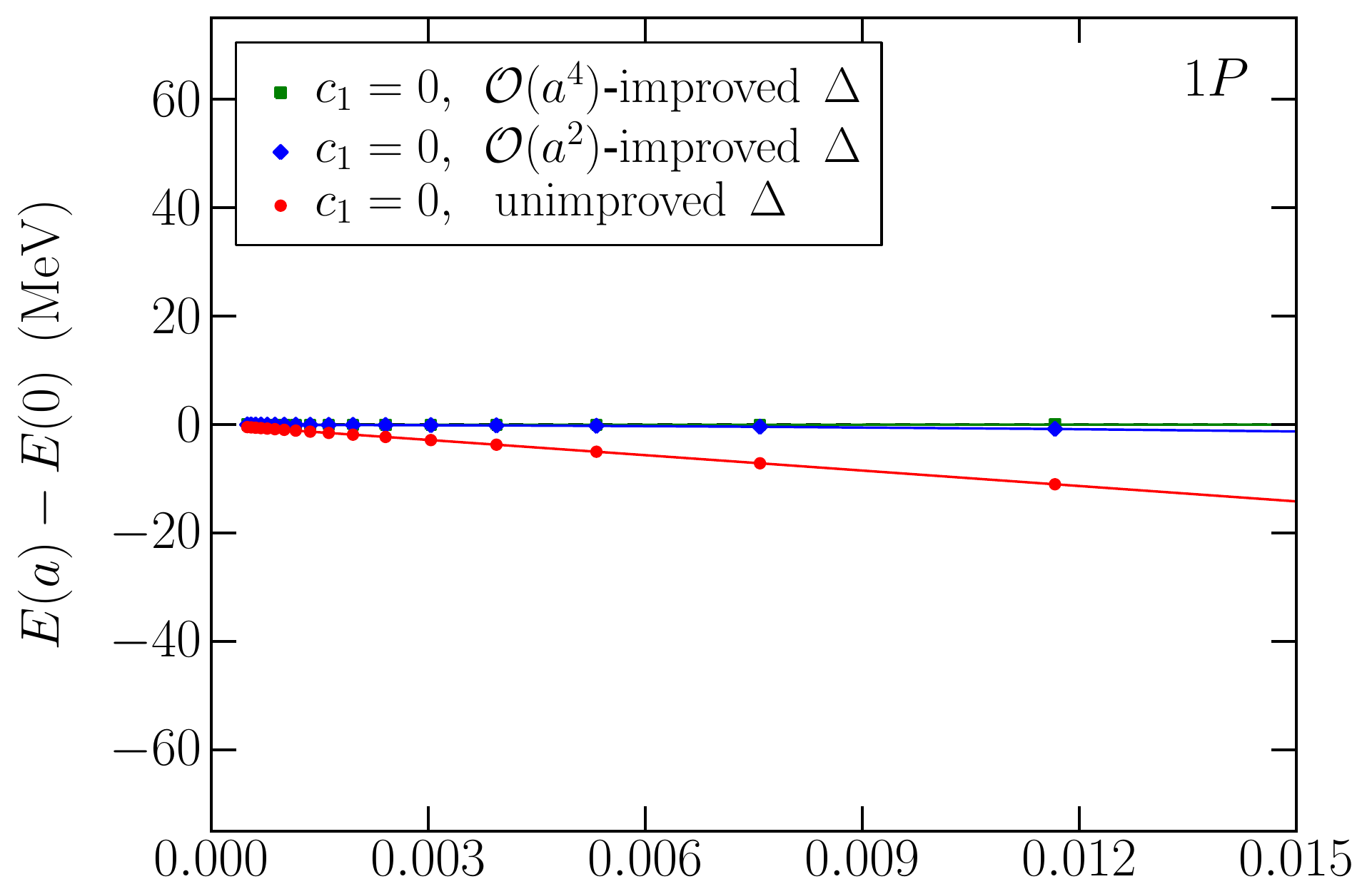}
 \includegraphics[width=0.46\linewidth]{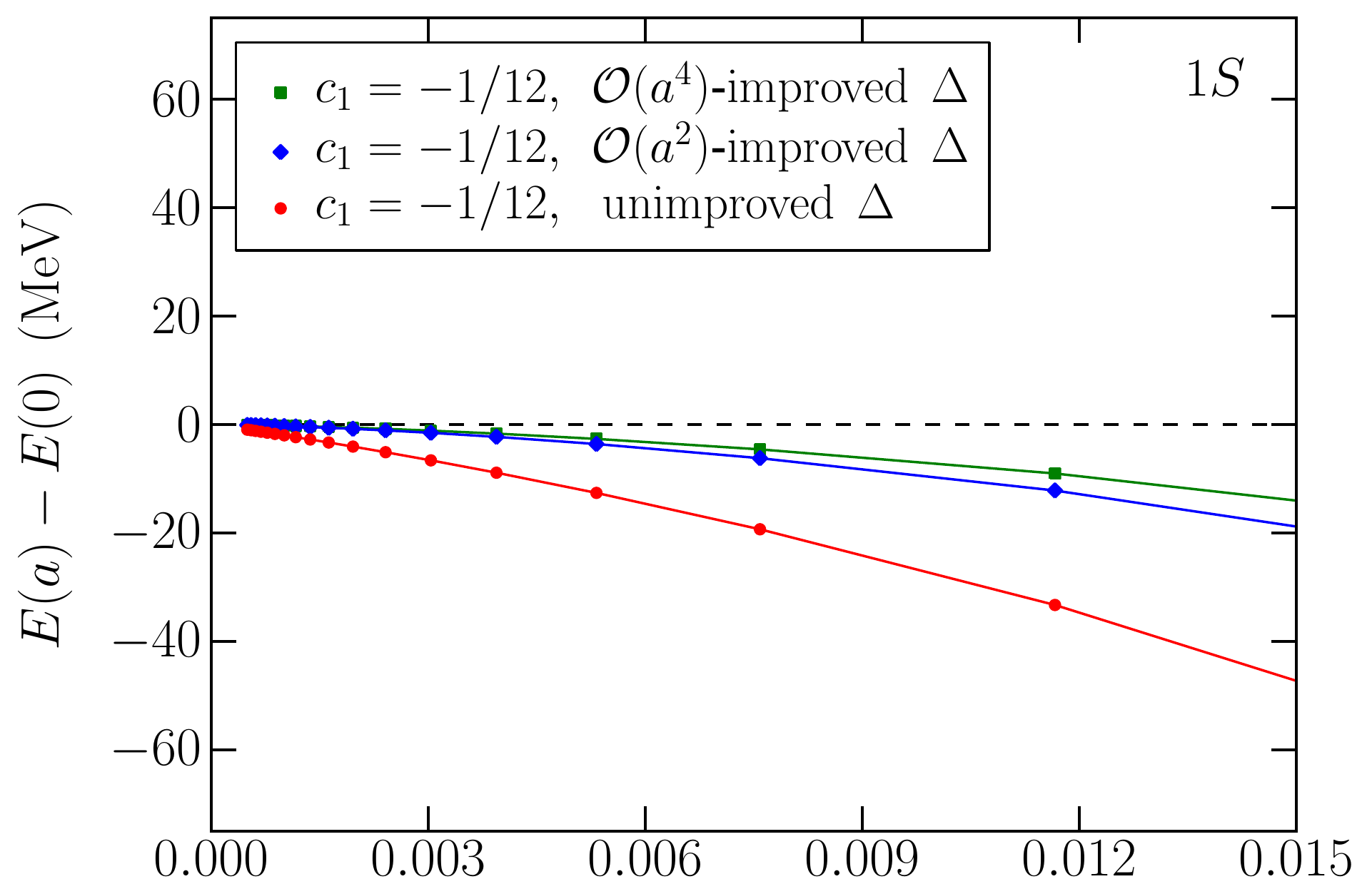}       \hfill \includegraphics[width=0.46\linewidth]{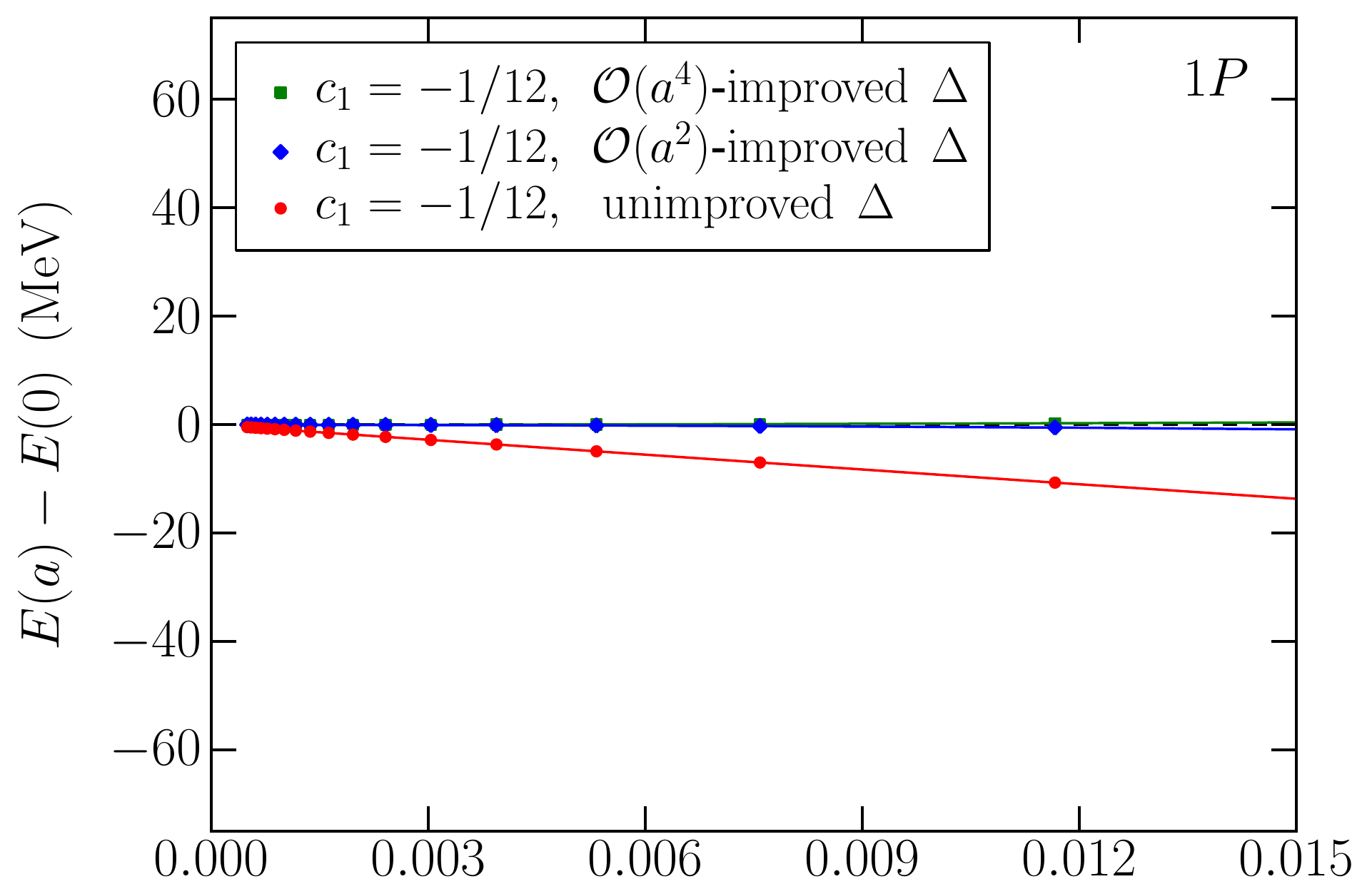}
 \includegraphics[width=0.46\linewidth]{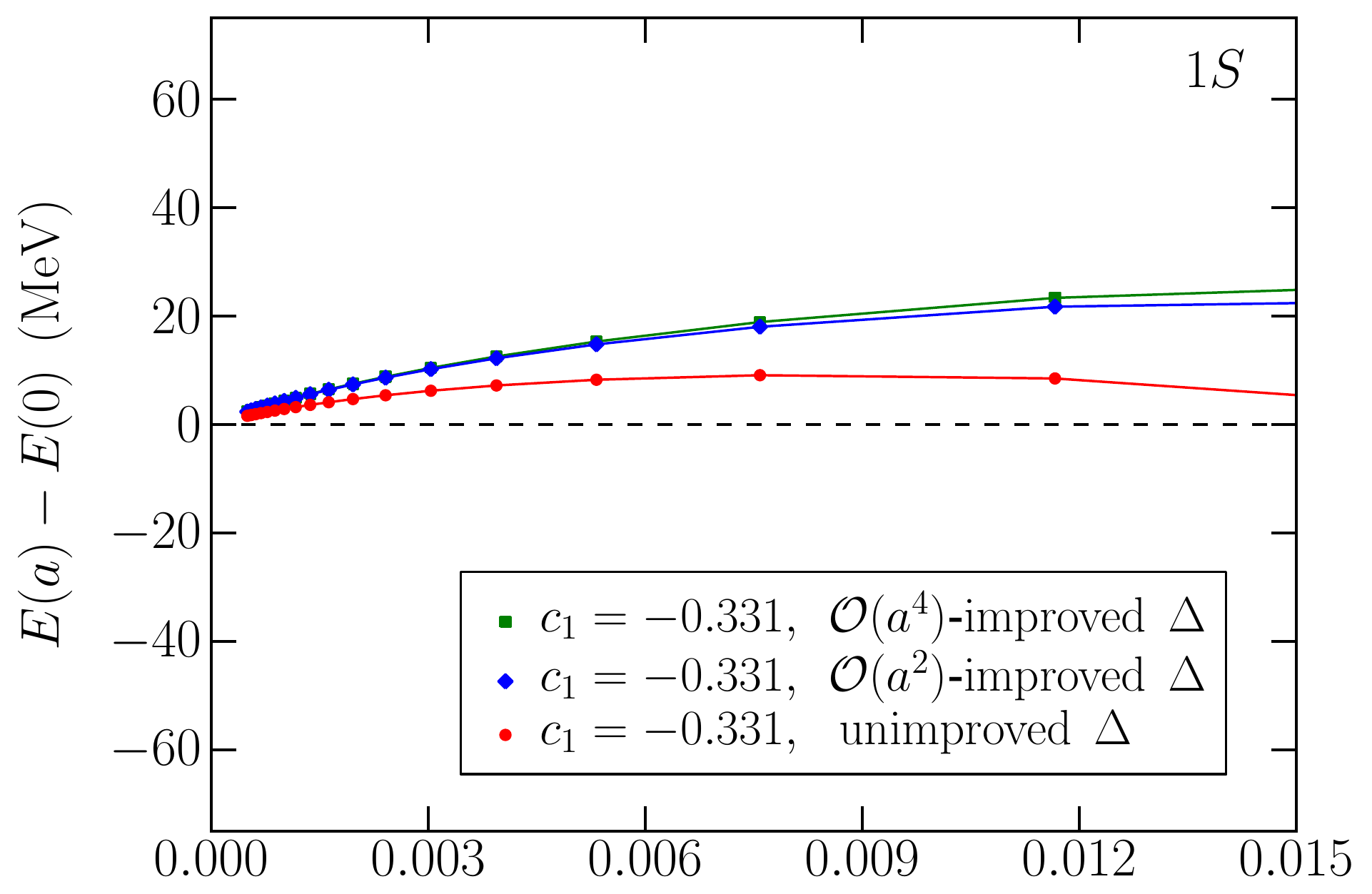}  \hfill \includegraphics[width=0.46\linewidth]{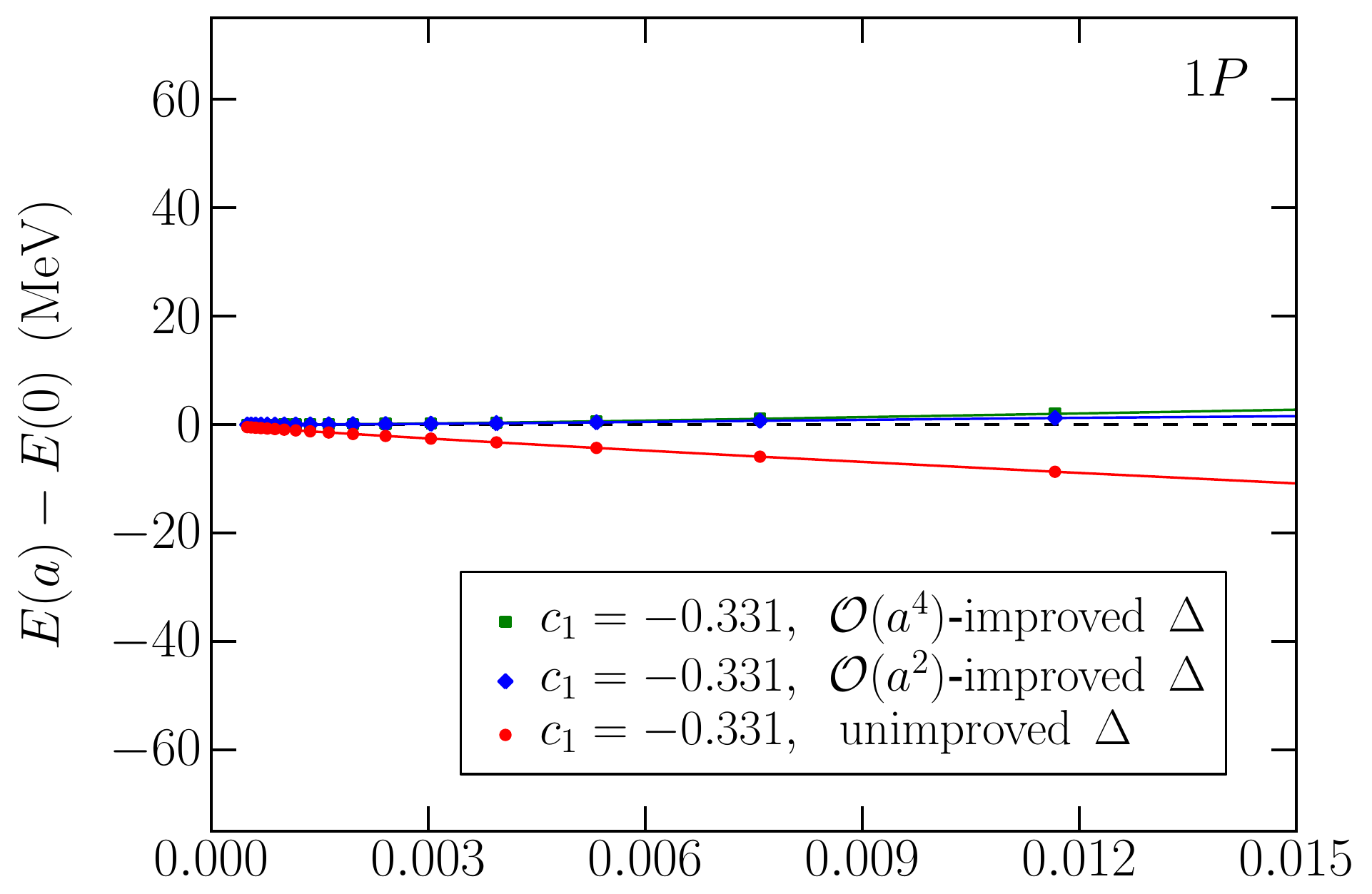}
 \includegraphics[width=0.46\linewidth]{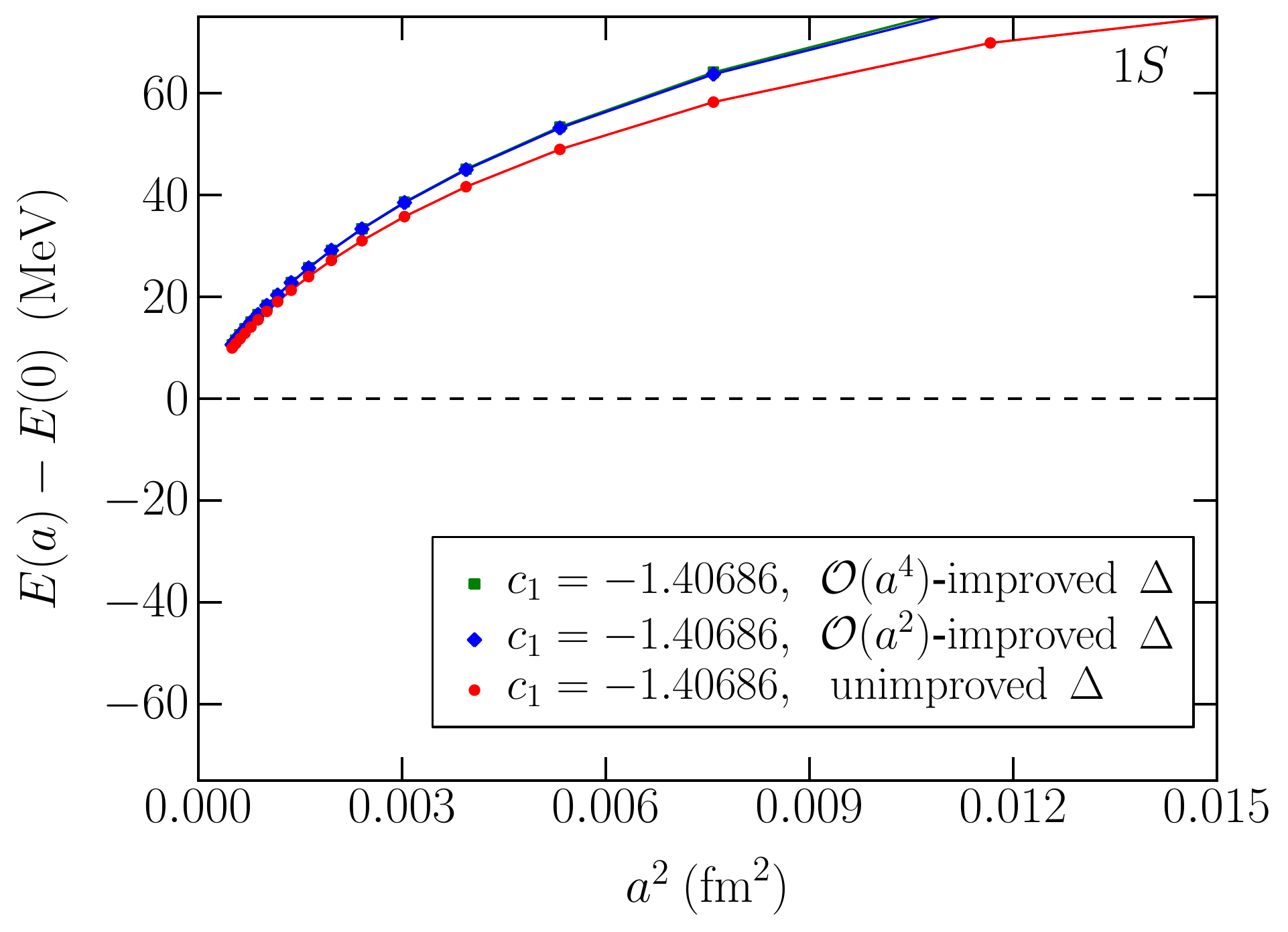}     \hfill \includegraphics[width=0.46\linewidth]{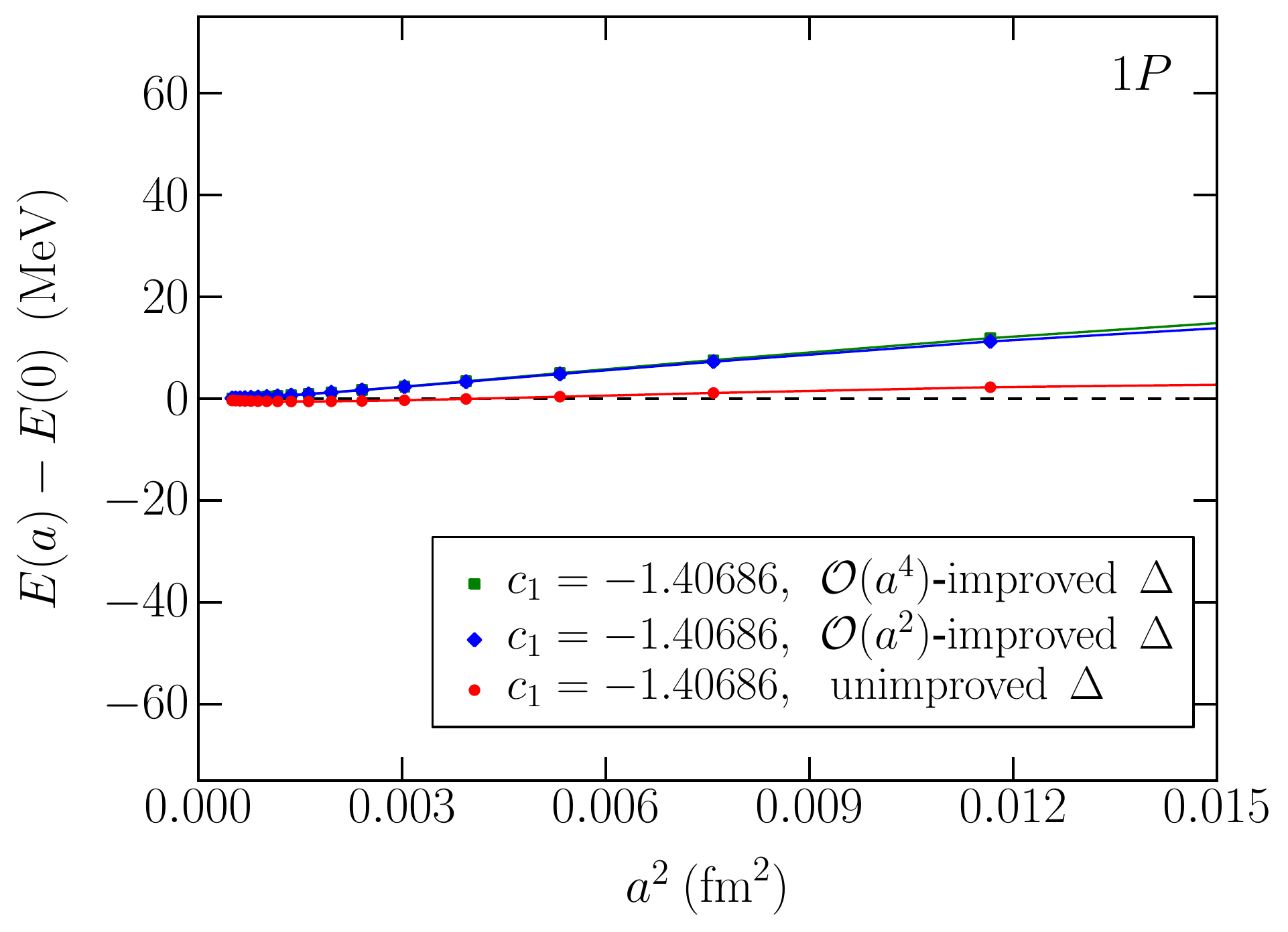}
\caption{\label{fig:Laplace_improvement}Shift in the lattice potential model $1S$ and $1P$ energies as a function of the lattice spacing,
for different levels of Symanzik improvement in the Laplace operator.}
\end{figure*}

Figure \ref{fig:Laplace_improvement} shows the deviations in the $1S$ and $1P$ energies from their continuum values
as a function of $a^2$, for the three different levels of Symanzik improvement in the lattice Laplacian defined in Eq.~(\ref{eq:Laplace}).
As can be seen in the figure, the difference between the results from the $\mathcal{O}(a^2)$- and $\mathcal{O}(a^4)$-improved Laplace operators is small,
much smaller than the difference between the results from the unimproved and $\mathcal{O}(a^2)$-improved
Laplace operators. This indicates that the remaining discretization errors in the $\mathcal{O}(a^4)$-improved Laplacian are negligible.
However, for the $S$-wave state a significant shift in the energy from its continuum value remains at finite lattice spacing.
This error can be interpreted as the tree-level gluonic contribution to the discretization errors, stemming from the use of the lattice potential (\ref{eq:Vlat}).
As expected, at small lattice spacings, the $\mathcal{O}(a^2)$-improved L\"uscher-Weisz action shows significantly smaller
tree-level discretization errors than the other actions.

For the $P$-wave states, the gluonic discretization errors are much smaller than for the $S$ wave states.
This is expected because the dominant correction in the potential arises at the origin, where the wave function
vanishes for all states other than the $S$-wave states. Only the DBW2 action leads to significant tree-level
gluon discretization errors in the $1P$ energy.

\begin{figure*}[h!]
 \includegraphics[width=0.45\linewidth]{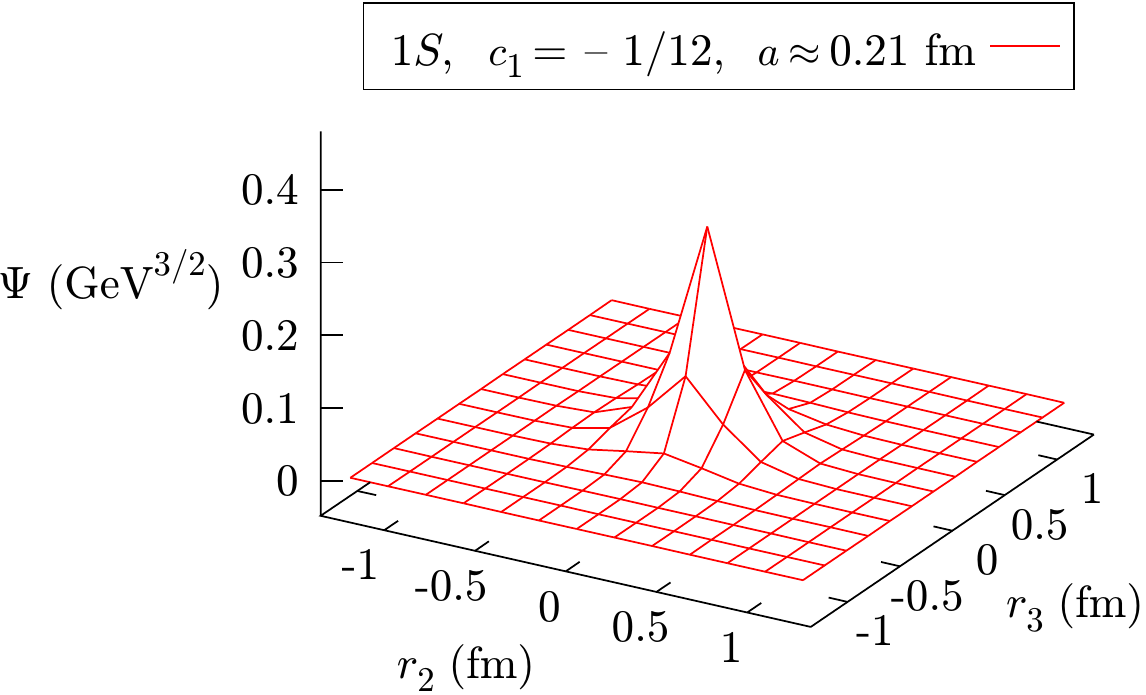}  \hfill \includegraphics[width=0.45\linewidth]{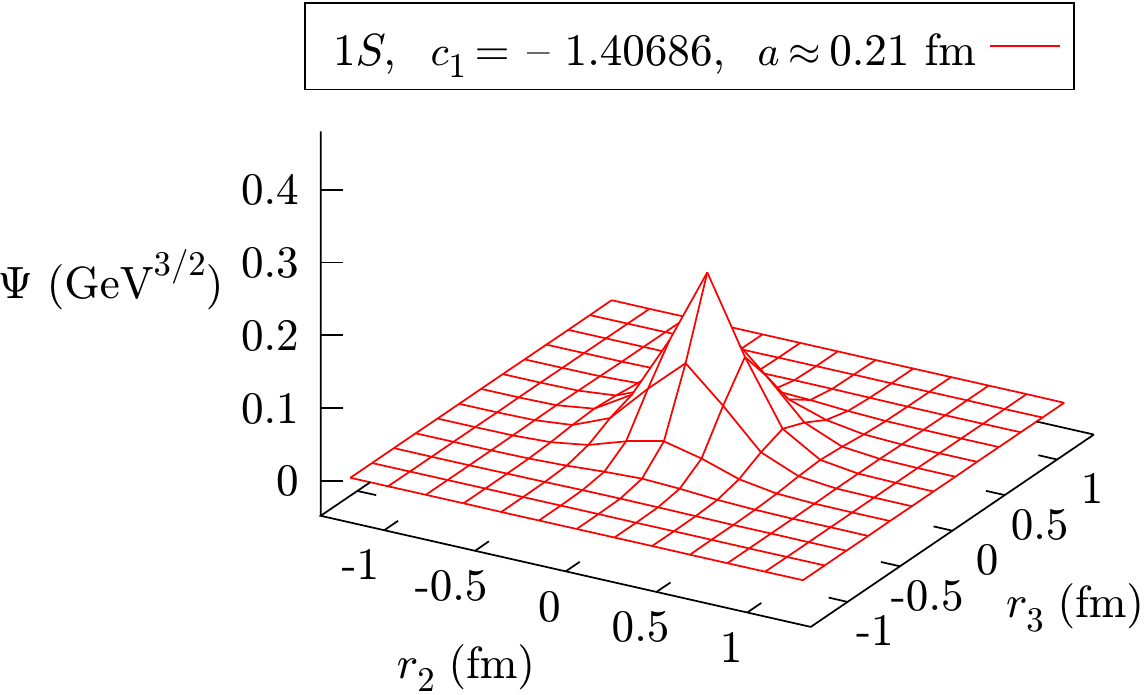}

\vspace{3ex}

 \includegraphics[width=0.45\linewidth]{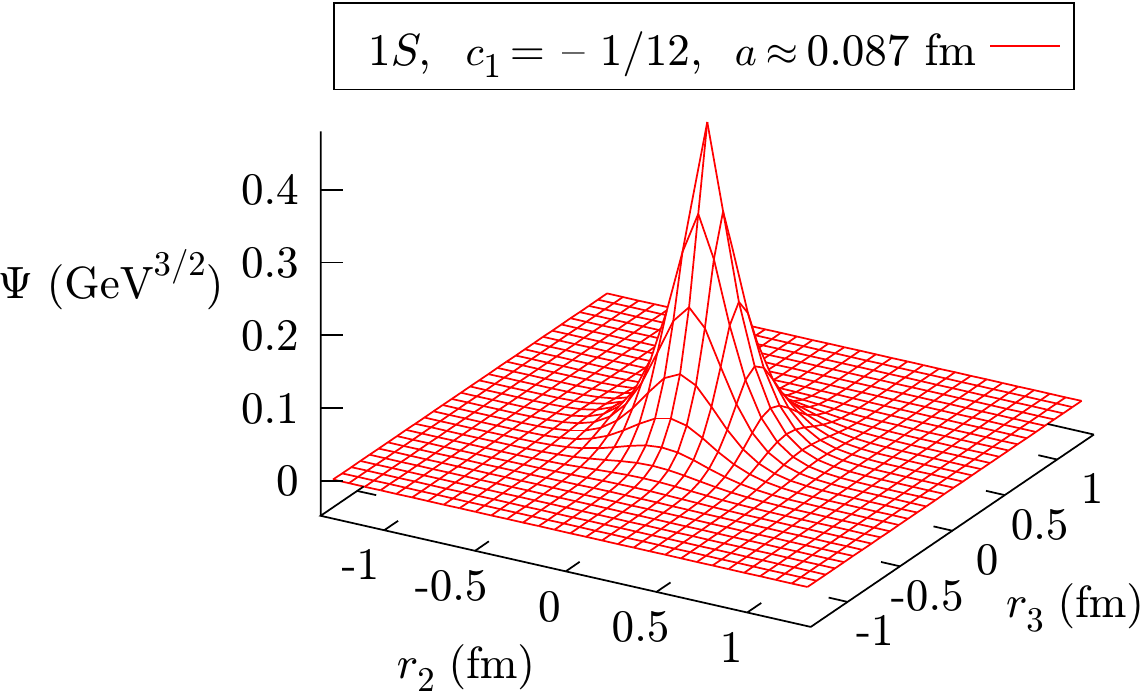}    \hfill \includegraphics[width=0.45\linewidth]{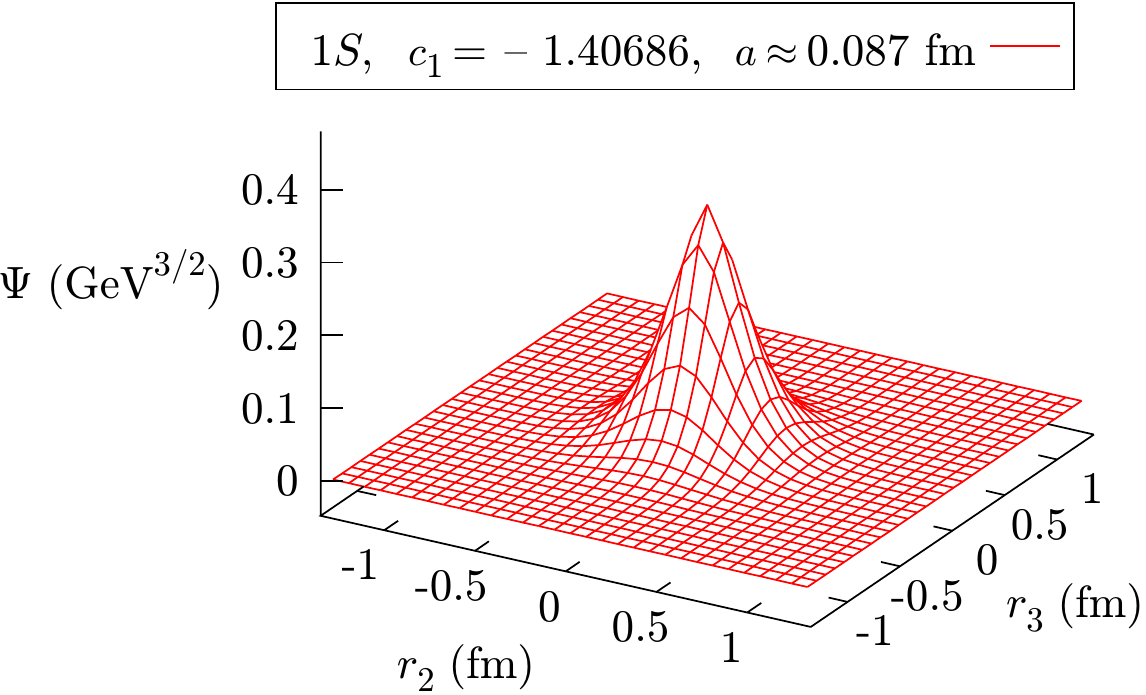}

\vspace{3ex}

 \includegraphics[width=0.45\linewidth]{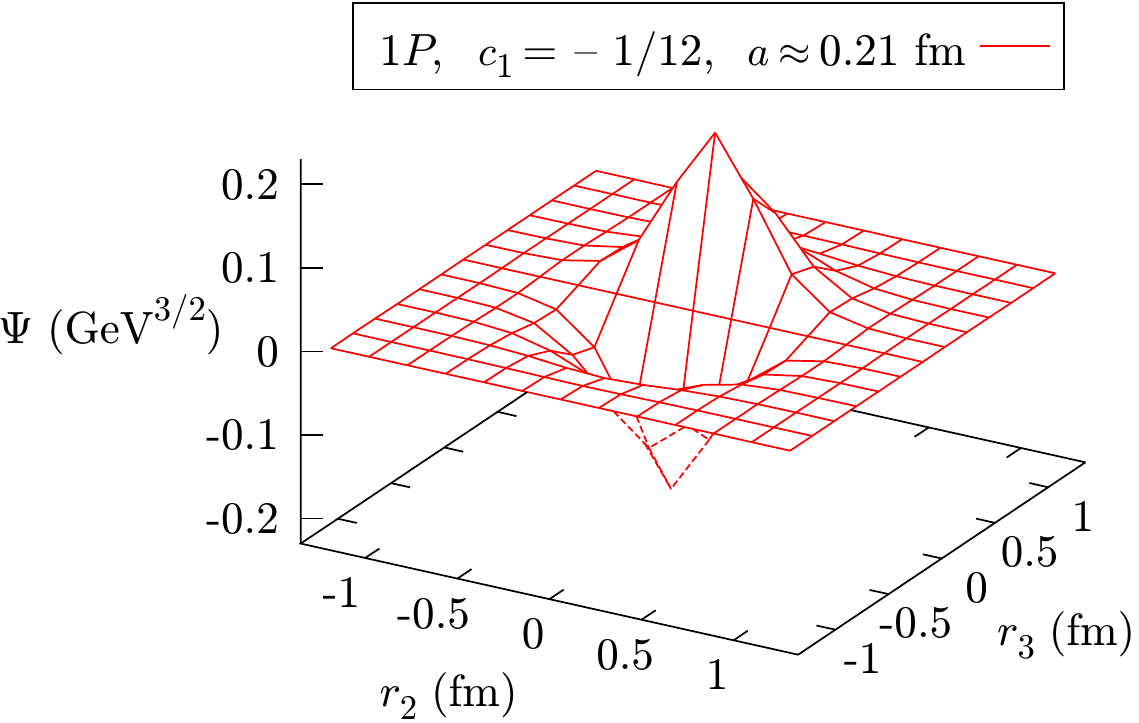}  \hfill \includegraphics[width=0.45\linewidth]{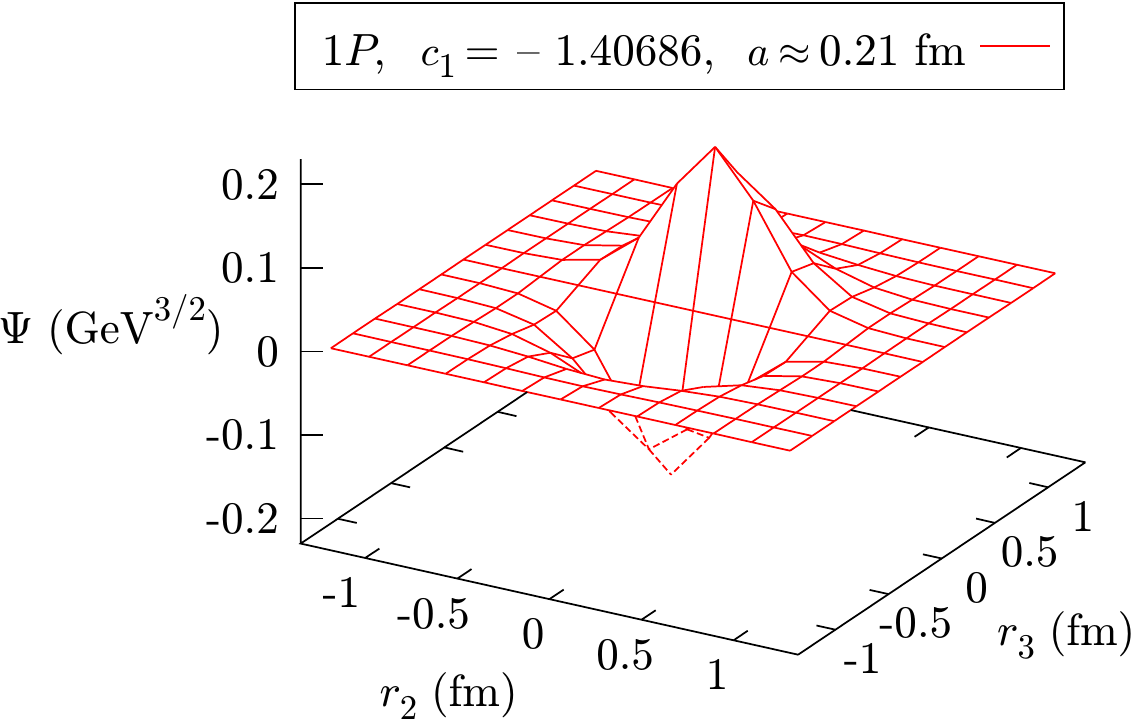}

\vspace{3ex}

 \includegraphics[width=0.45\linewidth]{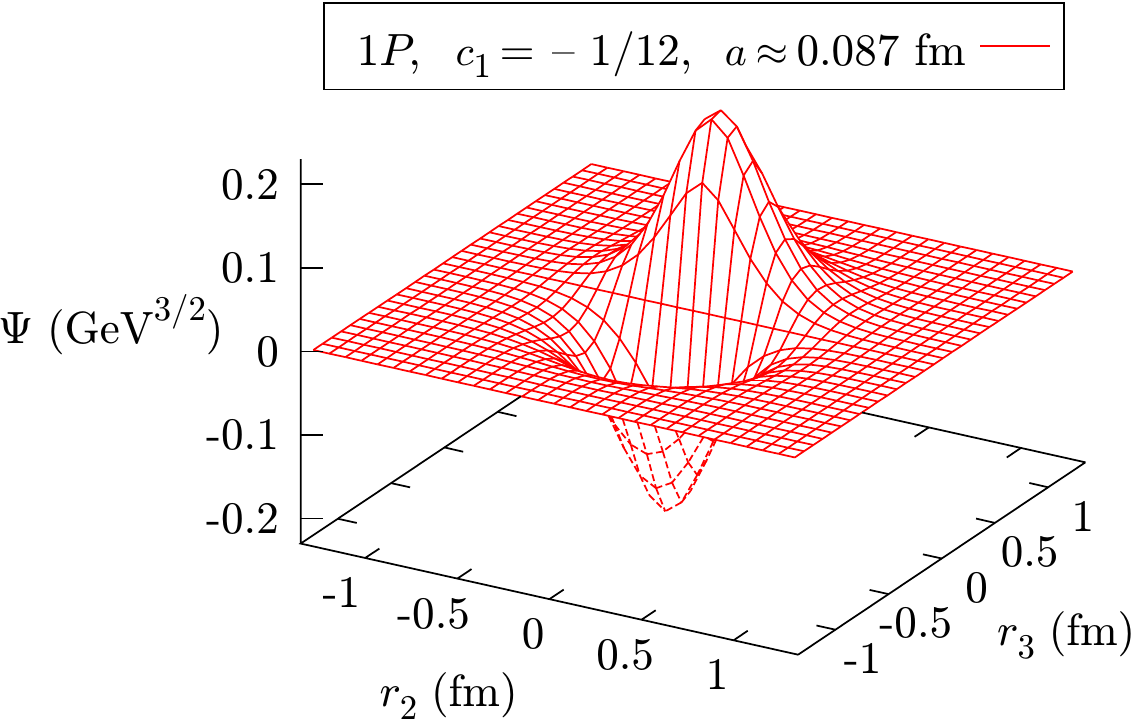}    \hfill \includegraphics[width=0.45\linewidth]{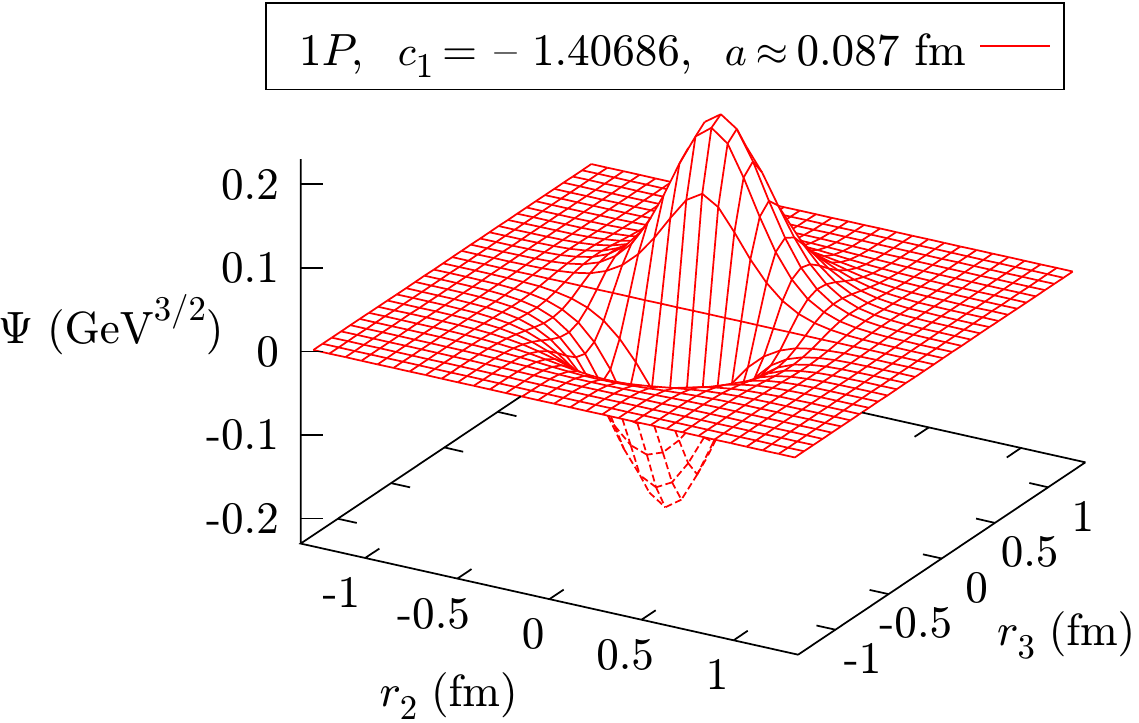}
\caption{\label{fig:wave_funcs}Lattice potential model wave functions $\Psi(r_1, r_2, r_3)$ at $r_1=0$ for the $1S$ and $1P$ states.
Data are shown for the tree-level L\"uscher-Weisz gluon action ($c_1=-1/12$) and the DBW2 gluon action ($c_1=-1.40686$), at two different lattice spacings,
using the $\mathcal{O}(a^4)$-improved Laplacian in all cases.}
\end{figure*}

Examples of $1S$ and $1P$ eigenfunctions of the Hamiltonian (\ref{eq:H_pot}) with the $\mathcal{O}(a^4)$-improved Laplacian are shown in Fig.~\ref{fig:wave_funcs}.
The large negative coefficient $c_1=-1.40686$ of the DBW2 action leads to a visible distortion of the $1S$ wave function compared to the tree-level
L\"uscher-Weisz action with $c_1=-1/12$, while the $1P$ state is only weakly affected by the choice of $c_1$.
The broadening of the $1S$ wave function is expected because a negative coefficient $c_1$ shifts the potential at short distances upwards. In particular, at $\bs{r}=0$, one has
\begin{equation}
\frac{3a}{4g^2} \:V^{\rm lat,0}(0) \approx  \left\{\begin{array}{ll}   -0.25273\:, & \hspace{4ex}c_1=0, \\
                                                                       -0.21903\:, & \hspace{4ex}c_1=-1/12, \\
                                                                       -0.16437\:, & \hspace{4ex}c_1=-0.331, \\
                                                                       -0.09384\:, & \hspace{4ex}c_1=-1.40686.
\end{array} \right.
\end{equation}
However, note that the broadening of the lattice $1S$ wave function caused by a negative value of $c_1$ does not necessarily mean that the hyperfine
splitting is reduced. This will be discussed further in Sec.~\ref{sec:gluon_errors_NP}.

\begin{figure*}[h!]
 \includegraphics[width=0.46\linewidth]{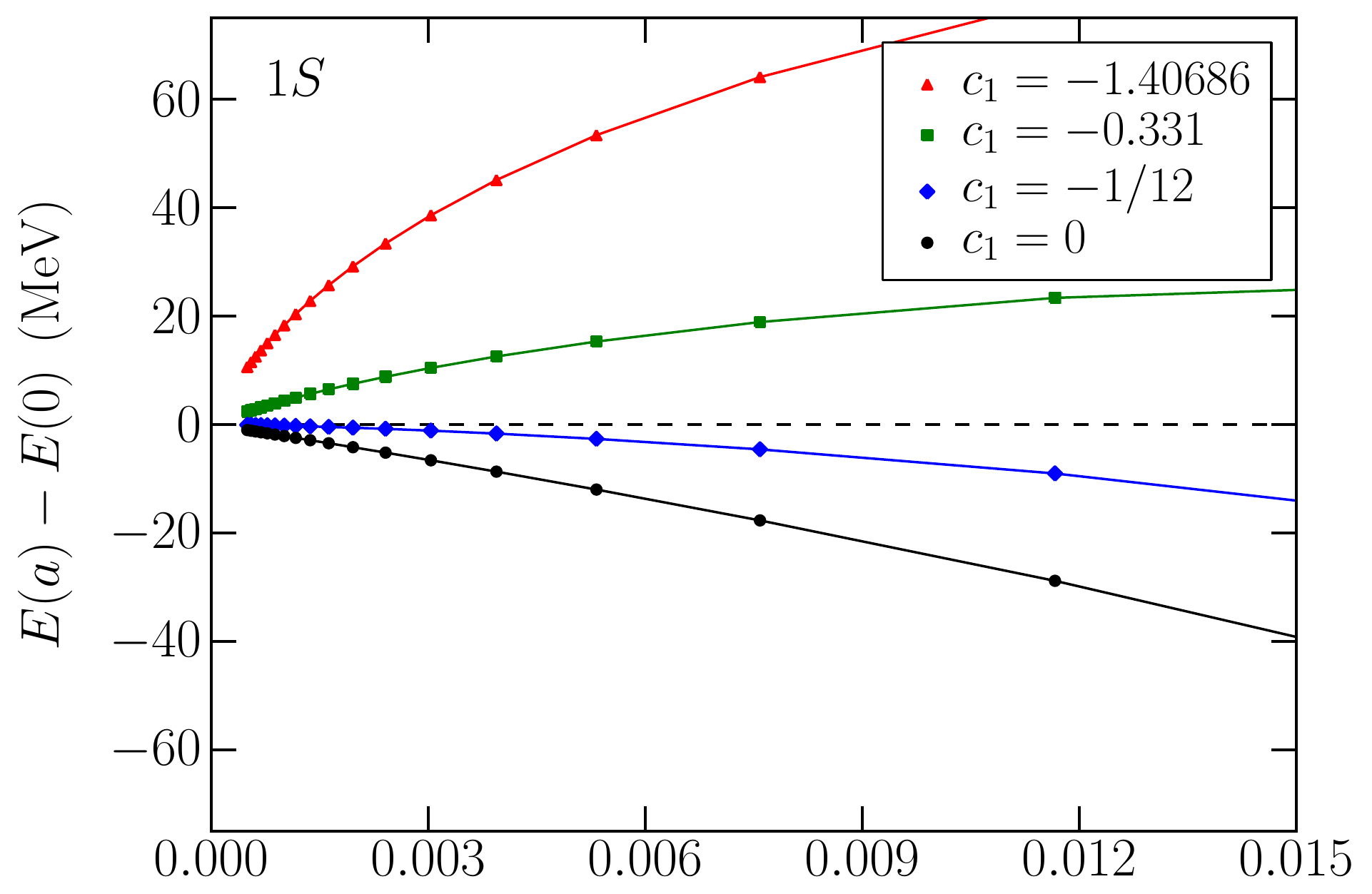}     \hfill \includegraphics[width=0.46\linewidth]{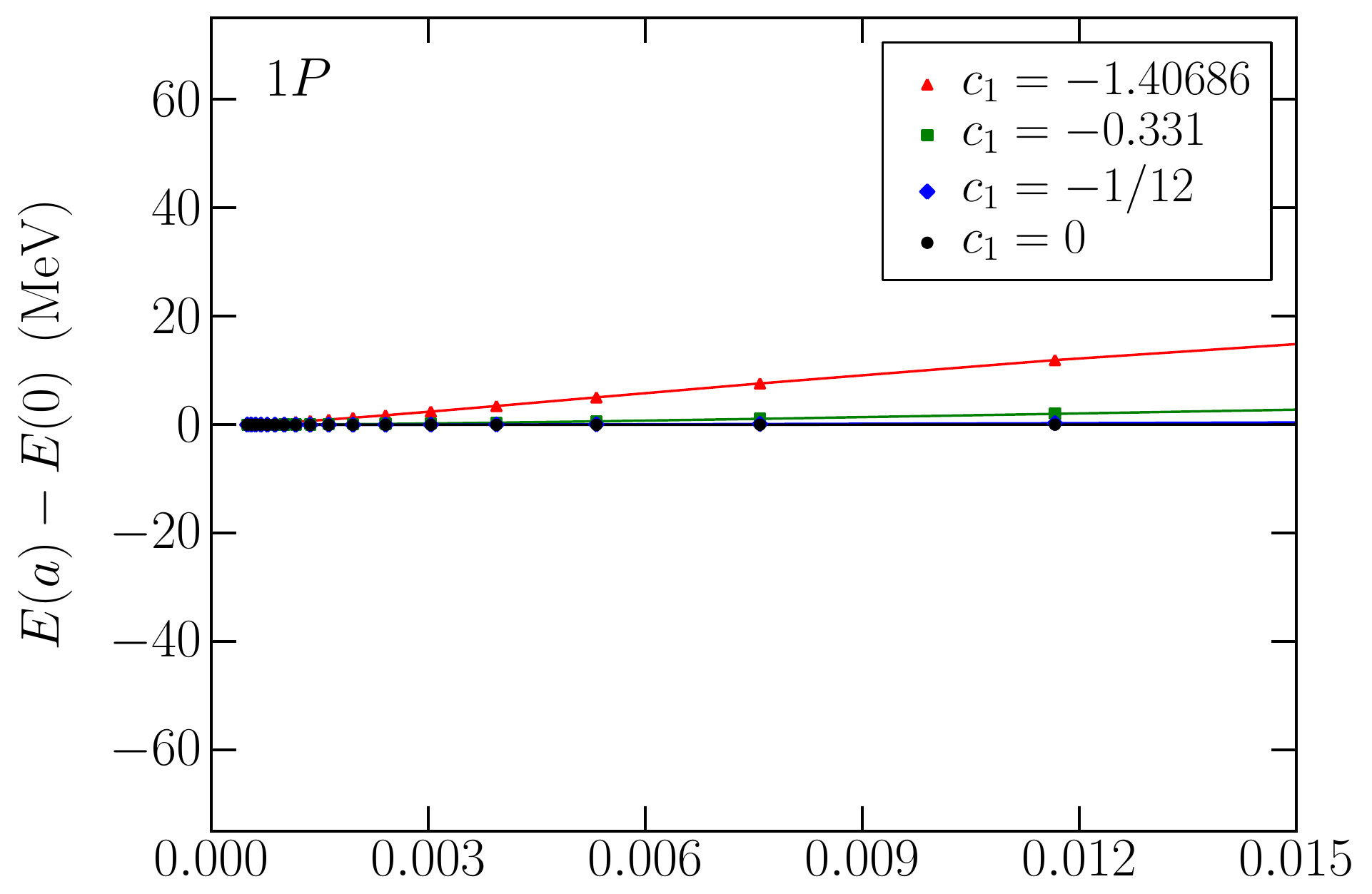}
 \includegraphics[width=0.46\linewidth]{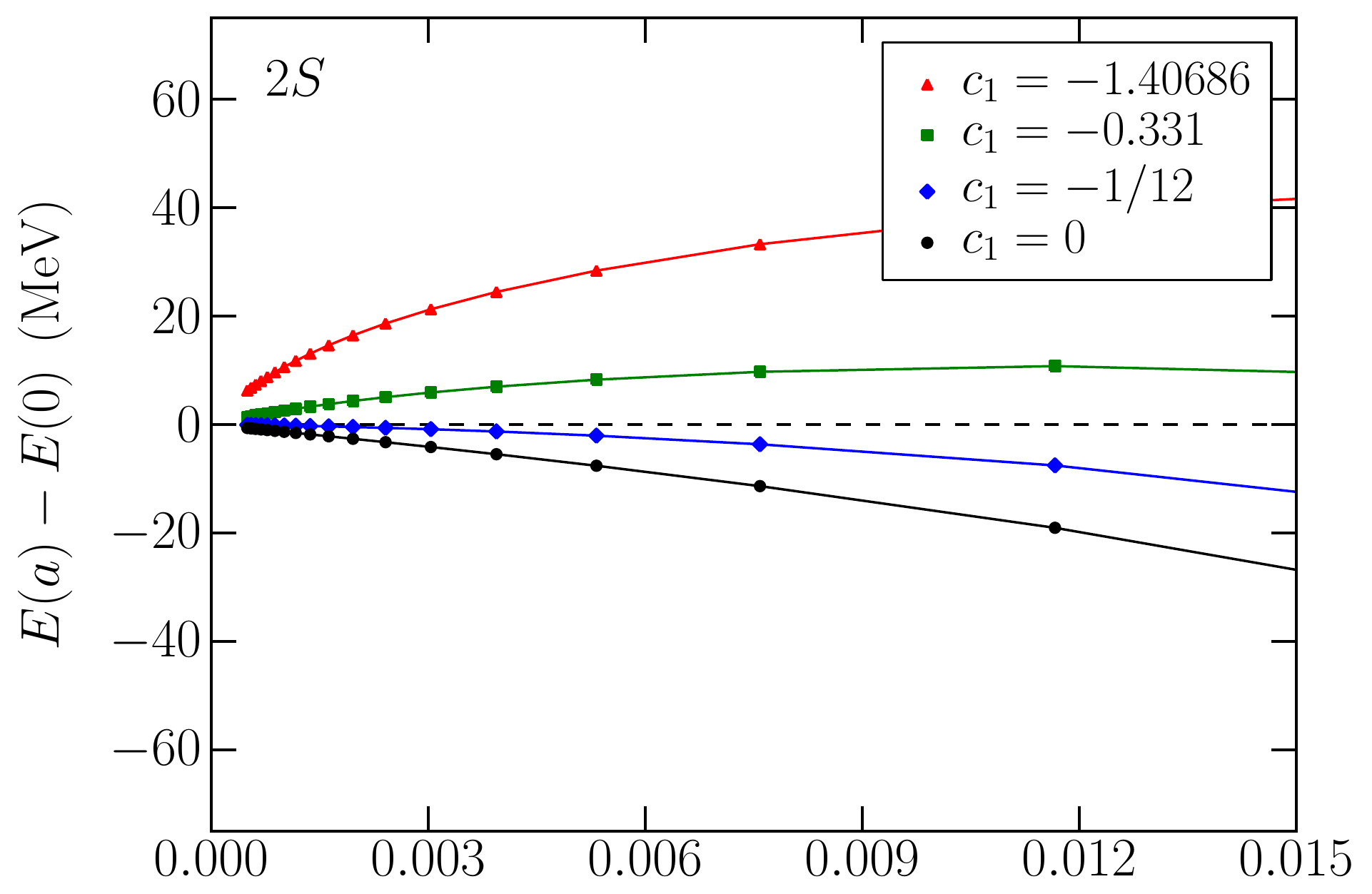}     \hfill \includegraphics[width=0.46\linewidth]{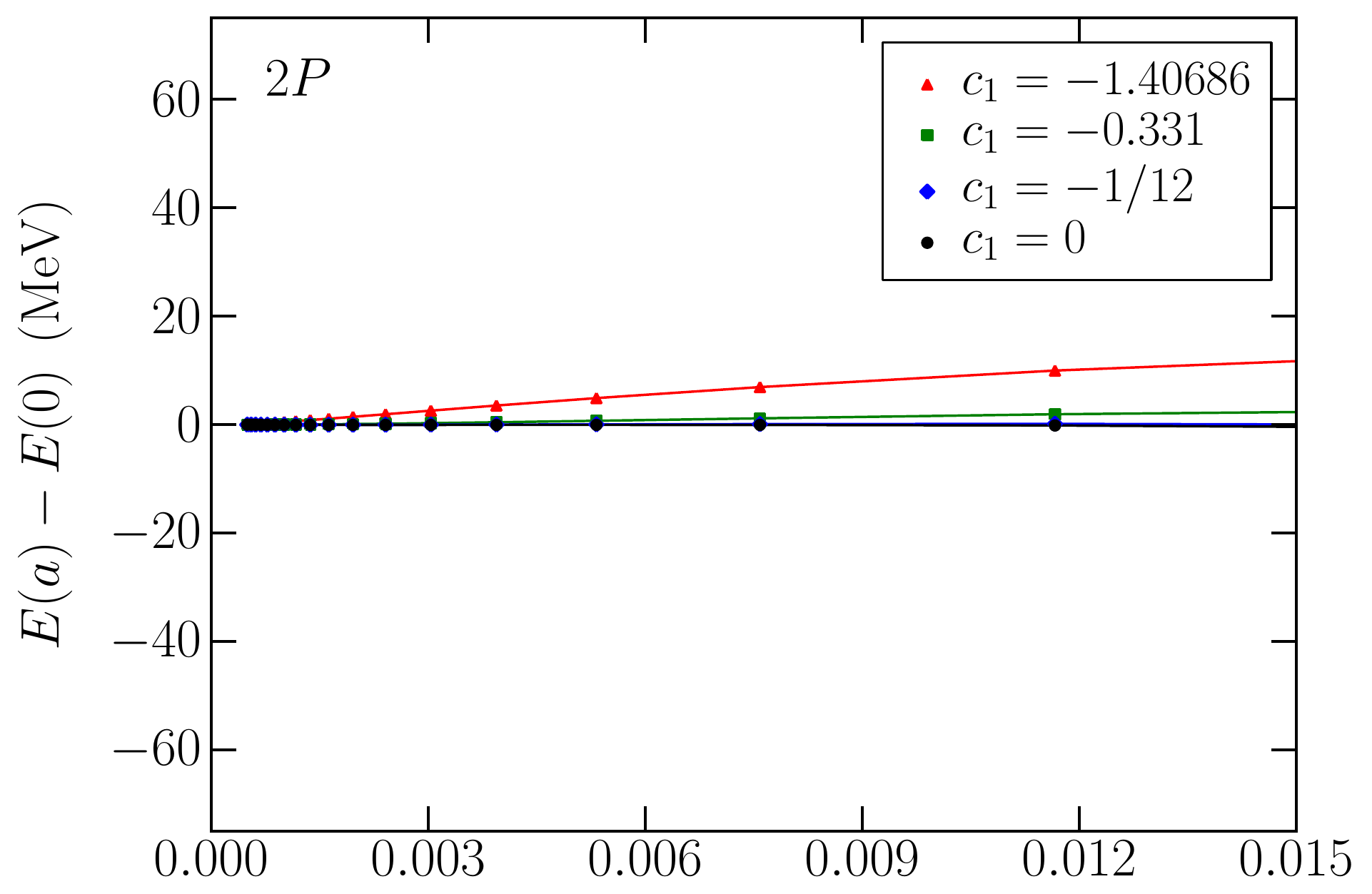}
 \includegraphics[width=0.46\linewidth]{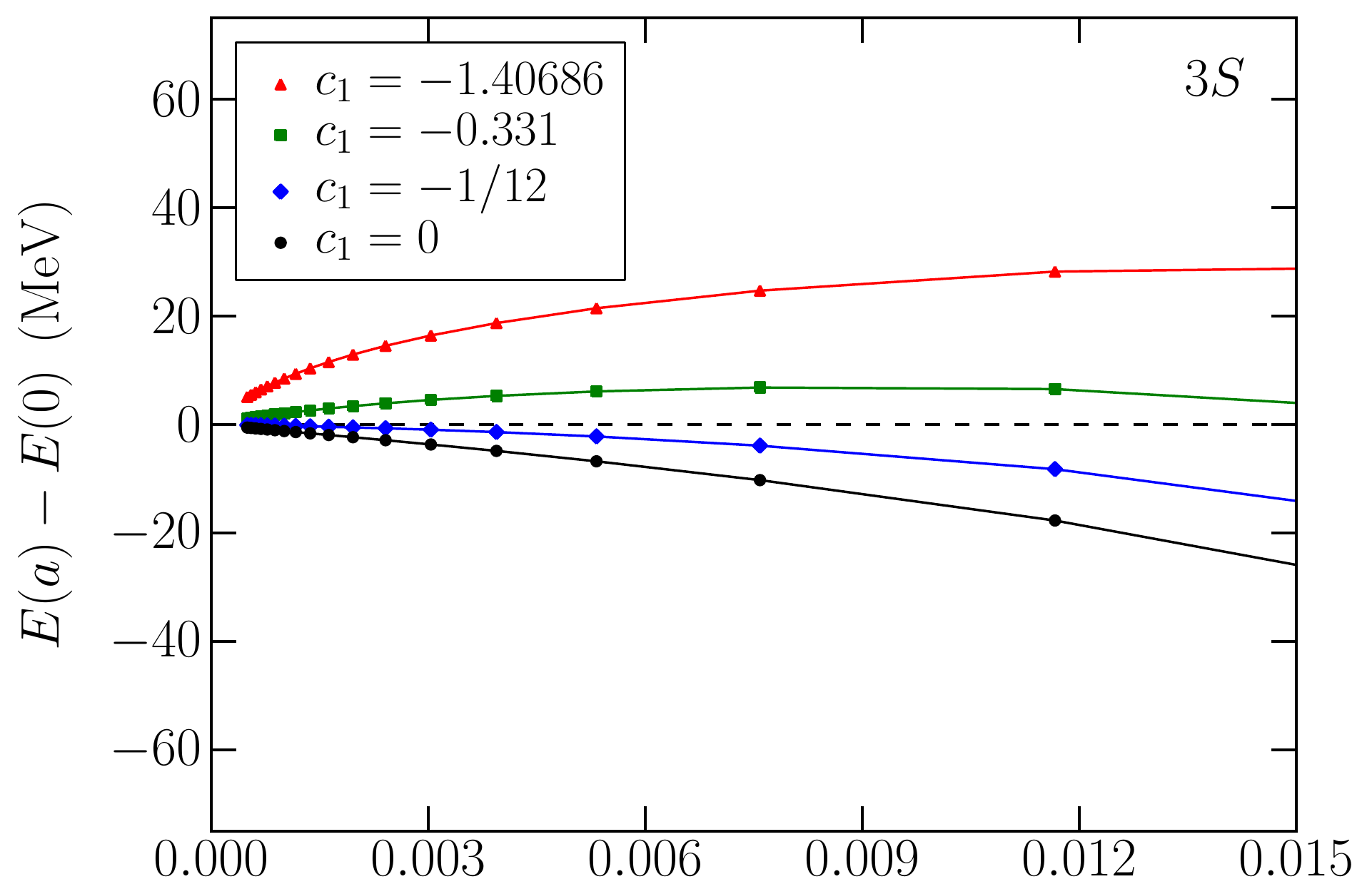}     \hfill \includegraphics[width=0.46\linewidth]{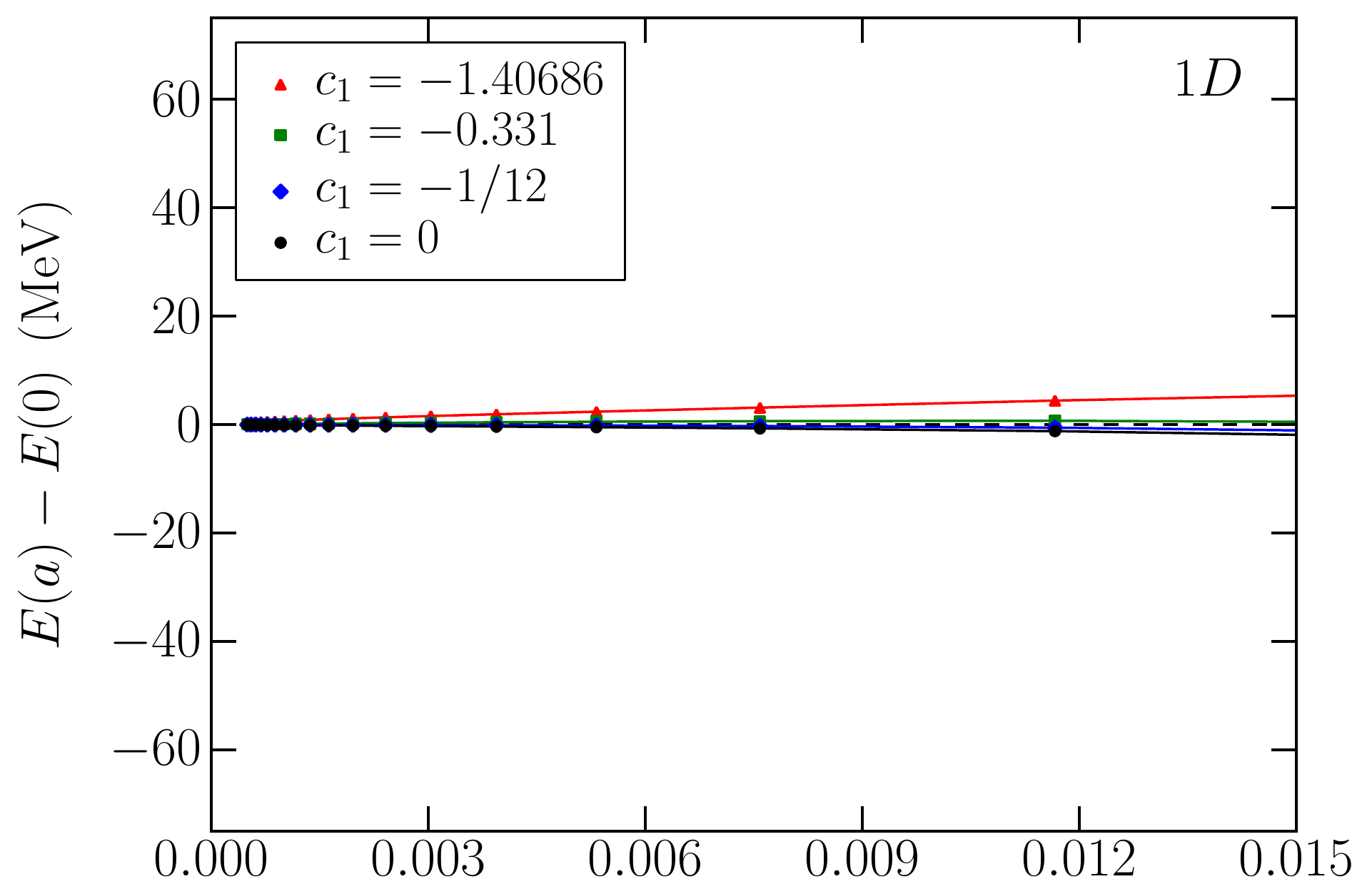}
 \includegraphics[width=0.46\linewidth]{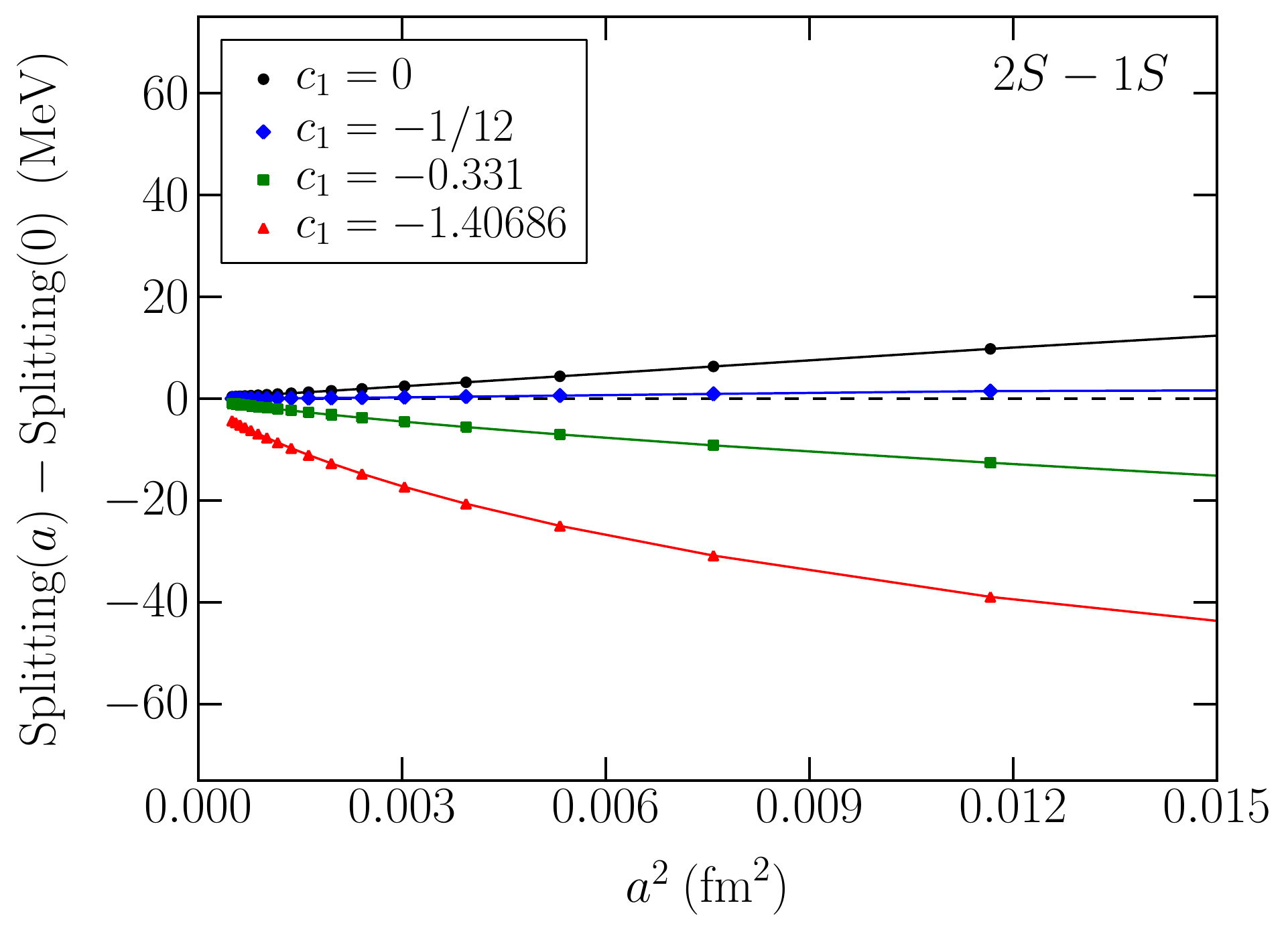}  \hfill \includegraphics[width=0.46\linewidth]{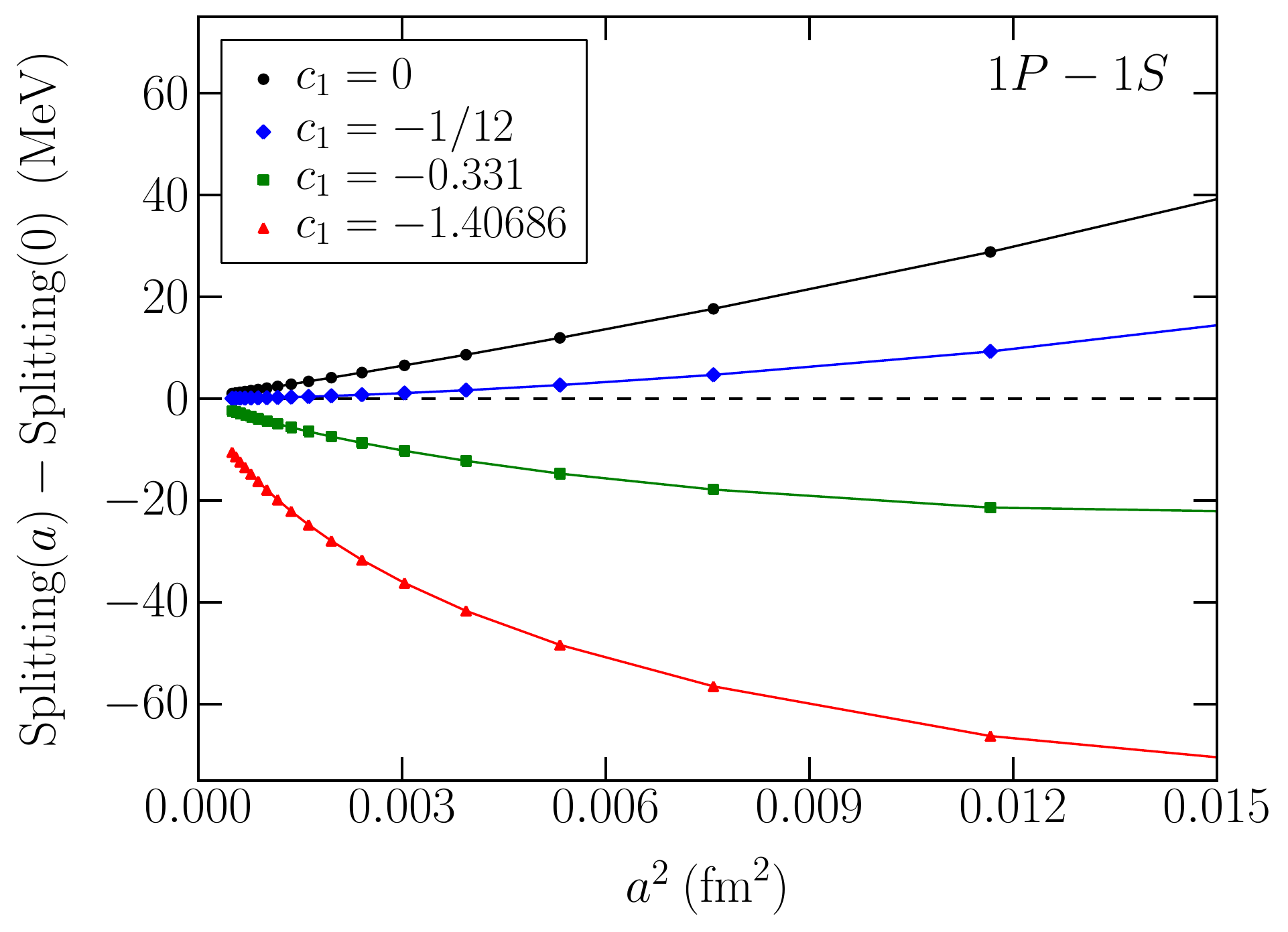}
\caption{\label{fig:gluon_errors}Shift in the lattice potential model energy levels and splittings as a function of the lattice spacing. All data shown
in this figure were generated with the $\mathcal{O}(a^4)$-improved Laplacian, so that the shifts are dominated by gluon discretization errors.}
\end{figure*}

The results for the shifts in the energies of the $1S$, $2S$, $3S$, $1P$, $2P$ and $1D$ states, obtained with the $\mathcal{O}(a^4)$-improved Laplacian
and the four different gluon actions, are summarized in Fig.~\ref{fig:gluon_errors}. This figure also shows the $2S-1S$ and $1P-1S$ splittings,
demonstrating that the $2S-1S$ splitting has smaller tree-level gluon discretization errors and is therefore better suited for
setting the lattice scale. For the Iwasaki action, the gluonic tree-level discretization errors in the $2S-1S$ splitting
are found to be about 2.6\% at $a=0.11$ fm and 1.6\% at $a=0.08$ fm, respectively. Note that the $2P-1P$ splitting is nearly free
of gluonic discretization errors and therefore appears to be a good alternative choice for the scale setting. However, in the actual
lattice QCD calculation the $2P-1P$ splitting has much larger \emph{statistical} errors than the $2S-1S$ splitting (see Sec.~\ref{sec:radial_orbital_lattice}).

\clearpage

\begin{figure*}[h!]
 \includegraphics[width=0.46\linewidth]{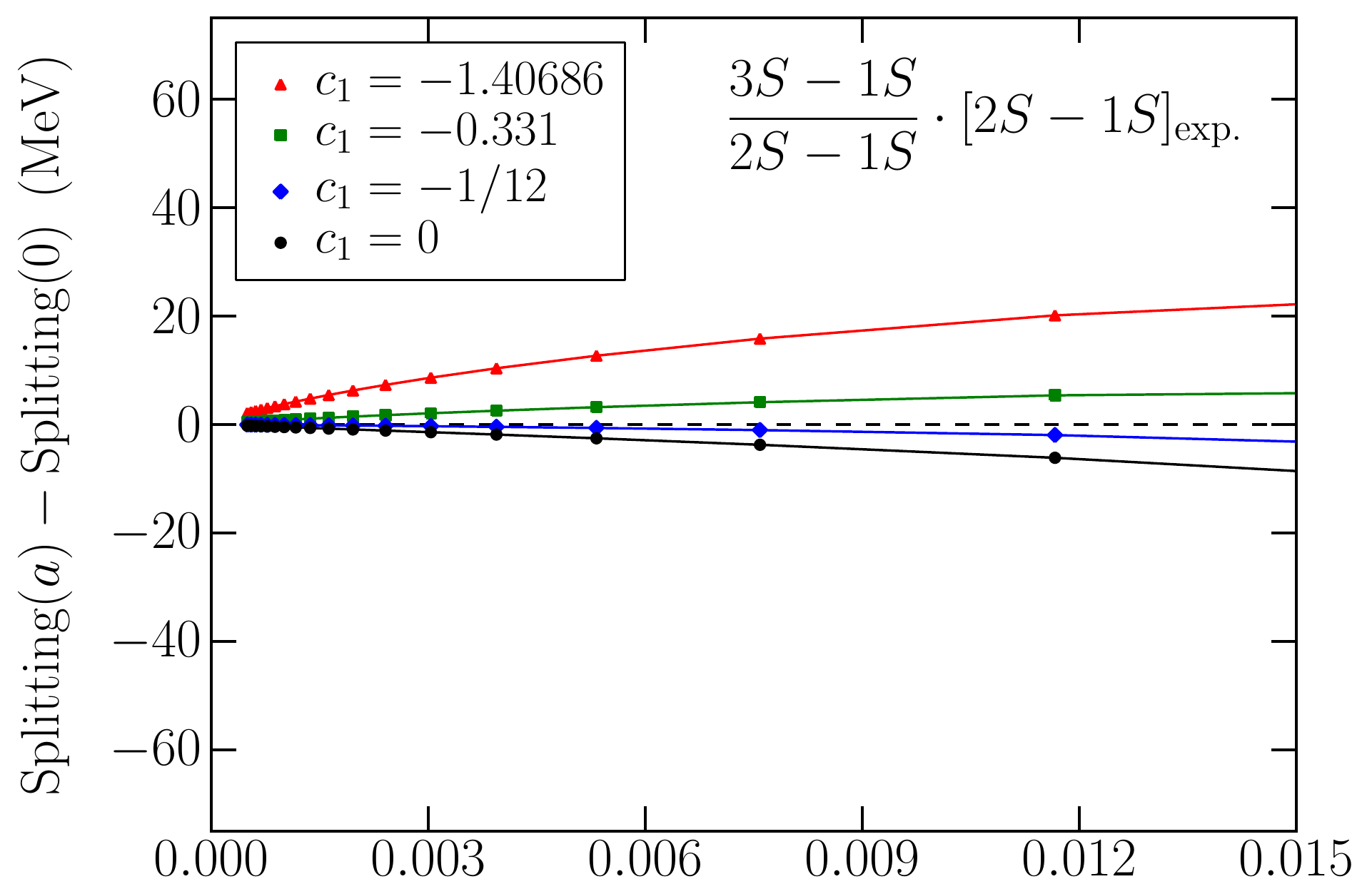}  \hfill \includegraphics[width=0.46\linewidth]{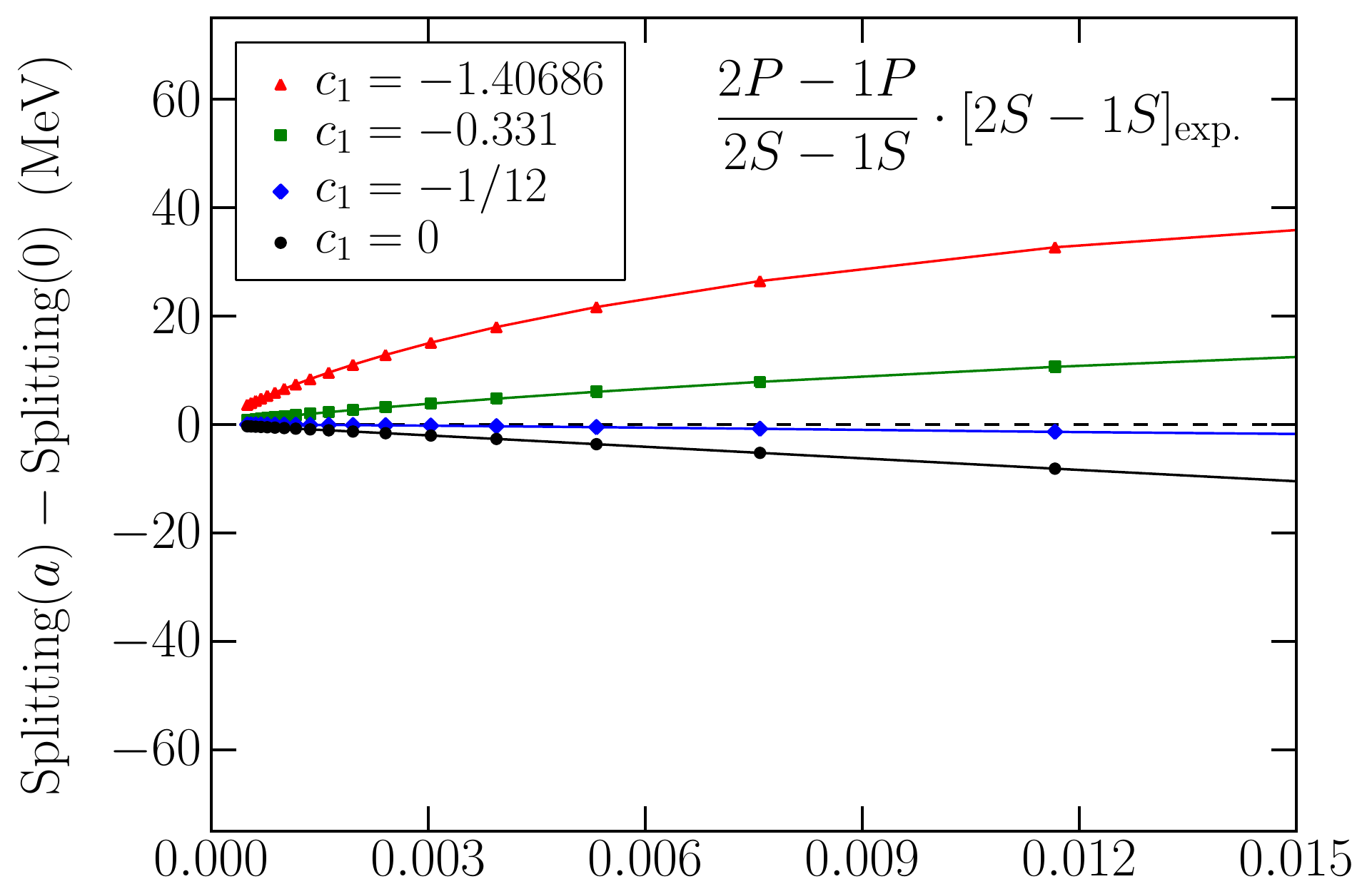}
 \includegraphics[width=0.46\linewidth]{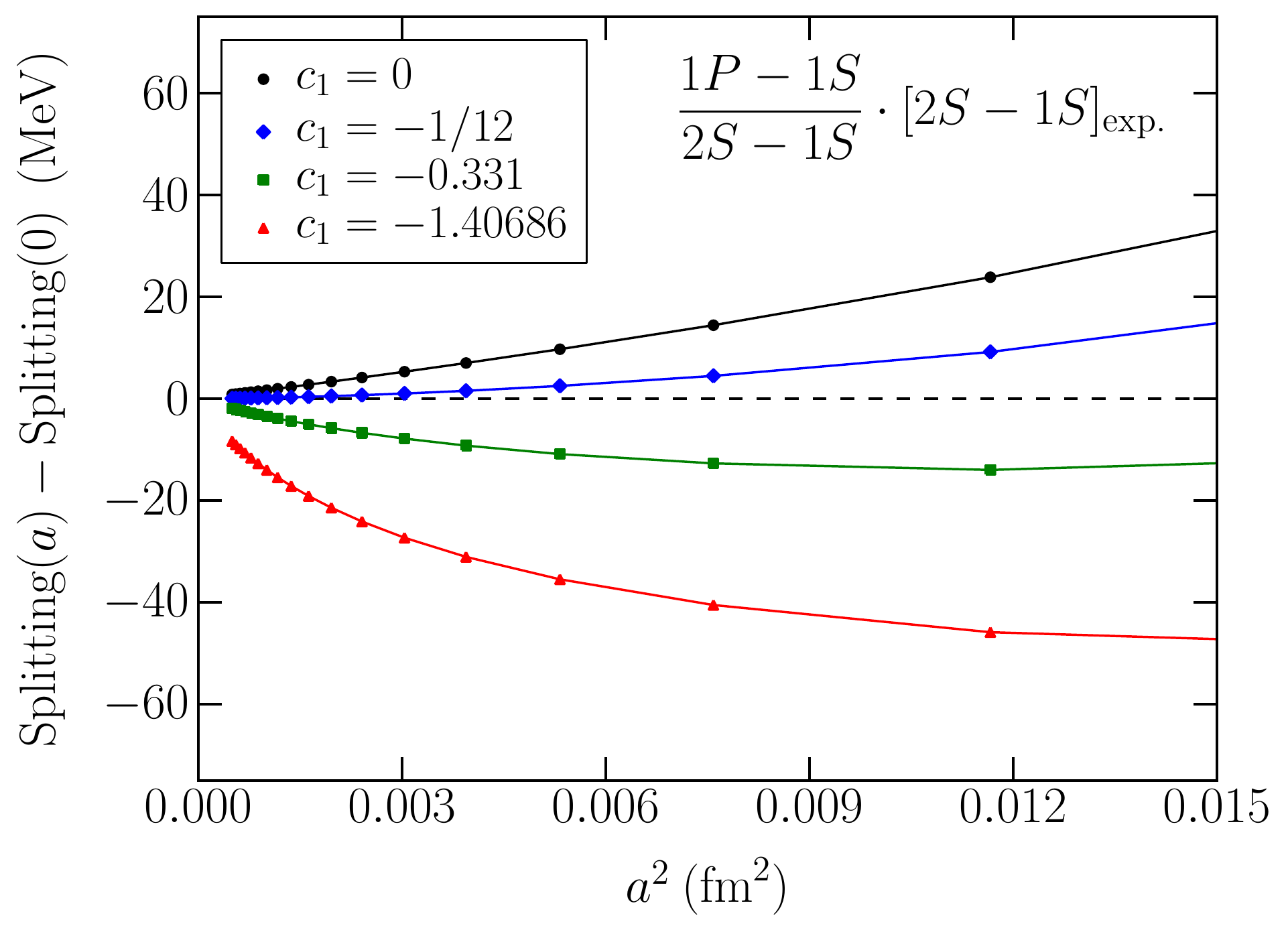}  \hfill \includegraphics[width=0.46\linewidth]{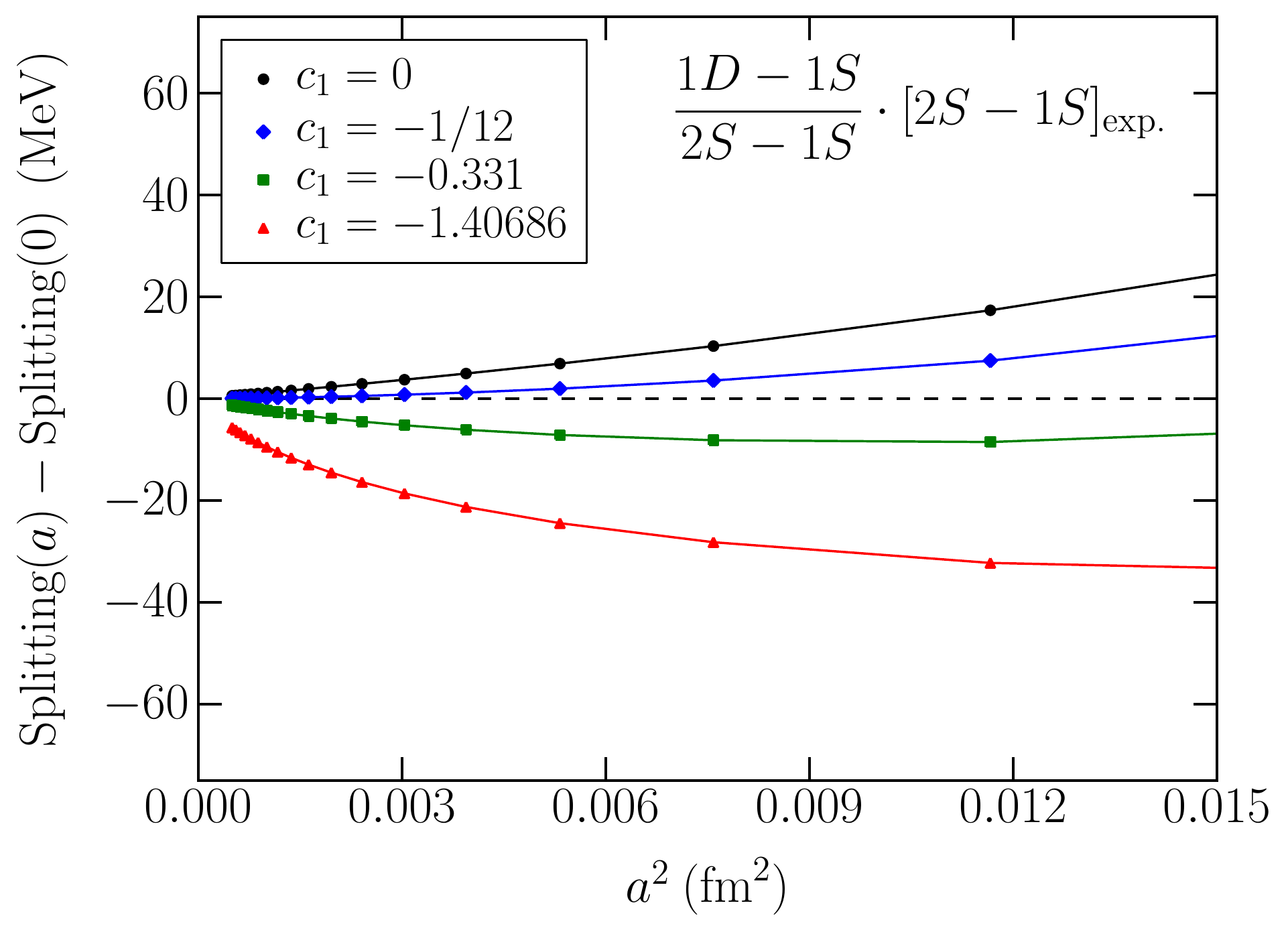}
\caption{\label{fig:gluon_errors_2S1S_scale}Shift in the lattice potential model energy splittings, rescaled using the $2S-1S$ splitting, as a function of the lattice spacing.
All data shown in this figure were generated with the $\mathcal{O}(a^4)$-improved Laplacian, so that the shifts are dominated by gluon discretization errors.}
\end{figure*}

Figure \ref{fig:gluon_errors_2S1S_scale} shows the gluonic discretization errors in the $3S-1S$, $1P-1S$, $2P-1P$, and $1D-1S$
splittings for the case that they are calculated using the $2S-1S$ splitting to set the scale. In the $2P-1P$ splitting,
previously nearly free of gluonic discretization errors, this process introduces new gluon errors. However, in the other
splittings shown in Fig.~\ref{fig:gluon_errors_2S1S_scale}, the scale setting with the $2S-1S$ splitting leads to a partial
cancellation of gluonic discretization errors. The results for the relative errors obtained with the Iwasaki
action at $a=0.11$ fm and $a=0.08$ fm are given in Table \ref{tab:gluonic_discr_errors}.

\begin{table}[h!]
\begin{tabular}{ccccc}
\hline\hline
 & \hspace{4ex} & error ($a=0.11$ fm) & \hspace{4ex} & error ($a=0.08$ fm) \\
\hline
$3S-1S$       &&  $\phantom{-}0.6$\%    &&  $\phantom{-}0.4$\%   \\
$1P-1S$       &&  $-3.4$\%   &&  $-2.8$\%   \\
$2P-1P$       &&  $\phantom{-}2.6$\%    &&  $\phantom{-}1.6$\%   \\
$1D-1S$       &&  $-1.2$\%   &&  $-1.1$\%   \\
\hline\hline
\end{tabular}
\caption{\label{tab:gluonic_discr_errors}Estimates of tree-level gluon discretization errors in radial and orbital bottomonium energy splittings
computed with the Iwasaki action and using the $2S-1S$ splitting to set the scale. A negative sign indicates a negative deviation from the continuum value.}
\end{table}

\vspace{-3ex}

\subsection{\label{sec:gluon_errors_NP}Results from MILC ensembles and gluon discretization errors in spin splittings}

In order to study the influence of the gauge action on the bottomonium energy splittings nonperturbatively, the calculations 
presented in the main part of this paper were repeated on two ensembles of lattice gauge fields generated by the MILC
collaboration \cite{Bazavov:2009bb}. These ensembles make use of the tadpole-improved one-loop L\"uscher-Weisz action \cite{Alford:1995hw}
for the gluons, which is based on order-$a^2$ Symanzik improvement rather than renormalization-group improvement.
\begin{table*}[ht!]
\begin{ruledtabular}
\begin{tabular}{ccllcccccc}
 $L^3\times T$ & $\beta$ & $a m_l$ & $a m_s$  & $a m_b$ & $u_{0L}$ & $n_{\rm conf}$ & $a$ (fm) & $m_{\pi}^{(\rm RMS)}$ (GeV) & $a m_b^{\rm (phys.)}$\\
\hline
 $24^3\times 64$  & $6.76$ & $0.005$ & $0.05$    & $2.3,\:2.64,\:2.7$    & $0.8362$ & $2099$ & $0.1198(11)\nb$    & $0.460$ &  $2.664(24)$ \\
 $28^3\times 96$  & $7.09$ & $0.0062$ & $0.031$  & $1.75,\:1.86,\:2.05$  & $0.8541$ & $1910$ & $0.08613(83)$   & $0.416$ &  $1.858(19)$ \\
\end{tabular}
\end{ruledtabular}
\caption{\label{tab:MILC_lattices}Parameters for the calculations on the MILC ensembles. The bare gauge couplings are given as $\beta=10/g^2$.
For the MILC ensembles, there are taste splittings between the different pions \protect\cite{Bazavov:2009bb},
and the root-mean-square masses taken from \protect\cite{Bazavov:2009tw,Bazavov:2009ir} are given.}
\end{table*}
The action includes the plaquette and rectangle terms, and in addition a third term (``parallelogram'').
Their coefficients $\beta_{\rm pl}$,  $\beta_{\rm rt}$, and $\beta_{\rm pg}$ were computed using one-loop perturbation theory,
but without the effects of sea quarks. These effects were later calculated and found to be significant \cite{Hao:2007iz}. Therefore, on the $(2+1)$-flavor MILC ensembles,
the gluon action is expected to have $\mathcal{O}(\alpha_s a^2)$ errors.

The parameters of the MILC ensembles used here are given in Table \ref{tab:MILC_lattices}. The values for the lattice
spacing and $a m_b^{\rm (phys.)}$ were computed using exactly the same methods as for the RBC/UKQCD ensembles, to minimize any
possible bias. The sea quarks in the MILC ensembles are implemented with the rooted staggered AsqTad action
\cite{Orginos:1998ue,Lepage:1998vj,Orginos:1999cr}. This leads to an effective averaging over multiple tastes
of sea pions \cite{Bazavov:2009bb}, and therefore the appropriate pion mass to consider for bottomonium is
the root-mean square (RMS) pion mass.
To facilitate the comparison, the results from the RBC/UKQCD ensembles were therefore interpolated/extrapolated
to match the RMS pion masses of the MILC ensembles. The lattice spacings of the coarse and fine MILC ensembles
also match the lattice spacings of the corresponding RBC/UKQCD ensembles. As discussed in Sec.~\ref{sec:radial_orbital_results},
the lattice spacing of the RBC/UKQCD ensembles changes slightly when the sea quark mass is changed, because the bare
gauge coupling is kept constant. It turns out that this shift makes the agreement of the lattice spacings even better after interpolation to match the MILC pion masses.
At the matching points, any significant difference between the results from the RBC/UKQCD and MILC ensembles
would indicate different systematics associated with the gluon and sea quark actions. In the following it is assumed
that the effect of changing the sea quark action is negligible for bottomonium.

The radial and orbital energy splittings are compared in Table \ref{tab:radial_orbital_splittings_RBC_vs_MILC_v4}. As can be seen
there, with the exception of the statistically most precise $1P-1S$ splitting, all results from the MILC ensembles agree with those from the RBC/UKQCD
ensembles within the statistical errors. At the coarse lattice spacing, the $1P-1S$ splitting from the MILC ensemble is found to be
about 9 MeV (1.4 standard deviations) higher than that from the RBC/UKQCD ensemble. At the fine lattice spacing, the $1P-1S$ splittings
from the MILC and RBC/UKQCD ensembles fully agree with each other within the statistical error of 7 MeV.
In contrast, the tree-level estimates in Fig.~\ref{fig:gluon_errors_2S1S_scale} would suggest a difference between the splittings from the
Iwasaki and tree-level L\"uscher-Weisz actions of 27 MeV at $a=0.12$ fm and 18 MeV at $a=0.09$ fm.
Note however that at these lattice spacings, the tadpole-improved one-loop L\"uscher-Weisz action has a value of $\beta_{\rm rt}/\beta_{\rm pl}$ that
is not as far away from the Iwasaki action as in the tree-level case \cite{Alford:1995hw}. Thus, nonperturbatively the errors caused by the Iwasaki action are likely
to be smaller than the tree-level estimates.

\begin{table*}[h!]
\begin{ruledtabular}
\begin{tabular}{lllll}
& \multicolumn{2}{l}{$a \approx 0.12$ fm, $m_\pi\approx460$ MeV} & \multicolumn{2}{l}{$a \approx 0.09$ fm, $m_\pi\approx416$ MeV} \\
                                                                                               & RBC &  MILC  & RBC  & MILC  \\
\hline
$\Upsilon(3S)-\Upsilon(1S)$       & $0.916(28)$  & $0.915(30)$  & $0.926(23)$  & $0.934(31)$  \\
$\overline{1^3P}-\Upsilon(1S)$    & $0.4426(41)$ & $0.4510(45)$ & $0.4349(45)$ & $0.4367(58)$  \\
$\overline{1^3P}-\overline{1S}$   & $0.4556(41)$ & $0.4645(46)$ & $0.4490(46)$ & $0.4515(60)$  \\
$\overline{2^3P}-\overline{1^3P}$ & $0.3932(92)$ & $0.385(12)$  & $0.396(13)$  & $0.386(11)$  \\
$\overline{2^3P}-\Upsilon(1S)$    & $0.836(11)$  & $0.831(12)$  & $0.829(14)$  & $0.823(15)$  \\
$\overline{2^3P}-\overline{1S}$   & $0.849(11)$  & $0.845(13)$  & $0.843(14)$  & $0.837(15)$  \\
$\Upsilon_2(1D)-\Upsilon(1S)$     & $0.7130(70)$ & $0.7143(70)$ & $0.7091(81)$ & $0.713(12)$  \\
\end{tabular}
\end{ruledtabular}
\caption{\label{tab:radial_orbital_splittings_RBC_vs_MILC_v4}Comparison of results
from the RBC/UKQCD and MILC ensembles: radial and orbital energy splittings in GeV, computed with the $v^4$ action.}
\end{table*}

The spin splittings from the RBC/UKQCD and MILC ensembles are compared in Table \ref{tab:spindep_splittings_RBC_vs_MILC_v4}
(for the order-$v^4$ NRQCD action) and Table \ref{tab:spindep_splittings_RBC_vs_MILC_v6} (for the order-$v^6$ NRQCD action).
As can be seen there, with the exception of the directly calculated $1S$ hyperfine splitting, all results from the MILC ensembles
are in agreement with the corresponding results from the RBC/UKQCD ensembles within the statistical errors. At the fine lattice spacing,
the $1S$ hyperfine splitting from the MILC ensemble is found to be about $5$\% (1.9 standard deviations)
higher than the $1S$ hyperfine splitting from the RBC/UKQCD ensemble.

\begin{table*}[h!]
\begin{ruledtabular}
\begin{tabular}{lllll}
& \multicolumn{2}{l}{$a \approx 0.12$ fm, $m_\pi\approx460$ MeV} & \multicolumn{2}{l}{$a \approx 0.09$ fm, $m_\pi\approx416$ MeV} \\
                                                                                               & RBC &  MILC  & RBC  & MILC  \\
\hline
$\Upsilon(1S)-\eta_b(1S)$                                                                      & $52.29(99)$  & $53.79(95)$  & $56.2(1.1)$  & $59.3(1.2)$  \\
$\Upsilon(2S)-\eta_b(2S)$                                                                      & $23.3(2.1)$  & $25.0(2.2)$  & $28.1(2.1)$  & $31.6(3.2)$  \\
$\chi_{b2}(1P)-\chi_{b1}(1P)$                                                                  & $20.3(1.1)$  & $21.5(1.3)$  & $20.4(1.5)$  & $21.8(1.8)$  \\
$\chi_{b1}(1P)-\chi_{b0}(1P)$                                                                  & $27.1(1.0)$  & $27.3(1.1)$  & $29.8(1.3)$  & $29.8(1.6)$  \\
$\overline{1^3P}-h_b(1P)$                                                                      & $2.94(66)$   & $2.90(70)$   & $1.93(84)$   & $2.1(1.2)$  \\
$-2\chi_{b0}(1P)-3\chi_{b1}(1P)+5\chi_{b2}(1P)$                                                & $155.3(6.4)$ & $162.1(7.5)$ & $161.6(8.8)$ & $168.7(9.6)$  \\
$-2\chi_{b0}(1P)+3\chi_{b1}(1P)-\chi_{b2}(1P)$                                                 & $33.9(2.1)$  & $33.1(2.3)$  & $39.4(2.7)$  & $38.0(3.8)$  \\
\\[-2ex]
$\displaystyle \frac{\Upsilon(2S)-\eta_b(2S)}{\Upsilon(1S)-\eta_b(1S)}$                        & $0.447(39)$  & $0.466(41)$  & $0.501(38)$  & $0.532(52)$  \\
\\[-2ex]
$\displaystyle \frac{\Upsilon(1S)-\eta_b(1S)}{-2\chi_{b0}(1P)+3\chi_{b1}(1P)-\chi_{b2}(1P)}$   & $1.517(88)$  & $1.63(11)$   & $1.416(98)$  & $1.56(15)$  \\
\\[-2ex]
$\displaystyle \frac{\Upsilon(2S)-\eta_b(2S)}{-2\chi_{b0}(1P)+3\chi_{b1}(1P)-\chi_{b2}(1P)}$   & $0.692(75)$  & $0.755(83)$  & $0.720(73)$  & $0.83(12)$  \\
\end{tabular}
\end{ruledtabular}
\caption{\label{tab:spindep_splittings_RBC_vs_MILC_v4}Comparison of results
from the RBC/UKQCD and MILC ensembles: spin splittings, computed with the $v^4$ action. All results in MeV, except for the dimensionless ratios.}
\end{table*}

\begin{table*}[h!]
\begin{ruledtabular}
\begin{tabular}{lllll}
& \multicolumn{2}{l}{$a \approx 0.12$ fm, $m_\pi\approx460$ MeV} & \multicolumn{2}{l}{$a \approx 0.09$ fm, $m_\pi\approx416$ MeV} \\
                                                                                               & RBC &  MILC  & RBC  & MILC  \\
\hline
$\Upsilon(1S)-\eta_b(1S)$                                                                      & $44.39(84)$  & $45.64(81)$  & $46.49(91)$  & $48.64(96)$  \\
$\Upsilon(2S)-\eta_b(2S)$                                                                      & $17.3(1.8)$  & $18.8(1.9)$  & $20.9(1.9)$  & $22.3(2.4)$  \\
$\chi_{b2}(1P)-\chi_{b1}(1P)$                                                                  & $16.55(99)$  & $17.7(1.2)$  & $16.5(1.3)$  & $17.5(1.7)$  \\
$\chi_{b1}(1P)-\chi_{b0}(1P)$                                                                  & $24.06(91)$  & $24.3(1.0)$  & $25.9(1.1)$  & $26.5(1.5)$  \\
$\overline{1^3P}-h_b(1P)$                                                                      & $2.18(60)$   & $2.13(62)$   & $1.27(74)$   & $1.21(96)$  \\
$-2\chi_{b0}(1P)-3\chi_{b1}(1P)+5\chi_{b2}(1P)$                                                & $130.6(5.9)$ & $137.0(6.9)$ & $134.3(7.8)$ & $140.9(8.8)$  \\
$-2\chi_{b0}(1P)+3\chi_{b1}(1P)-\chi_{b2}(1P)$                                                 & $31.5(1.9)$  & $30.9(2.2)$  & $35.6(2.4)$  & $35.7(3.7)$  \\
\\[-2ex]
$\displaystyle \frac{\Upsilon(2S)-\eta_b(2S)}{\Upsilon(1S)-\eta_b(1S)}$                        & $0.392(40)$  & $0.413(42)$  & $0.450(41)$  & $0.458(47)$  \\
\\[-2ex]
$\displaystyle \frac{\Upsilon(1S)-\eta_b(1S)}{-2\chi_{b0}(1P)+3\chi_{b1}(1P)-\chi_{b2}(1P)}$   & $1.396(81)$  & $1.48(10)$   & $1.304(88)$  & $1.36(14)$  \\
\\[-2ex]
$\displaystyle \frac{\Upsilon(2S)-\eta_b(2S)}{-2\chi_{b0}(1P)+3\chi_{b1}(1P)-\chi_{b2}(1P)}$   & $0.552(66)$  & $0.607(75)$  & $0.593(67)$  & $0.626(90)$  \\
\end{tabular}
\end{ruledtabular}
\caption{\label{tab:spindep_splittings_RBC_vs_MILC_v6}Comparison of results
from the RBC/UKQCD and MILC ensembles: spin splittings, computed with the $v^6$ action. All results in MeV, except for the dimensionless ratios.}
\end{table*}

Recall from Sec.~\ref{sec:gluon_errors_tree_results} that a negative coefficient $c_1$ in the gluon action
leads to a broadening of the lattice $1S$ wave function. In the continuum, the leading-order hyperfine splitting is proportional to $|\psi(0)|^2$, and one might therefore expect naively
that a negative coefficient $c_1$ reduces the hyperfine splitting \cite{DeGrand:2002vu}. If this picture was correct, for example
the simple plaquette action ($c_1=0$) would give a significantly larger hyperfine splitting than the Iwasaki action ($c_1=-0.331$).
In Ref.~\cite{Manke:2000dg}, the authors compared their results for the $1S$ hyperfine splitting, computed using the
Iwasaki action, to the results from \cite{Eicker:1998vx} that used the same NRQCD action and the same number of sea quark flavors,
but the plaquette gluon action. The lattice spacing was $a\approx0.10$ fm in both cases. At $a=0.10$ fm, the lattice potential model from
Sec.~\ref{sec:gluon_errors_tree} gives $|\psi^{(c_1=0)}(0)|^2/|\psi^{(c_1=-0.331)}(0)|^2\approx 1.5$.
In contrast, the results for the $1S$ hyperfine splitting from the two groups were in agreement within the statistical error of about 5\%.
Clearly, the simple continuum picture for the hyperfine splitting does not apply here.
On the lattice, the spin-dependent potential responsible for the $S$-wave hyperfine splitting will have non-zero values also at $\bs{r}\neq 0$. It is expected to be
a complicated function that depends both on the lattice gluon propagator and on the discretization
of the chromomagnetic field strength in the NRQCD action.

Given the comparison of results for the $1S$ hyperfine splitting from the Iwasaki action with results from two other gluon actions as discussed above,
it seems reasonable to assume a gluonic discretization error of 5\% for the hyperfine splittings
calculated with the Iwasaki action at $a\approx0.08$ fm. The same error estimate is used for the ratios of hyperfine and tensor splittings.
For the ratio of the $2S$ and $1S$ hyperfine splittings, a partial cancellation of gluonic discretization errors is expected (as in the $2S-1S$ splitting in
Fig.~\ref{fig:gluon_errors}), and therefore a $2.5$\% gluon error is estimated for this ratio at $a\approx0.08$ fm.

\begin{figure*}[h!]
 \includegraphics[width=0.415\linewidth]{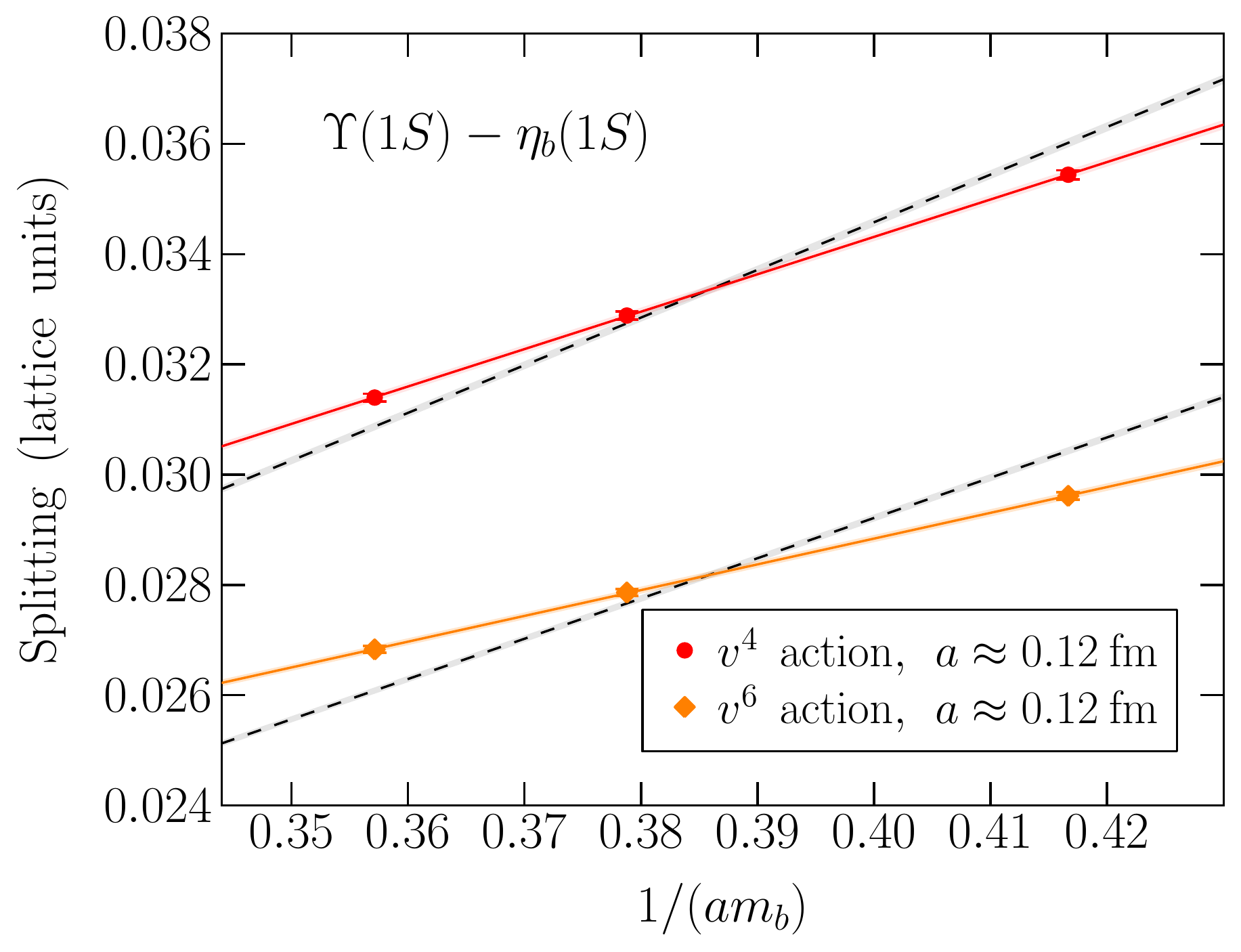}  \hfill \includegraphics[width=0.415\linewidth]{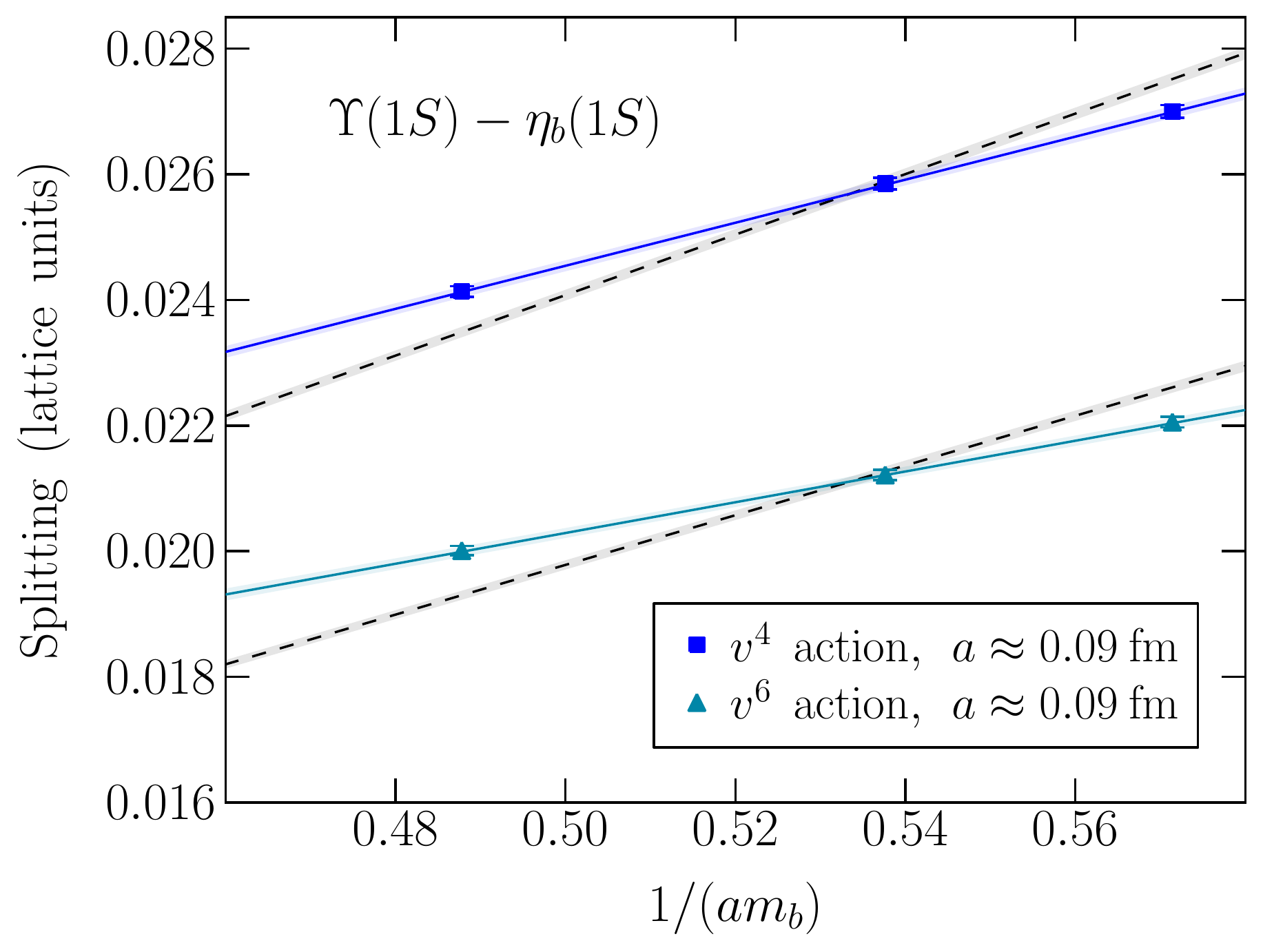}
\caption{\label{fig:spin_dep_mb_dep_MILC} Heavy-quark mass dependence of the $1S$ hyperfine splittings on the MILC ensembles (see Fig.~\ref{fig:spin_dep_mb_dep} for the data from the RBC/UKQCD ensembles).}
\end{figure*}

Finally, recall from Fig.~\ref{fig:spin_dep_mb_dep} that the $1S$ hyperfine splitting computed on the RBC/UKQCD ensembles shows
an $a m_b$-dependence that is slightly different from the simple proportionality to $1/(am_b)$ seen in the other spin splittings. In
Fig.~\ref{fig:spin_dep_mb_dep_MILC}, the $1S$ hyperfine splitting in lattice units computed on the coarse and fine MILC ensembles is plotted
as a function of $1/(a m_b)$. As can be seen there, the behavior is very similar to that found on the RBC/UKQCD ensembles.

\end{document}